\documentclass{aa}  
\usepackage[]{aalongtable}
\usepackage{scalefnt}
\usepackage{graphicx}
\usepackage{txfonts}
\def\kms   {km~s$^{-1}$}
\def\loggf {log~$gf$}
\def\Teff  {$T_\mathrm{eff}$}
\def\logg  {log~$g$}
\begin{document}

   \title{High-resolution abundance analysis of very metal-poor r-I stars
\thanks{Observations obtained with the VLT, at the European Southern 
Observatory, Paranal, Chile, under
proposal 080.D-0194(A) (PI:V. Hill)}}


\author{
C.~Siqueira Mello\inst{1}
\and
V.~Hill\inst{2}
\and
B.~Barbuy\inst{1}
\and
M.~Spite\inst{3}
\and
F.~Spite\inst{3}
\and
T.C.~Beers\inst{4}
\and
E.~Caffau\inst{3}
\and
P.~Bonifacio\inst{3}
\and
R.~Cayrel\inst{3}
\and
P.~Fran\c cois\inst{5}
\and 
H.~Schatz\inst{6}
\and
S.~Wanajo\inst{7,8}
}
\offprints{C. Siqueira Mello Jr. (cesar.mello@usp.br).}

\institute{
IAG, Universidade de S\~ao Paulo, Rua do Mat\~ao 1226,
Cidade Universit\'aria, S\~ao Paulo 05508-900, Brazil
\and
Universit\'e de Sophia-Antipolis,
 Observatoire de la C\^ote d'Azur, CNRS UMR 6202, BP4229, 06304 Nice Cedex 4, France
\and
GEPI, Observatoire de Paris,  CNRS, UMR 8111, F-92195 Meudon Cedex, France
\and
National Optical Astronomy Observatory, Tucson, Arizona 85719, USA 
and JINA: Joint Institute for Nuclear Physics; email: beers@noao.edu
\and
GEPI, Observatoire de Paris, CNRS, UMR 8111, 61 Av. de l'Observatoire, 75014 Paris, France
\and
{National Superconducting Cyclotron Laboratory, Department of Physics and Astronomy 
and Joint Institute for Nuclear Astrophysics, Michigan State University, East Lansing, 
MI 48824, USA; email: schatz@nscl.msu.edu}
\and
National Astronomical Observatory of Japan, 2-21-1 Osawa, Mitaka, Tokyo 181-8588, 
Japan; email: shinya.wanajo@nao.ac.jp
\and
RIKEN, iTHES Research Group, 2-1 Hirosawa, Wako, Saitama, 351-0198, Japan
}
\date{Received 18 March 2014; accepted 28 March 2014}

\titlerunning{}
\abstract
{Moderately r-process-enriched stars (r-I; $+0.3 \le \hbox{[Eu/Fe]} \le +1.0$) are at least four times as common as those that are greatly enriched 
in r-process elements (r-II; $\hbox{[Eu/Fe]} > +1.0$), and the abundances in their atmospheres are important tools for obtaining a better 
understanding of the nucleosynthesis processes responsible for the origin of the elements beyond the iron peak.}
{The main aim of this work is to derive abundances for a sample of seven metal-poor stars with $-3.4$~$\leq\hbox{[Fe/H]}\leq$~$-2.4$ 
classified as r-I stars, to understand the role of these stars for constraining the
astrophysical nucleosynthesis event(s) that is(are) responsible for the production of the r-process, and to investigate whether they differ,
in any significant way, from the r-II stars.}
{We carried out a detailed abundance analysis based on high-resolution spectra obtained with the VLT/UVES spectrograph, 
using spectra in the wavelength ranges 3400~-~4500~{\rm \AA}, 6800~-~8200~{\rm \AA}, and 8700~-~10,000~{\rm \AA}, 
with resolving power R $\sim$~40\,000 (blue arm) and R $\sim$~55\,000 (red arm). 
The OSMARCS LTE 1D model atmosphere grid was employed, along with the 
spectrum synthesis code Turbospectrum.}
{We have derived abundances of the light elements Li, C, and N, the $\alpha$-elements Mg, Si, S, Ca, and Ti, the odd-Z 
elements Al, K, and Sc, the iron-peak elements V, Cr, Mn, Fe, Co, and Ni, and the trans-iron elements from the first peak 
(Sr, Y, Zr, Mo, Ru, and Pd), the second peak (Ba, La, Ce, Pr, Nd, Sm, Eu, Gd, Tb, Dy, Ho, Er, Tm, and Yb), 
the third peak (Os and Ir, as upper limits), and the actinides (Th) regions. The results are compared with values for these elements for r-II and
``normal'' very and extremely metal-poor stars reported in the literature, ages based on 
radioactive chronometry are explored using different models, and a number of conclusions about the r-process and the 
r-I stars are presented. Hydrodynamical models were used for some elements, and general behaviors for the 3D 
corrections were presented. Although the abundance ratios of the second r-process peak elements (usually 
associated with the main r-process) are nearly identical for r-I and
r-II stars, the first r-process peak abundance
ratios (probably associated with the weak r-process) are more enhanced in r-I stars than in r-II stars, suggesting that differing
nucleosynthesis pathways were followed by stars belonging to these two different classifications.}
{}

   \maketitle
%

\section{Introduction}

The heavy elements (those beyond the iron-peak elements) are primarily formed by the capture of
neutrons. This process takes place through the slow (s) or the rapid
(r) process, where the addition of a neutron is considered slow or
rapid relative to the time for $\beta$ decay to take place (which
differs from isotope to isotope). The main s-process is thought to occur
on a relatively long timescale in asymptotic giant branch (AGB) stars
(see, e.g., Herwig 2005, Sneden et al. 2008). The r-process instead
occurs on a rather short timescale, typical of supernovae explosions
(e.g., Winteler et al. 2012; Wanajo 2013) or other brief events, such as
neutron star mergers (e.g., Goriely et al. 2011; Korobkin et al. 2012;
Rosswog et al. 2014; Wanajo et al. 2014). In the early evolution of the
Galaxy, matter may have been enriched in heavy elements through the
r-process alone, as first suggested by Truran (1981) (see also Roederer
et al. 2010, 2014, and references therein). To date, full understanding
of r-process nucleosynthesis and the astrophysical sites that account
for its operation remain unclear (e.g., Wanajo \& Ishimaru 2006; Kratz
et al. 2007; Langanke \& Thielemann 2013). The best sources of
information to constrain this process are the abundances of very and
extremely metal-poor stars, since they contain the nucleosythesis
products from early generations of stars and are essentially unaltered
by later production events.

Beers \& Christlieb (2005) classified metal-poor stars in terms of their
metallicities and the enhancements of carbon and the r- and s-process
elements. Metal-poor stars that are enhanced in the neutron-capture
elements provide a unique opportunity to study the r-process in the early Galaxy {\it in situ}, 
even for extremely metal-poor stars
($\hbox{[Fe/H]} < -3.0$). In the absence of enhancement of the
neutron-capture elements, most species would be too weak to be
detected in high-resolution spectra. Two classes of metal-poor stars
were defined by Beers \& Christlieb (2005) according to their
enhancement in r-process elements -- the moderately r-process-enhanced stars, with
$+0.3\leq\hbox{[Eu/Fe]}$\footnote{We adopt the notation
[A/B]~=~log(n$_{A}$/n$_{B}$)$_{star}$~-~log(n$_{A}$/n$_{B}$)$_{\odot}$,
where {\it n} is the number density of atoms.}~$\leq+1.0$ and
$\hbox{[Ba/Eu]}<0$, designated as r-I stars, and the highly
r-process-enhanced stars, with $\hbox{[Eu/Fe]}>+1.0$ and
$\hbox{[Ba/Eu]}<0$, designated as r-II stars. According to Beers \&
Christlieb (2005), the r-I stars appear to be, on the whole, at least 
several times as common as their more extreme counterparts, the
r-II stars.

In the context of the Hamburg/ESO R-process Enhanced Star survey (HERES)
(Christlieb et al. 2004; Barklem et al. 2005, hereafter B05), 253
metal-poor halo stars in the metallicity range $-3.8<\hbox{[Fe/H]}<-1.5$
were studied. The spectra were obtained with the VLT/UVES spectrograph,
using a slit width of 2'', yielding a resolving power R~$\sim$~20\,000,
and a typical signal-to-noise ratio S/N~$\sim$~50 per pixel, covering the wavelength region
3760~-~4980\,{\rm \AA}. Based on these observations, B05 identified 8
r-II and 35 r-I stars, showing that the r-I stars are, in fact, more
than four times as common as the r-II stars. 

Barklem et al. (2005) (and others before) suggested that the pattern of the
neutron-capture elements of r-II stars, including the reference stars
CS~22892-052 (McWilliam et al. 1995; Sneden et al. 1996, 2000, 2009) and
CS~31082-001 (Cayrel et al. 2001; Hill et al. 2002), closely followed
the scaled solar system r-process abundances for elements beyond barium
(except for Th and U). In stars of lower r-element enrichment, such as
the r-I stars, B05 noted that the lighter element abundances (in
particular Sr, Y, Zr) are generally higher than expected from the solar
pattern. Moreover, r-II stars were found to occupy a narrow metallicity
range, centered on $\hbox{[Fe/H]}\sim-2.8$, with a small scatter ($\sim$
0.16 dex). The r-I stars, on the other hand, were found to occur across
a wide metallicity range. From the data of B05, the abundance ratios
from [C/Fe] to [Zn/Fe] appeared to be the same in the r-II, r-I stars,
and normal metal-poor stars (those without CNO or neutron-capture
enhancements). These results have been strengthened, with only a few
exceptions, by high-resolution spectroscopic observations of additional
metal-poor stars in recent years. 

Based on the analysis of a sample of metal-poor stars with
$-3.4$~$\leq\hbox{[Fe/H]}\leq$~$-2.4$ classified as r-I stars, the
purpose of the present paper is to perform a detailed comparison of the
abundance ratios of elements for the r-I and r-II stars, and thereby
gain a better understanding of the nature of the r-process and the
likely astrophysical site(s) with which it might be associated. In Sect.
2, the observations and reductions are reported. Sect. 3 describes
the abundance determinations. In Sect. 4, the results are provided,
followed by a discussion in Sect. 5. Conclusions are given in Sect. 6.

\section {Observations and reductions}

Seven r-I stars were selected from the list of B05, with
$\hbox{[Fe/H]}\leq-2.3$, since, at higher metallicity, enrichment of the
interstellar medium from which these stars were born with s-process
elements contributed by AGB stars becomes more of a problem (see, e.g.,
Fran\c{c}ois et al. 2007). We also discarded the carbon-enhanced
metal-poor (CEMP) stars, since many of these stars have been enriched in
C and heavy s-process elements, because of mass-transfer from an AGB companion. 
The atmospheric parameters of the selected stars (from
B05) are given in Table~\ref{atm_barklem}, along with their reported 
[Eu/Fe], [Ba/Eu], and [C/Fe], the abundance ratios upon which their original 
classification was based.

The seven r-I stars were observed with the VLT/UVES spectrograph in November
6~-~10, 2007. The blue and red arms were centered on 3900\,{\rm \AA} and
8600\,{\rm \AA}, and the spectra were obtained with a 1'' slit, a
resolving power of R $\sim$~40\, 000 and R $\sim$~55\,000, respectively,
with about five pixels per resolution element. Spectra in the wavelength
ranges 3400~-~4500~{\rm \AA}, 6800~-~8200~{\rm \AA}, and
8700~-~10,000~{\rm \AA} were obtained. The log of observations is given
in Table \ref{log}. Standard data reductions were performed, employing
the UVES pipeline in the ESO~-~GASGANO environment, including flatfield
correction, bias and dark subtraction, cosmic-ray removal, spectral
extraction, and wavelength calibration with comparison arc-line spectra
taken before or after each exposure.

In the frame of the Large Program (LP) ``First Stars'' (PI: R. Cayrel),
some 35 very metal-poor giants were analyzed based on very high-quality
high-resolution spectra. The abundance patterns from C to Zn are
presented in Cayrel et al. (2004), and also (for very metal-poor
main-sequence turnoff stars) in Bonifacio et al. (2009). Fran\c{c}ois et
al. (2007) derived the abundance patterns of the neutron-capture elements
for many of these stars. Among the giants, six r-I stars and three r-II stars
(CS~31082-001, CS~22953-003, and CS~22892-052) were studied in detail and 
were added to the present sample of r-I stars for the purpose of
discussion. The relative abundances of the heavy elements for
CS~31082-001 come from Hill et al. (2002), Barbuy et al. (2011), and
Siqueira-Mello et al. (2013), for CS~22953-003 they were taken from Fran\c{c}ois
et al. (2007), and for CS~22892-052 from Sneden et al. (1996,
2000, 2009). 

For the star CS~30315-029, we also used a spectrum obtained during the 
LP ``First Stars,'' in the regions centered on 3960\,{\rm \AA} and
5730\, {\rm \AA}, to measure the Ba lines. For the other six stars, 
we derived Ba abundances from the HERES spectra.

\begin{table*}
\caption{Atmospheric parameters, radial velocities, [Eu/Fe], [Ba/Eu], and [C/Fe] taken from Barklem et al. (2005). 
Our set of stars were chosen with {\bf $\rm+0.3 \le [Eu/Fe] \le
+1.0$} and $\rm[Ba/Eu]<0$.}             
\label{atm_barklem}      
\centering                          
\begin{tabular}{cccccrrrr}        
\hline\hline                 
\noalign{\smallskip}
\hbox{Star} & \hbox{\Teff} & \hbox{log$g$} & \hbox{[Fe/H]} & \hbox{$\xi$} & \hbox{V$_{\rm r}$} & \hbox{[Eu/Fe]} & \hbox{[Ba/Eu]} & \hbox{[C/Fe]}\\ 
\noalign{\smallskip}
\hline                        
\noalign{\smallskip}
& \hbox{(K)} & \hbox{[cgs]} & \hbox{} & \hbox{(\kms)} & \hbox{(\kms)} & \hbox{} & \hbox{} & \hbox{} \\
\noalign{\smallskip}
\hline                        
\noalign{\smallskip}
\hbox{CS 30315-029} & 4541 & 1.07 & $-$3.33 & 2.06 & $-$169.2 & $+$0.73 & $-$0.31 & $-$0.52 \\
\hbox{HE 0057-4541} & 5083 & 2.55 & $-$2.32 & 1.67 &     13.4 & $+$0.62 & $-$0.74 & $+$0.09 \\
\hbox{HE 0105-6141} & 5218 & 2.83 & $-$2.55 & 1.66 &      3.8 & $+$0.68 & $-$0.50 & $+$0.11 \\
\hbox{HE 0240-0807} & 4729 & 1.54 & $-$2.68 & 1.96 &  $-$98.8 & $+$0.73 & $-$0.53 & $-$0.43 \\
\hbox{HE 0516-3820} & 5342 & 3.05 & $-$2.33 & 1.48 &    153.5 & $+$0.67 & $-$0.77 & $+$0.30 \\
\hbox{HE 0524-2055} & 4739 & 1.57 & $-$2.58 & 1.95 &    255.3 & $+$0.49 & $-$0.42 & $-$0.33 \\
\hbox{HE 2229-4153} & 5138 & 2.47 & $-$2.62 & 1.79 & $-$139.6 & $+$0.45 & $-$0.73 & $+$0.28 \\
\noalign{\smallskip}
\hline                                   
\end{tabular}
\end{table*}

\subsection{Radial velocities}

For each spectrum, Table \ref{log} lists the measured geocentric ($\rm
RV_G$) and barycentric ($\rm RV_B$) radial velocities, along with the
mean $\rm RV_B$ for each star. These measurements were made with the blue spectra; 
a precision of about 1\,\kms was achieved. 

Comparison to the values given by B05 shows good agreement, within the
error bars. Consequently, our sample stars do not present any indication
of binarity.

\subsection{Measurement of equivalent widths}

The spectra were normalized, corrected for radial-velocity shifts, and
combined to produce the final average data. We measured the equivalent 
widths (EW) of a set of weak iron and titanium lines in their neutral
and ionized states. The EWs were measured with a semi-automatic code,
which traces the continuum and uses a Gaussian profile to fit the
absorption lines. The central wavelength of each line was left as a free
parameter, and the full width at half maximum (FWHM) was computed as well. 
For blends on the blue or the red wing, this part of the line
can be excluded from the computations. The EWs of the Fe and Ti lines
used in this work are given in Table \ref{EW_measurements} along with
the corresponding \loggf~values. 

\begin{figure}
\centering
\resizebox{80mm}{!}{\includegraphics[angle=0]{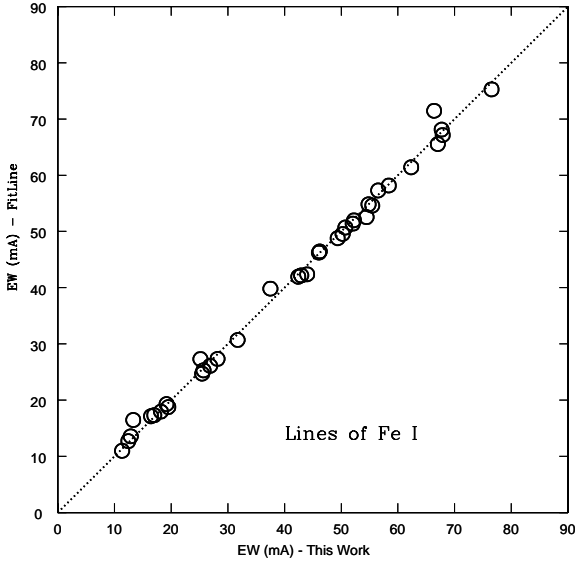}}
\caption{Comparison of EWs measured for a set of Fe~I lines in
CS~30315-029 using the code developed in this work and EWs obtained 
with the code FitLine.}
\label{EW_compara}
\end{figure}

To check the reliability of the implemented code, the results
were compared with those obtained using Fitline (Fran\c{c}ois et al.
2003), an automated line-fitting procedure based on the algorithms of
Charbonneau (1995). Fig. \ref{EW_compara} shows a comparison of the EWs
obtained with the present method and with Fitline for CS~30315-029, and
demonstrates excellent agreement.
 
\begin{table*}
\caption{Log of observations: coordinates, date of observation, exposure time, air mass at the beginning and at the 
end of the observation, geocentric and barycentric radial velocities.}             
\label{log}      
\centering                          
\begin{tabular}{cccccccrrr}        
\hline\hline                 
\noalign{\smallskip}
\hbox{Target} & \hbox{$\alpha$(J2000)} & \hbox{$\delta$(J2000)} &
\hbox{Date} &\hbox{Exp.} &\hbox{Airmass} &\hbox{Airmass} & \hbox{RV$\rm_{G}$} & 
\hbox{RV$\rm_{B}$}& \hbox{Mean $\rm RV_B$} \\
\noalign{\smallskip}
\hline
\noalign{\smallskip}
\hbox{} & \hbox{} & \hbox{} & & \hbox{(sec)} & \hbox{Start} & \hbox{End} & \hbox{(\kms)} & \hbox{(\kms)} & \hbox{(\kms)} \\
\noalign{\smallskip}
\hline
\noalign{\smallskip}
\hbox{CS 30315-029} & 23:34:26.5 & $-$26:42:19 & 06.11.07 & 3600 & 1.063 & 1.006 & $-$144.95 & $-$169.43 & $-$170.0\\
\hbox{           } &            &              & 08.11.07 & 3600 & 1.063 & 1.123 & $-$145.39 & $-$170.36 &         \\
\hbox{           } &            &              & 08.11.07 & 3600 & 1.127 & 1.322 & $-$145.23 & $-$170.31 &         \\
\hbox{HE 0057-4541} & 00:59:59.2 & $-$45:24:54 & 07.11.07 & 3600 & 1.143 & 1.080 &     30.93 &     14.15 &     13.7\\
\hbox{           } &            &              & 07.11.07 & 3600 & 1.072 & 1.081 &     30.79 &     13.90 &         \\
\hbox{           } &            &              & 07.11.07 & 3600 & 1.081 & 1.145 &     30.74 &     13.76 &         \\
\hbox{           } &            &              & 07.11.07 & 3600 & 1.146 & 1.281 &     30.89 &     13.82 &         \\
\hbox{           } &            &              & 08.11.07 & 3600 & 1.156 & 1.299 &     30.62 &     13.34 &         \\
\hbox{           } &            &              & 08.11.07 & 3600 & 1.303 & 1.564 &     30.67 &     13.31 &         \\
\hbox{HE 0105-6141} & 01:07:38.0 & $-$61:25:17 & 09.11.07 & 4500 & 1.425 & 1.683 &     20.09 &      5.33 &     5.3 \\
\hbox{HE 0240-0807} & 02:42:57.6 & $-$07:54:35 & 07.11.07 & 3600 & 1.095 & 1.233 &  $-$96.02 & $-$100.77 & $-$101.8\\
\hbox{           } &            &              & 08.11.07 & 4500 & 1.232 & 1.631 &  $-$95.96 & $-$101.31 &         \\
\hbox{           } &            &              & 10.11.07 & 3600 & 1.641 & 1.290 &  $-$96.44 & $-$102.06 &         \\
\hbox{           } &            &              & 10.11.07 & 3600 & 1.287 & 1.120 &  $-$96.42 & $-$102.14 &         \\
\hbox{           } &            &              & 10.11.07 & 3600 & 1.117 & 1.050 &  $-$96.17 & $-$102.01 &         \\
\hbox{           } &            &              & 10.11.07 & 3600 & 1.045 & 1.069 &  $-$96.14 & $-$102.15 &         \\
\hbox{           } &            &              & 10.11.07 & 3600 & 1.070 & 1.175 &  $-$95.89 & $-$102.03 &         \\
\hbox{           } &            &              & 10.11.07 & 3600 & 1.178 & 1.408 &  $-$95.78 & $-$102.04 &         \\
\hbox{HE 0516-3820} & 05:18:12.9 & $-$38:17:33 & 09.11.07 & 3600 & 1.056 & 1.144 &    148.20 &    154.38 &   154.4 \\
\hbox{HE 0524-2055} & 05:27:04.4 & $-$20:52:42 & 07.11.07 & 3600 & 1.007 & 1.013 &    244.19 &    256.32 &   255.4 \\
\hbox{           } &            &              & 07.11.07 & 3600 & 1.014 & 1.087 &    243.53 &    255.55 &         \\
\hbox{           } &            &              & 08.11.07 & 3300 & 1.030 & 1.116 &    243.65 &    255.31 &         \\
\hbox{           } &            &              & 10.11.07 & 3600 & 1.007 & 1.064 &    243.32 &    254.42 &         \\
\hbox{HE 2229-4153} & 22:32:49.0 & $-$41:38:25 & 08.11.07 & 2700 & 1.076 & 1.137 & $-$113.97 & $-$139.92 & $-$138.5\\
\hbox{           } &            &              & 09.11.07 & 2700 & 1.046 & 1.066 & $-$111.74 & $-$137.66 &         \\
\hbox{           } &            &              & 09.11.07 & 2700 & 1.046 & 1.055 & $-$111.94 & $-$137.90 &         \\
\noalign{\smallskip}
\hline                                   
\end{tabular}
\end{table*}

\begin{table*}
\caption{Identifications, magnitudes, and reddening.}             
\label{flux}      
\scalefont{0.78}
\centering                          
\begin{tabular}{cccccccccc}        
\hline\hline                 
\noalign{\smallskip}
\hbox{Star} & \hbox{2MASS ID} & \hbox{($V$)*} & \hbox{($B-V$)*} & \hbox{($V-R_{C}$)*} & 
\hbox{($V-I_{C}$)*} & \hbox{($J$)**} & \hbox{($H$)**} & \hbox{($K_{S}$)**} & \hbox{E($B-V$)***}\\
\noalign{\smallskip}
\hline
\noalign{\smallskip}
\hbox{CS 30315-029} & \hbox{23342669-2642140} & 13.661$\pm$0.004 & 0.915$\pm$0.007 & 0.569$\pm$0.004 & 1.143$\pm$0.004 & 11.780$\pm$0.020 & 11.209$\pm$0.021 & 11.124$\pm$0.023 & 0.020\\
\hbox{HE 0057-4541} & \hbox{00595927-4524534} & 14.829$\pm$0.005 & 0.699$\pm$0.009 & 0.441$\pm$0.007 & 0.890$\pm$0.007 & 13.376$\pm$0.021 & 12.970$\pm$0.026 & 12.877$\pm$0.031 & 0.016\\
\hbox{HE 0105-6141} & \hbox{01073782-6125176} & 13.516$\pm$0.004 & 0.645$\pm$0.006 & 0.403$\pm$0.005 & 0.856$\pm$0.006 & 12.161$\pm$0.023 & 11.758$\pm$0.022 & 11.663$\pm$0.025 & 0.020\\
\hbox{HE 0240-0807} & \hbox{02425772-0754354} & 14.971$\pm$0.005 & 0.896$\pm$0.012 & 0.524$\pm$0.007 & 1.082$\pm$0.008 & 13.213$\pm$0.022 & 12.707$\pm$0.032 & 12.625$\pm$0.031 & 0.025\\
\hbox{HE 0516-3820} & \hbox{05181291-3817326} & 14.377$\pm$0.007 & 0.615$\pm$0.013 & 0.401$\pm$0.009 & 0.839$\pm$0.009 & 12.937$\pm$0.023 & 12.507$\pm$0.029 & 12.469$\pm$0.027 & 0.026\\
\hbox{HE 0524-2055} & \hbox{05270444-2052420} & 14.013$\pm$0.004 & 0.878$\pm$0.007 & 0.526$\pm$0.005 & 1.076$\pm$0.005 & 12.256$\pm$0.026 & 11.747$\pm$0.026 & 11.623$\pm$0.019 & 0.045\\
\hbox{HE 2229-4153} & \hbox{22324904-4138252} & 13.322$\pm$0.003 & 0.676$\pm$0.005 & 0.420$\pm$0.004 & 0.875$\pm$0.006 & 11.937$\pm$0.025 & 11.497$\pm$0.021 & 11.456$\pm$0.022 & 0.012\\
\noalign{\smallskip}
\hline
\end{tabular}
\tablebib{*: Broadband UBVR$_{C}$I$_{C}$ (subscript ``C'' indicates the Cousins system) 
from HK and Hamburg/ESO surveys (Beers et al. 2007); 
**: 2MASS (Cutri et al. 2003); 
***: Infrared Processing and Analysis Center (IRSA, Schlegel et al. 1998).}
\end{table*}

\begin{table*}
\caption{Temperatures and errors derived using the calibrations by Alonso et al. (1999) 
for several colors, and the final temperature adopted for each star. The error on the 
adopted temperature does not take into account the uncertainty on the reddening.}             
\label{temp}      
\centering                          
\begin{tabular}{crcccccc}        
\hline\hline                 
\noalign{\smallskip}
\hbox{Star} & \hbox{\Teff($B-V$)} & \hbox{\Teff($V-I$)} & \hbox{\Teff($V-R$)} & \hbox{\Teff($J-H$)} & \hbox{\Teff($J-K$)} & \hbox{\Teff($V-K$)} & 
\hbox{Adopted \Teff}\\
\noalign{\smallskip}
\hline
\noalign{\smallskip}
\hbox{CS 30315-029} & 4703$\pm$96  & 4523$\pm$125 & 4959$\pm$150 & 4383$\pm$170 & 4622$\pm$125 & 4621$\pm$25 & 4570$\pm$53\\
\hbox{HE 0057-4541} & 5104$\pm$167 & 5069$\pm$125 & 5372$\pm$150 & 5158$\pm$170 & 5153$\pm$125 & 5237$\pm$40 & 5144$\pm$60\\
\hbox{HE 0105-6141} & 5273$\pm$167 & 5176$\pm$125 & 5638$\pm$150 & 5156$\pm$170 & 5167$\pm$125 & 5398$\pm$40 & 5234$\pm$60\\
\hbox{HE 0240-0807} & 4689$\pm$96  & 4656$\pm$125 & 5064$\pm$150 & 4707$\pm$170 & 4847$\pm$125 & 4802$\pm$40 & 4740$\pm$53\\
\hbox{HE 0516-3820} & 5400$\pm$167 & 5244$\pm$125 & 5633$\pm$150 & 5067$\pm$170 & 5302$\pm$125 & 5333$\pm$40 & 5269$\pm$60\\
\hbox{HE 0524-2055} & 4736$\pm$96  & 4725$\pm$125 & 5096$\pm$150 & 4736$\pm$170 & 4734$\pm$125 & 4813$\pm$40 & 4749$\pm$53\\
\hbox{HE 2229-4153} & 5146$\pm$167 & 5097$\pm$125 & 5529$\pm$150 & 4966$\pm$170 & 5218$\pm$125 & 5354$\pm$40 & 5156$\pm$60\\
\noalign{\smallskip}
\hline
\end{tabular}
\end{table*}

\section{Abundance derivation}

The OSMARCS 1D model atmosphere grid in local thermodynamic equilibrium (LTE) 
was employed (Gustafsson et al. 2003, 2008). We used the spectrum synthesis code Turbospectrum (Alvarez
\& Plez 1998), which includes proper treatment of diffusion in the UV, 
treatment of scattering in the blue and UV domain, molecular
dissociative equilibrium, and collisional broadening by H, He, and
H$_{2}$, following Anstee \& O'Mara (1995), Barklem \& O'Mara (1997),
and Barklem et al. (1998). The calculations used the Turbospectrum
molecular line lists (Alvarez \& Plez 1998), and atomic line lists from
the VALD2 compilation (Kupka et al. 1999).

\subsection{Atmospheric parameters}

\subsubsection{Effective temperature derivation}

The effective temperatures were previously derived by B05, based on
2MASS ($J-H$) and ($J-K_{S}$) colors, and calibrations by Alonso et al.
(1999). We calculated photometric temperatures based on several colors:
($B-V$), ($V-I$), ($V-R$), ($J-H$), ($J-K$) and ($V-K$).
Table~\ref{flux} lists $BVR_{C}$$I_{C}$ (Beers et al. 2007) and 2MASS
$JHK_{S}$ (Skrutskie et al. 2006) colors and magnitudes for the sample
stars. Calibrations by Alonso et al. (1999) were applied, with reddening
$E(B-V)$ computed with the Galactic Reddening and Extinction Calculator
from the Infrared Processing and Analysis Center (IRSA)
\footnote{http://irsa.ipac.caltech.edu/applications/DUST/}, which is
based on Schlegel et al. (1998). The extinction laws given by Dean et
al. (1978) and Rieke \& Lebofsky (1985) were adopted.

The relations established by Fernie (1983) between the photometric
systems of Johnson and Cousins were used, and the results agree well 
with those obtained from Bessell (1979). Colors from 2MASS
were transformed into the Caltech (CIT) system, and from this into the TCS
(Telescopio Carlos S\'anchez) system, following the relations established by
Carpenter (2001) and Alonso et al. (1998) (infrared flux method, IRFM).
The transformations were also performed with the ESO photometric system
in place of CIT, without significant differences in the final
temperatures.

Table \ref{temp} lists the temperatures deduced from the different
colors, as well as the adopted temperatures for each star. The adopted
temperatures do not take into account the temperatures estimated from
($V-R$), since they appear to be systematically higher than the other
individual estimates. Table \ref{temp} provides the errors suggested by
Alonso et al. (1999). Cayrel et al. (2004) adopted an error on E($B-V$)
toward halo stars of about 0.02 mag, corresponding to a temperature
error of about 60 K. This leads to a total error in the final
temperature estimate of $\Delta$\Teff$~\sim$~100~K, which agrees well 
with B05.

\subsubsection{Derivation of gravity and microturbulence velocity}

An iterative method was adopted to derive estimates of the surface
gravity, \logg, and microturbulence velocity, $\xi$:

\begin{itemize}
\item {By fixing the photometric \Teff, together with [Fe/H] and $\xi$ from B05, we calculated the ionization 
equilibrium for a series of models. We then chose the \logg~that minimizes 
([\ion{Ti}{I}/H]-[\ion{Ti}{II}/H])+([\ion{Fe}{I}/H]-[\ion{Fe}{II}/H]).}
\item {Adopting the newly derived \logg~and [Fe/H] values, $\xi$ was chosen to minimize the dependence of 
[\ion{Fe}{I}/H] on the equivalent width of the line.}
\item {Using the newly derived $\xi$, [Fe/H], and \logg~were recomputed until stable values were obtained.}
\end{itemize}
 
\begin{figure}
\centering
\resizebox{80mm}{!}{\includegraphics[angle=0]{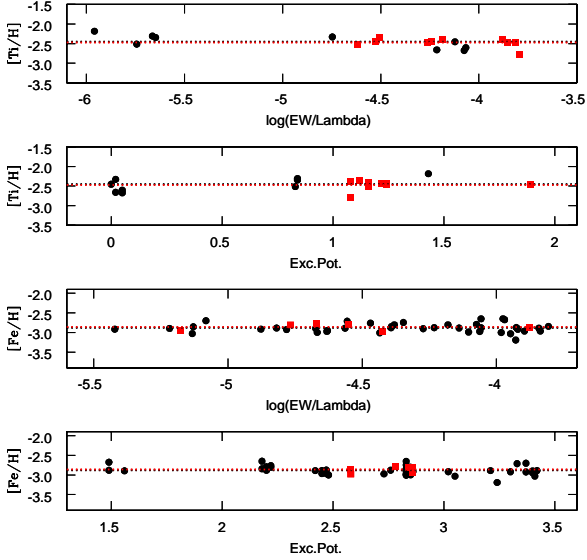}}
\caption{Excitation and ionization equilibria of Ti and Fe lines, resulting from 
the new set of atmospheric parameters for HE~0240-0807. The black dots are 
the abundances obtained from \ion{Ti}{I} and \ion{Fe}{I} lines, and the red squares are 
those from \ion{Ti}{II} and \ion{Fe}{II} lines; the black dotted lines represent the 
average metallicity from the neutral species, and the red dotted lines represent the 
average metallicity from the ionized species.}
\label{avalia_HE02}
\end{figure}

Fig. \ref{avalia_HE02} shows an example of the dependence of
[\ion{Fe}{I}/H], [\ion{Fe}{II}/H], [\ion{Ti}{I}/H], and [\ion{Ti}{II}/H]
on $\rm \log(EW/\lambda)$, and on the excitation potential of the lines 
for HE~0240-0807. With the adopted parameters, 
the \ion{Fe}{I} and \ion{Fe}{II} lines (black and red dots,
respectively) lead to the same iron abundance (the same is observed for
the \ion{Ti}{I} and \ion{Ti}{II} lines). Our choice of
\Teff~(photometric temperature) is also supported by the fulfilled ionization
and excitation equilibria for [Fe/H] and [Ti/H], as shown in Fig.
\ref{avalia_HE02}. In this check of consistency, the \ion{Fe}{I} lines
with an excitation potential lower than 1.4~eV were excluded, because
they are severely affected by NLTE effects (see Cayrel et al. 2004).  

The adopted parameters are given in Table \ref{atm_novos}. Typical
uncertainties are $\Delta$\logg~=~0.1 dex and $\Delta\xi$~=~0.2
\kms~(e.g. Cayrel et al. 2004). As pointed out by Cayrel et al. (2004),
the \logg~values may be affected by NLTE effects (overionization), and
by uncertainties in the oscillator strengths of the Fe and Ti lines.
Comparing these results with the parameters from B05, the average difference (and
dispersion as the standard deviation) in the temperatures is
$\Delta\hbox{\Teff}=10(24)$ K. For surface gravity, metallicity, and
microturbulence, the results are $\Delta\hbox{\logg}=-0.08~(0.23)$
[cgs], $\Delta\hbox{[Fe/H]}=-0.10~(0.07)$ dex and $\Delta\xi=0.06~(0.17)
$ \kms~. The results show a good consistency between the two sets of
parameters.

\begin{table}
\caption{Adopted atmospheric parameters.}             
\label{atm_novos}      
\centering                          
\begin{tabular}{cccccccc}        
\hline\hline                 
\noalign{\smallskip}
\hbox{Star} & \hbox{\Teff} & \hbox{\logg} & \hbox{[Fe/H]} & \hbox{$\xi$} \\
\noalign{\smallskip}
\hline
\noalign{\smallskip}
\hbox{} & \hbox{(K)} & \hbox{[cgs]} & \hbox{} & \hbox{(\kms)} \\
\noalign{\smallskip}
\hline
\noalign{\smallskip}
\hbox{CS 30315-029} & 4570 & 0.99 & $-$3.40 & 2.22\\
\hbox{HE 0057-4541} & 5144 & 2.73 & $-$2.40 & 1.79\\
\hbox{HE 0105-6141} & 5234 & 2.91 & $-$2.57 & 1.42\\
\hbox{HE 0240-0807} & 4740 & 1.50 & $-$2.86 & 2.27\\
\hbox{HE 0516-3820} & 5269 & 2.17 & $-$2.51 & 1.48\\
\hbox{HE 0524-2055} & 4749 & 1.53 & $-$2.77 & 2.20\\
\hbox{HE 2229-4153} & 5156 & 2.67 & $-$2.62 & 1.63\\
\noalign{\smallskip}
\hline
\end{tabular}
\end{table}

\subsection{Derived abundances}

The elemental abundances were derived using equivalent widths for lines
of titanium and iron. A line-by-line fitting was carried out for the
other elements, with special attention given to the molecular bands (CH,
CN, NH), to the lines affected by hyperfine splitting, or that were severely
blended, and to all the lines of the heavy elements. 

Synthetic spectra were convolved with Gaussian profiles that take into
account, in combination, the effects of macroturbulence, rotational, and
instrumental broadening. This analysis is very sensitive to the
determination of the continuum. We determined a local 
continuum around each line before computing the $\chi^{2}$ between
computed and observed spectra, using the synthetic spectra as reference.

The position and intensity of the telluric lines were
computed by means of the line-by-line radiative transfer model code 
(LBLRTM; Clough et al. 2005), using the high-resolution transmission molecular 
absorption database (HITRAN; Rothman et al. 2009). We used the same transmission, T$_{0}$($\lambda$), computed for a
standard atmosphere by Chen et al. (2013) for the altitude of the
Observatory, modified to take into account the airmass variations from
star to star. Stellar lines that are severely blended by telluric lines
were excluded.

The final abundance estimates for each element are the mean of
abundances from all individual lines; the internal (observational)
uncertainty corresponds to the standard deviation of the distribution of
estimates. Table
\ref{finalAbund} presents the LTE abundances A(X)\footnote{We adopted the
notation A(X)~=~log~(X)~=~log~n(X)/n(H)~+~12, with n~=~number density of
atoms} and the [X/Fe] abundance ratios of all studied elements for each
star of the sample. The complete line list as well as abundance
values from individual lines can be found in Table \ref{linelist}.

\begin{table*}
\caption{LTE abundances.}             
\label{finalAbund}      
\scalefont{0.7}
\centering                          
\begin{tabular}{crrrrrrrrrrrrrr}        
\hline\hline                 
\noalign{\smallskip}
\hbox{Element} & \hbox{A(X)} & \hbox{[X/Fe]} & \hbox{A(X)} & \hbox{[X/Fe]} & \hbox{A(X)} & \hbox{[X/Fe]} 
& \hbox{A(X)} & \hbox{[X/Fe]} & \hbox{A(X)} & \hbox{[X/Fe]} & \hbox{A(X)} & \hbox{[X/Fe]} & \hbox{A(X)} & \hbox{[X/Fe]} \\
\noalign{\smallskip}
\hline
\noalign{\smallskip}
& \multicolumn{2}{c}{CS~30315-029} & \multicolumn{2}{c}{HE~0057-4541} & \multicolumn{2}{c}{HE~0105-6141} 
& \multicolumn{2}{c}{HE~0240-0807} & \multicolumn{2}{c}{HE~0516-3820} & \multicolumn{2}{c}{HE~0524-2055} 
& \multicolumn{2}{c}{HE~2229-4153} \\
\noalign{\smallskip}
\hline
\noalign{\smallskip}
\hbox{Fe I*}    & 4.07     & $-$3.45  &  5.16    & $-$2.36  & 4.94     & $-$2.58  & 4.64     & $-$2.88  & 5.02     & $-$2.50  & 4.76     & $-$2.76  & 4.89     & $-$2.63  \\
\hbox{Fe II*}   & 4.08     & $-$3.44  &  5.16    & $-$2.36  & 4.96     & $-$2.56  & 4.66     & $-$2.86  & 5.02     & $-$2.50  & 4.76     & $-$2.76  & 4.89     & $-$2.63  \\
\hbox{Li I}     &  -----   &   -----  &  1.14    & $+$2.47  & 1.16     & $+$2.71  &  -----   &  -----   & 0.96     & $+$2.43  &  -----   &  -----   & 0.98     & $+$2.58  \\
\hbox{C}        & 5.00     & $-$0.06  &  6.58    & $+$0.44  & 6.10     & $+$0.17  & 5.20     & $-$0.43  & 6.35     & $+$0.35  & 5.84     & $+$0.10  & 6.40     & $+$0.53  \\
\hbox{N(CN)}    & 5.60     & $+$1.19  &  5.65    & $+$0.15  & -----    & -----    & 5.70     & $+$0.72  & 5.70     & $+$0.34  & 5.95     & $+$0.85  & -----    & -----   \\
\hbox{N(NH)}    & 5.95     & $+$1.54  &  5.70    & $+$0.20  & $<$5.20  & $<-$0.08 & 6.05     & $+$1.07  & 5.80     & $+$0.44  & 6.20     & $+$1.10  & 5.35     & $+$0.12  \\
\hbox{Mg}       & 4.75     & $+$0.65  &  5.65    & $+$0.47  & 5.42     & $+$0.45  & 5.23     & $+$0.57  & 5.55     & $+$0.51  & 5.60     & $+$0.82  & 5.55     & $+$0.64  \\
\hbox{Al}       & 2.68     & $-$0.35  &  3.50    & $-$0.61  & 3.20     & $-$0.70  & 3.26     & $-$0.34  & 3.30     & $-$0.67  & 3.40     & $-$0.31  & 3.25     & $-$0.59  \\
\hbox{Si}       & 4.80     & $+$0.72  &  5.94    & $+$0.78  & 5.64     & $+$0.69  & 5.30     & $+$0.65  & 5.45     & $+$0.43  & 5.37     & $+$0.61  & 5.48     & $+$0.59  \\
\hbox{S}        & 4.33     & $+$0.61  &  5.35    & $+$0.55  & 5.25     & $+$0.66  & 4.95     & $+$0.66  & 5.35     & $+$0.69  & 5.08     & $+$0.68  & 5.05     & $+$0.52  \\
\hbox{K}        & 2.60     & $+$0.93  &  3.30    & $+$0.55  & 2.92     & $+$0.38  & 2.95     & $+$0.71  & 3.18     & $+$0.57  & 3.27     & $+$0.92  & 3.25     & $+$0.77  \\
\hbox{Ca}       & 3.32     & $+$0.43  &  4.33    & $+$0.37  & 4.17     & $+$0.41  & 3.82     & $+$0.37  & 4.23     & $+$0.40  & 4.00     & $+$0.44  & 4.13     & $+$0.44  \\
\hbox{Sc}       & $-$0.36  & $-$0.02  &  0.84    & $+$0.10  & 0.77     & $+$0.24  & 0.38     & $+$0.16  & 0.75     & $+$0.15  & 0.42     & $+$0.09  & 0.62     & $+$0.15  \\
\hbox{Ti I}     & 1.86     & $+$0.40  &  2.95    & $+$0.42  & 2.81     & $+$0.49  & 2.45     & $+$0.43  & 2.74     & $+$0.34  & 2.54     & $+$0.40  & 2.67     & $+$0.41  \\
\hbox{Ti II}    & 1.80     & $+$0.35  &  2.96    & $+$0.42  & 2.83     & $+$0.51  & 2.43     & $+$0.41  & 2.77     & $+$0.38  & 2.53     & $+$0.39  & 2.66     & $+$0.39  \\
\hbox{V I}      & 0.32     & $-$0.24  &  1.57    & $-$0.07  & 1.32     & $-$0.10  & 1.12     & $-$0.01  & 1.40     & $-$0.10  & 1.22     & $-$0.02  & 1.35     & $-$0.01  \\
\hbox{V II}     & 0.53     & $-$0.03  &  1.87    & $+$0.23  & 1.65     & $+$0.22  & 1.37     & $+$0.24  & 1.69     & $+$0.19  & 1.34     & $+$0.10  & 1.51     & $+$0.14  \\
\hbox{Cr}       & 1.78     & $-$0.41  &  3.06    & $-$0.21  & 2.83     & $-$0.24  & 2.30     & $-$0.47  & 3.05     & $-$0.09  & 2.44     & $-$0.44  & 2.75     & $-$0.25  \\
\hbox{Mn}       & 1.58     & $-$0.35  &  2.78    & $-$0.23  & 2.37     & $-$0.43  & 2.07     & $-$0.43  & 2.73     & $-$0.14  & 2.26     & $-$0.35  & 2.42     & $-$0.32  \\
\hbox{Co}       & 1.74     & $+$0.26  &  2.86    & $+$0.30  & 2.80     & $+$0.45  & 2.29     & $+$0.24  & 2.81     & $+$0.40  & 2.45     & $+$0.29  & 2.63     & $+$0.34  \\
\hbox{Ni}       & 2.78     & $-$0.01  &  3.79    & $-$0.08  & 3.75     & $+$0.09  & 3.31     & $-$0.05  & 3.85     & $+$0.12  & 3.39     & $-$0.08  & 3.70     & $+$0.10  \\
\hbox{Sr}       & $-$0.84  & $-$0.32  &  0.80    & $+$0.24  & 0.64     & $+$0.29  & $-$0.08  & $-$0.13  & 0.65     & $+$0.23  & 0.15     & $-$0.01  & 0.57     & $+$0.28  \\
\hbox{Y}        & $-$1.45  & $-$0.22  & $-$0.30  & $-$0.14  & $-$0.39  & $-$0.03  & $-$0.72  & $-$0.06  & $-$0.14  & $+$0.16  & $-$0.71  & $-$0.16  & $-$0.47  & $-$0.05  \\
\hbox{Zr}       & $-$0.77  & $+$0.10  &  0.54    & $+$0.32  & 0.21     & $+$0.20  & $-$0.03  & $+$0.26  & 0.51     & $+$0.44  & 0.11     & $+$0.30  & 0.22     & $+$0.27  \\
\hbox{Mo}       & $<-$1.15 & $<+$0.37 & -----    &  -----   & -----    & -----    &  -----   &  -----   &  -----   &  -----   & $<-$0.35 & $<+$0.49 & -----    &  -----   \\
\hbox{Ru}       & $-$0.95  & $+$0.65  &  0.24    & $+$0.76  & -----    & -----    & $<-$0.20 & $<+$0.83 &  -----   &  -----   & $-$0.15  & $+$0.77  & $<-$0.13 & $<+$0.66 \\
\hbox{Pd}       & $<-$1.10 & $<$0.68  & -----    &  -----   & -----    & -----    &  -----   &  -----   &  -----   &  -----   &  -----   &  -----   & $<-$0.60 & $<+$0.37 \\
\hbox{Ba}       & $-$1.10  & $+$0.17  & $-$0.22  & $-$0.03  & $-$0.25  & $-$0.15  & $-$0.50  & $+$0.20  & $-$0.27  & $+$0.06  & $-$0.63  & $-$0.04  & $-$0.65  & $-$0.19  \\
\hbox{La}       & $-$2.06  & $+$0.24  & $-$1.00  & $+$0.22  & $-$1.15  & $+$0.28  & $-$1.44  & $+$0.29  & $-$1.18  & $+$0.18  & $-$1.57  & $+$0.05  & $-$1.58  & $-$0.08  \\
\hbox{Ce}       & $-$1.65  & $+$0.18  & $-$0.48  & $+$0.27  & $<-$0.85 & $<+$0.11 & $-$1.00  & $+$0.26  & $-$0.44  & $+$0.45  & $-$1.22  & $-$0.07  & $-$1.05  & $-$0.03  \\
\hbox{Pr}       & $-$2.13  & $+$0.56  & $<-$1.09 & $<$0.51  & $<-$0.85 & $<+$0.97 & $-$1.52  & $+$0.60  & $<-$0.88 & $<+$0.87 & $-$1.62  & $+$0.39  & $-$1.50  & $+$0.37  \\
\hbox{Nd}       & $-$1.56  & $+$0.44  & $-$0.56  & $+$0.35  & $-$0.53  & $+$0.59  & $-$0.93  & $+$0.50  & $-$0.46  & $+$0.60  & $-$1.03  & $+$0.28  & $-$1.00  & $+$0.19  \\
\hbox{Sm}       & $-$1.76  & $+$0.68  & $-$0.97  & $+$0.40  & $-$0.80  & $+$0.77  & $-$1.19  & $+$0.69  & $<-$0.46 & $<+$1.04 & $-$1.42  & $+$0.35  & $-$1.33  & $+$0.31  \\
\hbox{Eu}       & $-$2.24  & $+$0.68  & $-$1.27  & $+$0.58  & $-$1.53  & $+$0.52  & $-$1.58  & $+$0.78  & $-$1.34  & $+$0.64  & $-$1.75  & $+$0.50  & $-$1.72  & $+$0.39  \\
\hbox{Gd}       & $-$1.85  & $+$0.48  & $-$0.85  & $+$0.40  & $<-$0.90 & $<+$0.56 & $-$1.09  & $+$0.67  & $-$1.12  & $+$0.27  & $-$1.38  & $+$0.27  & $-$1.33  & $+$0.20  \\
\hbox{Tb}       & $-$2.40  & $+$0.76  & -----    &  -----   & -----    & -----    & $<-$1.78 & $<+$0.82 &  -----   &  -----   & $<-$1.95 & $<+$0.53 & -----    &  -----   \\
\hbox{Dy}       & $-$1.62  & $+$0.69  & $-$0.62  & $+$0.61  & $-$0.98  & $+$0.46  & $-$1.03  & $+$0.71  & $-$0.73  & $+$0.65  & $-$1.12  & $+$0.51  & $-$1.15  & $+$0.35  \\
\hbox{Ho}       & $-$2.38  & $+$0.56  & $-$1.32  & $+$0.53  & -----    & -----    & $-$1.63  & $+$0.74  & $<-$1.00 & $<+$0.99 & $-$1.80  & $+$0.45  & $-$1.90  & $+$0.22  \\
\hbox{Er}       & $-$1.75  & $+$0.73  & $-$0.78  & $+$0.63  & -----    & -----    & $-$1.22  & $+$0.69  & $<-$0.78 & $<+$0.76 & $-$1.27  & $+$0.53  & $-$1.17  & $+$0.50  \\
\hbox{Tm}       & $-$2.72  & $+$0.59  & $<-$1.46 & $<+$0.77 & -----    & -----    & $<-$1.86 & $<+$0.87 &  -----   &  -----   &  -----   &  -----   & $<-$1.93 & $<+$0.57 \\
\hbox{Yb}       & $-$1.98  & $+$0.60  & $-$0.95  & $+$0.55  & $-$1.22  & $+$0.49  & $-$1.25  & $+$0.76  & $-$0.97  & $+$0.67  & $-$1.58  & $+$0.32  & $-$1.53  & $+$0.24  \\
\hbox{Os}       & $<-$1.00 & $<+$1.08 & -----    &  -----   & -----    & -----    & $<-$0.43 & $<+$1.09 &  -----   &  -----   &  -----   &  -----   & -----    &  -----   \\
\hbox{Ir}       & $<-$0.74 & $<+$1.32 & $<$0.01  & $<+$0.99 & -----    & -----    & $<-$0.20 & $<+$1.29 &  -----   &  -----   & $<-$0.45 & $<+$0.93 & -----    &  -----   \\
\hbox{Th}       & $-$2.45  & $+$0.91  & $<-$1.40 & $<+$0.88 & -----    & -----    & $<-$1.92 & $<+$0.87 &  -----   &  -----   & $<-$2.12 & $<+$0.56 & -----    &  -----   \\
\noalign{\smallskip}
\hline
\end{tabular}
\tablebib{ *: [X/H] is used in place of [X/Fe].}
\end{table*}

\subsection{Uncertainties on the derived abundances}

We adopted typical errors in the atmospheric parameters:
$\Delta$\Teff~=~100~K, $\Delta$\logg~=~0.1~dex, and
$\Delta\xi$~=~0.2~\kms. It is possible to estimate the abundance
uncertainties that arise from each of these three sources independently.
However, the quadratic sum of the various sources of uncertainties is
probably not the best way to estimate the total error budget, since the
stellar parameters are not independent of each other, which adds significant 
covariance terms to this calculation. 

To solve this, we created a new atmospheric model with a 100~K lower
temperature, determining the corresponding surface gravity and
microturbulent velocity by the traditional method. The difference
between this model and the nominal model are expected to represent the total
error budget arising from the stellar parameters. The final error is the
sum of the uncertainty from the atmospheric parameters with the
observational error. Table \ref{error} shows the final values in
CS~30315-029 as an example.

\begin{table}
\caption{Observational and atmospheric errors in CS~30315-029, as well as the final uncertainties.}             
\label{error}      
\centering                          
\begin{tabular}{ccccc}        
\hline\hline                 
\noalign{\smallskip}
\hbox{Element} & \hbox{$\Delta_{obs}$} & \hbox{$\Delta_{atm}$} & \hbox{$\Delta_{final}$} & \hbox{$\Delta_{final[X/Fe]}$} \\
\noalign{\smallskip}
\hline
\noalign{\smallskip}
 & \hbox{(dex)} & \hbox{(dex)} & \hbox{(dex)} & \hbox{(dex)} \\
\noalign{\smallskip} 
\hline
\noalign{\smallskip}
\hbox{Fe~I}  & 0.09 & 0.09 & 0.13 & -- \\
\hbox{Fe~II} & 0.06 & 0.06 & 0.09 & -- \\
\hbox{C}     & 0.10 & 0.16 & 0.19 & 0.22 \\
\hbox{N}     & 0.10 & 0.10 & 0.14 & 0.21 \\
\hbox{Mg}    & 0.07 & 0.09 & 0.11 & 0.17 \\
\hbox{Al}    & 0.10 & 0.12 & 0.16 & 0.20 \\
\hbox{Si}    & 0.10 & 0.02 & 0.10 & 0.16 \\
\hbox{S}     & 0.10 & 0.12 & 0.16 & 0.20 \\
\hbox{K}     & 0.10 & 0.05 & 0.11 & 0.18 \\
\hbox{Ca}    & 0.07 & 0.02 & 0.07 & 0.14 \\
\hbox{Sc}    & 0.05 & 0.04 & 0.06 & 0.16 \\
\hbox{Ti~I}  & 0.11 & 0.14 & 0.17 & 0.21 \\
\hbox{Ti~II} & 0.07 & 0.09 & 0.12 & 0.17 \\
\hbox{V~I}   & 0.10 & 0.07 & 0.12 & 0.16 \\
\hbox{V~II}  & 0.06 & 0.07 & 0.09 & 0.14 \\
\hbox{Cr}    & 0.01 & 0.02 & 0.02 & 0.11 \\
\hbox{Mn}    & 0.10 & 0.07 & 0.12 & 0.17 \\
\hbox{Co}    & 0.10 & 0.04 & 0.11 & 0.17 \\
\hbox{Ni}    & 0.10 & 0.02 & 0.10 & 0.15 \\
\hbox{Sr}    & 0.10 & 0.06 & 0.12 & 0.17 \\
\hbox{Y}     & 0.02 & 0.07 & 0.07 & 0.13 \\
\hbox{Zr}    & 0.07 & 0.06 & 0.09 & 0.14 \\
\hbox{Ru}    & 0.10 & 0.10 & 0.14 & 0.19 \\
\hbox{Ba}    & 0.10 & 0.02 & 0.10 & 0.16 \\
\hbox{La}    & 0.05 & 0.07 & 0.09 & 0.14 \\
\hbox{Ce}    & 0.10 & 0.10 & 0.14 & 0.18 \\
\hbox{Pr}    & 0.03 & 0.08 & 0.09 & 0.14 \\
\hbox{Nd}    & 0.05 & 0.05 & 0.07 & 0.13 \\
\hbox{Sm}    & 0.08 & 0.07 & 0.10 & 0.16 \\
\hbox{Eu}    & 0.05 & 0.05 & 0.07 & 0.13 \\
\hbox{Gd}    & 0.10 & 0.05 & 0.11 & 0.16 \\
\hbox{Tb}    & 0.10 & 0.10 & 0.14 & 0.20 \\
\hbox{Dy}    & 0.07 & 0.12 & 0.14 & 0.18 \\
\hbox{Ho}    & 0.10 & 0.08 & 0.13 & 0.19 \\
\hbox{Er}    & 0.10 & 0.08 & 0.13 & 0.18 \\
\hbox{Tm}    & 0.10 & 0.07 & 0.12 & 0.17 \\
\hbox{Yb}    & 0.10 & 0.07 & 0.12 & 0.19 \\
\hbox{Th}    & 0.10 & 0.10 & 0.14 & 0.18 \\
\noalign{\smallskip}
\hline
\end{tabular}
\end{table}

\section{Results}

\subsection{Light and iron-peak elements}

\subsubsection{Lithium}

The atmospheres of cool giant stars are Li-depleted. This is expected,
since Li is destroyed at temperatures in excess of 2.5$\times$10$^{6}$K.
Such temperatures are reached in the inner atmospheric layers, which are mixed
by convection into the outer atmosphere of the star. 

\begin{figure}
\centering
\resizebox{80mm}{!}{\includegraphics[angle=0]{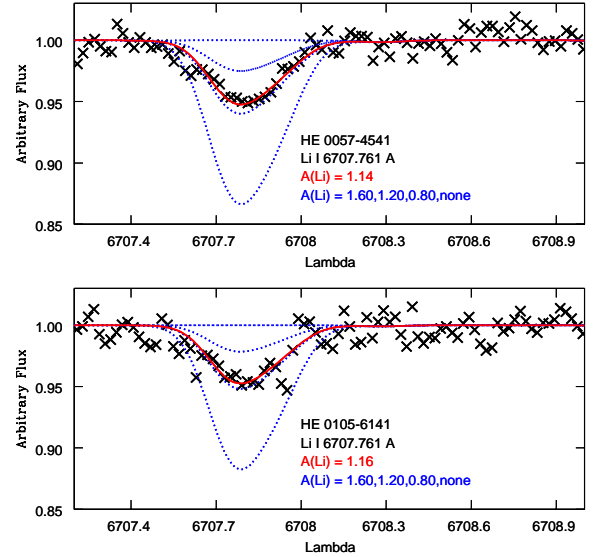}}
\caption{Li abundances for HE~0057-4541 (upper panel) and HE~0105-6141 (lower panel). 
Observations (crosses) are compared with synthetic spectra computed with different 
abundances (blue dotted lines), as well as with the adopted abundances (red solid lines).}
\label{Li_fig}
\end{figure}

HE~0057-4541 and HE~0105-6141 show very clear line profiles at 6707\,{\rm
\AA}. The final lithium abundances are A(Li)~=~$+$1.14 and A(Li)
~=~$+$1.16, respectively. Fig. \ref{Li_fig} shows the fit used for
both cases, where the blue dotted lines are the synthetic spectra
computed with the abundances indicated in the figure, and the red solid
lines are the abundances from the best fits. 

The lithium line profile of HE~0516-3820 is weak, but we derived an
abundance of A(Li)~=~$+$0.96. The line profile of HE~2229-4153 yields an
upper limit of A(Li)~$< +$0.98. As expected, all the values determined
are below the lithium plateau for warm dwarf halo stars, and correspond
to the Li abundance observed in giant stars after the first dredge-up,
but before the onset of extra mixing (Spite et al. 2005). 
A low NLTE correction of $\sim$0.05 dex on the Li abundance is given by 
Lind et al. (2009) for parameters similar to those of the present sample. 

For CS~30315-029 and HE~0524-2055, the lithium line is not visible (A(Li)
~$<$~0.0). These giants are cool and have very probably undergone 
extra-mixing events, as confirmed by their very low value of [C/N]:
[C/N]~=~$-$1.07 for CS~30315-029 and [C/N]~=~$-$0.75 for
HE~0524-2055 (see Spite et al. 2005). For HE~0240-0807, no spectra were
available in the lithium line region.

\subsubsection{Carbon and nitrogen}

\begin{figure}
\centering
\resizebox{80mm}{!}{\includegraphics[angle=0]{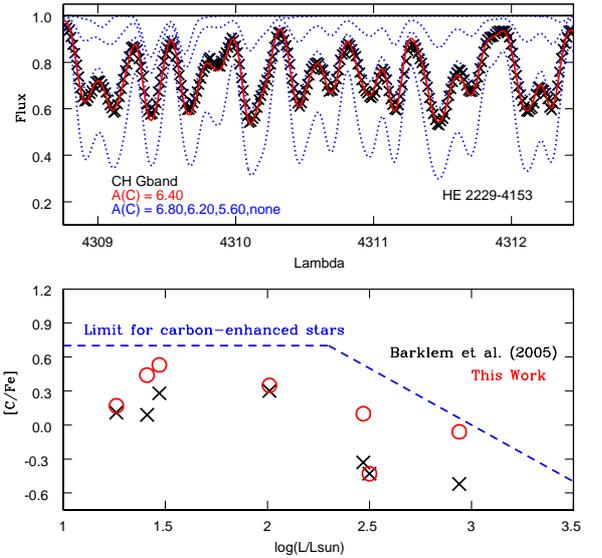}}
\caption{Upper panel: carbon abundance for HE~2229-4153, derived from the CH A$^{2}$$\Delta$-X$^{2}$$\Pi$ (0,0) 
transitions (G-band). Symbols are the same as in Fig. \ref{Li_fig}. 
Lower panel: comparison of the results obtained in this work (red circles) with 
those previously obtained by Barklem et al. (2005) (crosses). 
The dashed blue line corresponds to the limit for carbon-enhanced 
stars, as defined by Aoki et al. (2007).}
\label{Gband_fig}
\end{figure}

The carbon abundance was derived from the CH
A$^{2}$$\Delta$-X$^{2}$$\Pi$ (0,0) transitions (G-band), with lines
present in the range 4290~{\rm \AA}~-~4310~{\rm \AA}, and the bandhead
at $\sim$4310 {\rm \AA}. Fig. \ref{Gband_fig} shows an example of fit
(upper panel) for HE~2229-4153; and one can see that this region is
almost devoid of atomic lines. Beers \& Christlieb (2005) defined 
carbon-enhanced stars as having [C/Fe]~$>$~+1.0. However, in this work
we prefer the definition for the carbon-enhancement phenomenon in
metal-poor stars as presented by Aoki et al. (2007; [C/Fe]~$>$~+0.7),
which also takes into account that the surface abundance of
carbon in evolved stars will in general be lower than it was when the
star was in an earlier evolutionary stage as a consequence of mixing
events. The lower panel in Fig. \ref{Gband_fig} shows carbon abundances
obtained in this work (red circles) and the criterion for
carbon-enhanced stars as defined by Aoki et al. (2007) (dashed blue
line). The previous results from B05 are also presented for comparison
purposes (crosses). The new results, obtained from our high-resolution
spectra, confirm that none of the sample stars are carbon-enhanced. 

The nitrogen abundance was derived from the (0,0) bandhead of the blue
CN B$^{2}$$\Sigma$-X$^{2}$$\Sigma$ at 3883\,{\rm \AA}. However, this
line is weak and not detectable for HE~0105-6141 and HE~2229-4153. We
also used lines of the NH A$^{3}\Pi_{i-}$X$^{3}\Sigma_{-}$ band, with a
bandhead at 3360\,{\rm \AA}. The NH lines give the nitrogen abundance
with a dependence on the C abundance only through dissociative
equilibrium, but not directly in the line intensity (as for CN lines).

\begin{figure}
\centering
\resizebox{80mm}{!}{\includegraphics[angle=0]{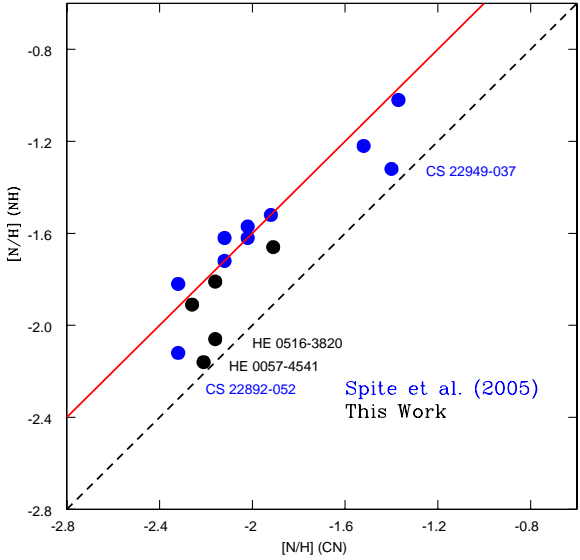}}
\caption{Abundance ratios for [N/H] obtained from the molecular bands of CN and 
NH in the present sample (black dots), compared with the results from 
Spite et al. (2005) (blue dots). The one-to-one relation is shown by a 
dashed black line. The solid red line indicates a correlation offset by 
0.4~dex.}
\label{fig_nitro}
\end{figure}

Figure \ref{fig_nitro} compares the results for our sample stars (black
dots), obtained with CN and NH, compared with the results from Spite et
al. (2005) (blue dots) for the LP ``First Stars''. Spite et al. (2005)
pointed out a systematic difference of 0.4~dex between the two
indicators, and the present results confirm this discrepancy, as shown
by the red solid line in Fig. \ref{fig_nitro}. The reason for this
difference is not clear, but it might be related to molecular constants 
and also to different 3D effects on the NH and CN lines. The molecular
lines are indeed formed very close to the surface, a region where 3D
effects can be important. The two carbon-rich stars from Spite et al.
(2005) are identified in Fig. \ref{fig_nitro}, CS~22949-037 ([C/Fe] =
$+$1.21) and CS~22892-052 ([C/Fe] = $+$0.93). 

\subsubsection{The $\alpha$-elements: Mg, Si, S, Ca, and Ti}

\begin{figure}
\centering
\resizebox{80mm}{!}{\includegraphics[angle=0]{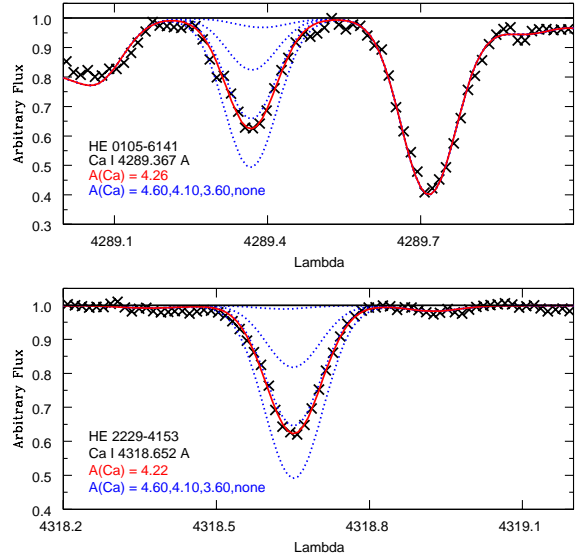}}
\caption{Ca abundances from the \ion{Ca}{I}~4289.37\,{\rm \AA} line for
HE~0105-6141 (upper panel), 
and from the \ion{Ca}{I}~4318.65\,{\rm \AA} line for HE~2229-4153 (lower panel). 
Symbols are the same as in Fig. \ref{Li_fig}.}
\label{Ca_fig}
\end{figure}

We derived abundances of magnesium, calcium, and titanium 
using lines of \ion{Mg}{I}, \ion{Ca}{I}, 
\ion{Ti}{I}, and \ion{Ti}{II}; Table \ref{linelist}  
reports the abundances for individual lines (see Table
\ref{EW_measurements} for Ti), and the averaged results are shown in
Table \ref{finalAbund}. Fig. \ref{Ca_fig} shows the fit for the 
\ion{Ca}{I}~4289.37\,{\rm \AA} line for HE~0105-6141 (upper panel) 
and for the \ion{Ca}{I}~4318.65~{\rm \AA} line for HE~2229-4153 (lower panel). 

Andrievsky et al. (2010) presented NLTE computations for Mg abundances in
stars observed in the framework of the LP ``First Stars'' and found an average correction of
$\Delta_{NLTE}=+0.19$~dex\footnote{$\Delta_{NLTE}$~=~A(X)
$_{NLTE}$~-~A(X)$_{LTE}$}. To estimate the 3D corrections on the
results, we used hydrodynamical models computed with the code CO5BOLD
(Freytag et al. 2002, 2012) with \Teff~=~5000~K, \logg~=~2.5~[cgs], and
two different metallicities, [Fe/H]~=~$-2$ and [Fe/H]~=~$-3$. The plane-parallel 
1D$_{LHD}$ model was used as reference model, computed with the
LHD code, which shares the microphysics and opacity with the CO5BOLD
code. The 3D correction is defined as A(3D)-A(1D$_{LHD}$) (see details
in Caffau et al. 2011). The line-formations were computed with 
Linfor3D\footnote{http://www.aip.de/Members/msteffen/linfor3d/files/linfor3d-manual} 
for the \ion{Mg}{I} lines studied in this work. The results obtained are
$\Delta_{3D}=+0.09$~dex and $\Delta_{3D}=-0.04$~dex for [Fe/H]~=~$-2$ and
[Fe/H]~=~$-3$, respectively. For further discussions about the NLTE computation and 3D
correction of the Ca abundances see Spite et al. (2012).

\begin{figure}
\centering
\resizebox{80mm}{!}{\includegraphics[angle=0]{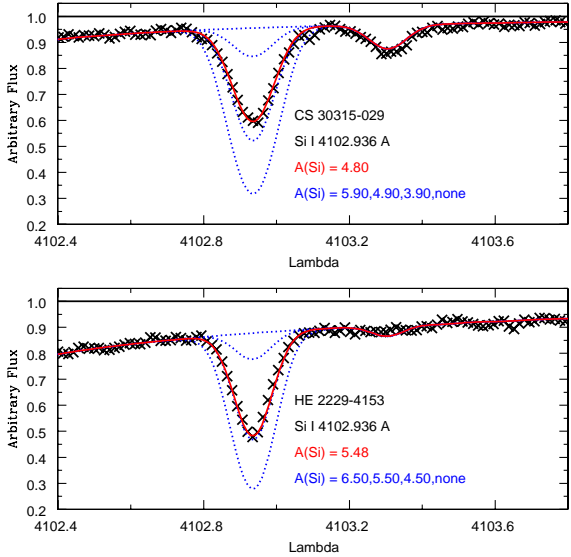}}
\caption{Si abundances from the \ion{Si}{I}~4102.94\,{\rm \AA} line for
CS~30315-029 (upper panel), 
and for HE~2229-4153 (lower panel). Symbols are the same as in Fig. \ref{Li_fig}.}
\label{Si_fig}
\end{figure}

In the visible region, the main indicators of silicon abundance are the
\ion{Si}{I} line at 3905.52\,{\rm \AA}, which in giants is too severely
blended by a CH line, and the \ion{Si}{I} line at 4102.94\,{\rm \AA},
located in the red wing of the H$\delta$ line. We therefore used
spectrum synthesis of this Si line, taking into account the hydrogen
line. Fig. \ref{Si_fig} shows the fits obtained for CS~30315-029 (upper
panel) and HE~2229-4153 (lower panel). 
Using the same approach as described for Mg to estimate the 3D effect 
on the Si abundances, the results obtained for the 4102.94\,{\rm \AA} line 
are $\Delta_{3D}=+0.04$~dex and $\Delta_{3D}=-0.09$~dex, with
[Fe/H]~=~$-2$ and [Fe/H]~=~$-3$, respectively.

\begin{figure}
\centering
\resizebox{80mm}{!}{\includegraphics[angle=0]{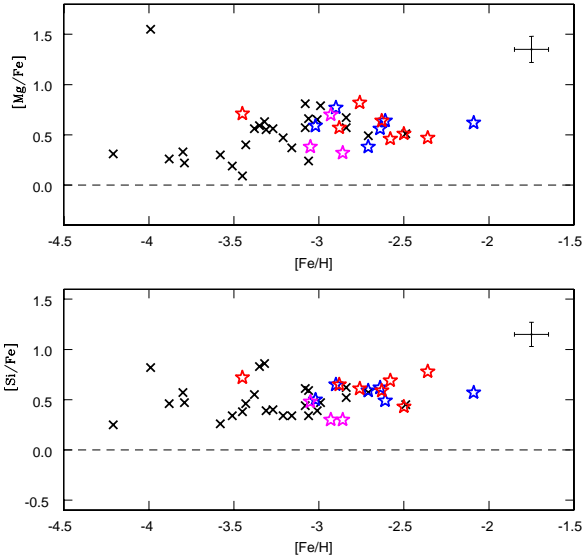}}
\caption{Comparison of the abundance ratios  for [Mg/Fe] (upper panel) and [Si/Fe] (lower panel), 
obtained in this work (red
stars) with the LTE results from Cayrel et al. (2004): 
black crosses for normal stars; magenta for r-II stars; and blue for r-I stars. The average of the 
error bars is indicated in the figure.}
\label{Mg_Si_fig}
\end{figure}

\begin{figure}
\centering
\resizebox{80mm}{!}{\includegraphics[angle=0]{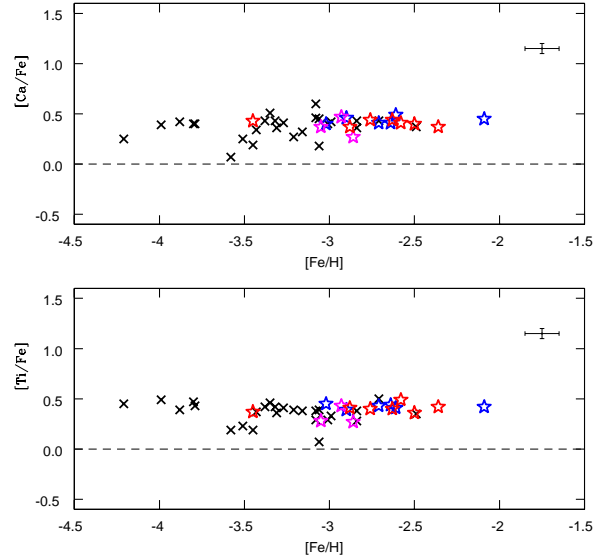}}
\caption{Comparison of the abundance ratios [Ca/Fe] (upper panel) and [Ti/Fe] (lower panel) 
obtained in this work with the LTE results from Cayrel et
al. (2004). Symbols are the same as in Fig. \ref{Mg_Si_fig}.}
\label{Ca_Ti_fig}
\end{figure}

Figures \ref{Mg_Si_fig} and \ref{Ca_Ti_fig} compare the LTE abundance 
trends of the $\alpha$-elements obtained in this work (red stars) and in the ESO LP
``First Stars'' (Cayrel et al. 2004; Bonifacio et al. 2009; black
crosses). The NLTE and 3D corrections are listed in the text for reference only. 
The r-I and r-II stars 
are represented by blue and magenta
star symbols, respectively. The $\alpha$-elements are enhanced for both
sets of stars and the trends are very similar. We remark in particular
that the scatter of [Mg/Fe] from star to star is larger than the scatter
of [Ca/Fe] and [Ti/Fe], as is generally observed. 

\begin{figure}
\centering
\resizebox{80mm}{!}{\includegraphics[angle=0]{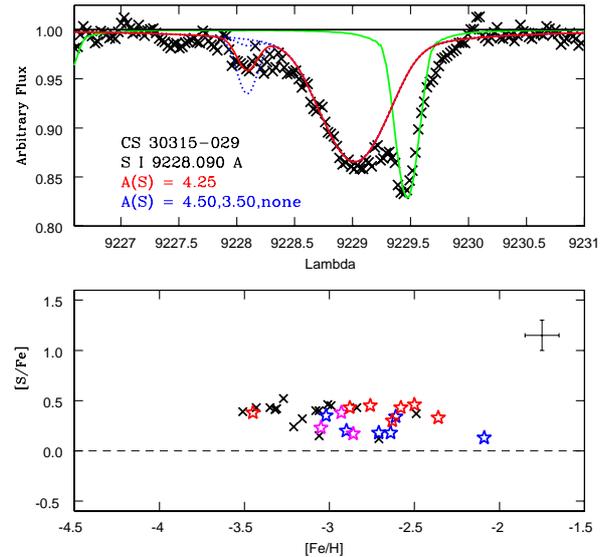}}
\caption{Upper panel: S abundances from the \ion{S}{I}~9228.09\,{\rm \AA}
line for the star 
CS~30315-029. The green solid lines are the synthetic telluric 
transmission. Lower panel: comparison of the abundance ratio [S/Fe] obtained in this work 
with the results from Spite et al. (2011). Symbols are the same as in Fig. \ref{Mg_Si_fig}.}
\label{S_fig}
\end{figure}

The abundance of sulfur was derived from the \ion{S}{I} lines at
9212.86\,{\rm \AA}, 9228.09\,{\rm \AA}, and 9237.54\,{\rm \AA}. This
wavelength region suffers from CCD fringes that affect the continuum
placement, and the definition of a local continuum is mandatory.
Another difficulty arises from the telluric H$_{2}$O lines present in the
region, which were taken into account according to the method described
in Sect. 3.2. Sulfur transitions that are severely blended by telluric lines were
not considered in the final result. In addition, the 
\ion{S}{I}~9228.09~{\rm \AA} line is located on the blue side of the Paschen
P$\zeta$ wing, which was included in the synthesis. Fig. \ref{S_fig}
shows the result for CS~30315-029 (upper panel). One can also
see a telluric line in the red wing of the P$\zeta$, represented by the
green solid line. 

Spite et al. (2011) presented S abundances for 33 metal-poor stars
observed in the framework of the LP ``First Stars,'' taking into account
NLTE calculations and 3D corrections. As an estimate, the average
correction $\Delta_{NLTE+3D}$~=~$-$0.23~dex from Spite et al. (2011) was
applied to the present results. Fig. \ref{S_fig} presents in the lower
panel the comparison of our final abundaces with the results obtained by
Spite et al. (2011). The formation of sulfur remains controversial, and the trend of 
[S/Fe] versus [Fe/H] in the early Galaxy is still debated. The result shows 
that this element presents an $\alpha$-element-like behavior, and the abundances 
derived in the present work follow the general trend.

It is important to note that the results do not come from a full 3D NLTE computation, 
and the NLTE and 3D corrections were obtained separately. It is well-known that the 
two effects combine in a clearly nonlinear way (see Gustafsson 2009 and references 
therein), but only diagnostic works have been done so far with the complete approach 
for few elements. In this sense, the corrections we present should be regarded as estimates.

\subsubsection{Odd-Z elements: Al, K, and Sc}

The published data from the HERES survey (B05) do not exhibit significant
differences between r-I and r-II stars for the abundances of the normal 
elements (i.e., the non neutron-capture elements).
Figs.~\ref{Mg_Si_fig} and \ref{Ca_Ti_fig} show that the same is true for
the ESO LP ``First Stars'' data. The only element that seems to differ
slightly between the stellar groups is the odd-Z element aluminum, thus
this element was measured in our new high-resolution spectra as well. In
addition, we decided to measure the odd-Z element potassium. For both S
and K, spectra in the near-infrared region are needed, which were
included in the observation program of the present work.

\begin{figure}
\centering
\resizebox{80mm}{!}{\includegraphics[angle=0]{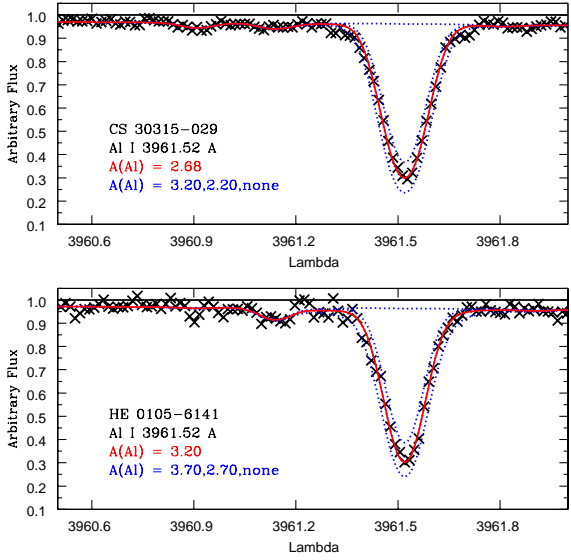}}
\caption{Al abundances from the \ion{Al}{I}~3961.52\,{\rm \AA} line for
CS~30315-029 (upper panel), 
and for HE~0105-6141 (lower panel). Symbols are the same as in Fig. \ref{Li_fig}.}
\label{Al_fig}
\end{figure}

The Al abundances were derived from the \ion{Al}{I}~3944.01\,{\rm \AA}
and \ion{Al}{I}~3961.52\,{\rm \AA} doublet lines. A blend with CH lines
for the \ion{Al}{I}~3944.01\,{\rm \AA} line was taken into account, but in some
stars the contribution of the molecular lines is strong, so we preferred
to adopt only the \ion{Al}{I}\,3961.52~{\rm \AA} line in the final
abundances. Fig. \ref{Al_fig} shows the results for CS~30315-029 (upper
panel) and HE~0105-6141 (lower panel).

\begin{figure}
\centering
\resizebox{80mm}{!}{\includegraphics[angle=0]{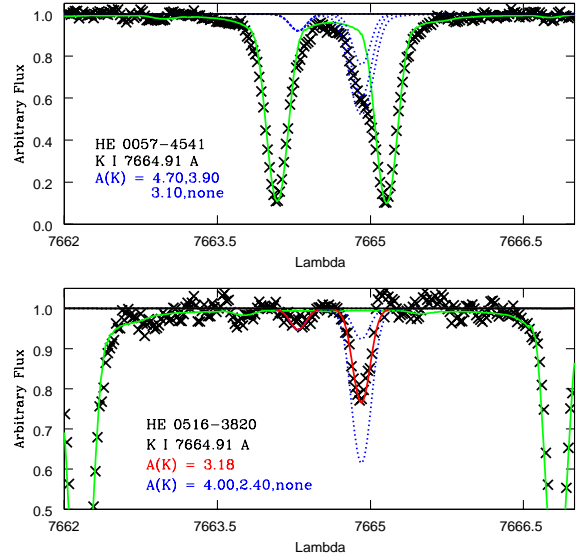}}
\caption{K abundances from the \ion{K}{I}~7664.91\,{\rm \AA} line for
HE~0057-4541 (upper panel), 
and for HE~0516-3820 (lower panel). Symbols are the same as in Fig. \ref{Li_fig}. The green solid lines are 
the synthetic telluric transmission features.}
\label{K_fig}
\end{figure}

The K abundances were derived from the \ion{K}{I}~7664.91\,{\rm \AA}
line and the \ion{K}{I}~7698.97\,{\rm \AA} doublet lines. Even
correcting for the telluric lines, for half of our sample stars the
\ion{K}{I}~7664.91\, {\rm \AA} line was severely compromised and was 
discarded, as shown in Fig.~\ref{K_fig} (upper panel) for HE~0057-4541.
For HE~0516-3820, it was possible to use the same line, and
the result is shown in Fig.~\ref{K_fig} (lower panel).

Given that the abundances were measured from resonance lines, which are
very sensitive to NLTE effects, it is important to correct for this
effect. Cayrel et al. (2004) adopted a constant correction for all the
stars as a rough estimate: for Al a correction of $+$0.65~dex from
Baum\"uller \& Gehren (1997) and Norris et al. (2001), and for K, a
correction of $-$0.35~dex from Ivanova \& Shimanskii (2000). 

Andrievsky et al. (2008) have computed the NLTE Al abundances in the LP
``First Stars'' sample. The NLTE corrections $\Delta_{NLTE}$ behavior 
with effective temperature, surface gravity, and metallicity of the model
is also provided (see Fig. 2 in their paper), from which we estimate
$\Delta_{NLTE}$ for the present sample.
The LTE Al abundances in Table \ref{finalAbund} must be corrected by
$\Delta_{NLTE}$~=~$+$0.20~dex,$+$0.65~dex,$+$0.75~dex,$+$0.25~dex,
$+$0.65~dex,$+$0.30~dex, and $+$0.75~dex for CS~30315-029, HE~0057-4541,
HE~0105-6141, HE~0240-0807, HE~0516-3820, HE~0524-2055, and
HE~2229-4153, respectively. 
Applying our computations to estimate the 3D corrections to Al 
abundances, the results obtained with the \ion{Al}{I}\,3961.52~{\rm \AA} 
line are $\Delta_{3D}=-0.02$~dex and $\Delta_{3D}=-0.33$~dex, with 
metallicities of [Fe/H]~=~$-2$ and [Fe/H]~=~$-3$, respectively, showing that 
this effect becomes strong in extremely metal-poor stars. 

Takeda et al. (2009) reanalyzed the K abundances from the EWs of the LP
``First Stars,'' showing that the NLTE corrections systematically
decrease with decreasing metallicity: $\Delta_{NLTE}$~$\sim$~$-$0.3~dex at
[Fe/H]~$\sim$~$-$2.5~dex and $\Delta_{NLTE}$~$\sim$~$-$0.2~dex at
[Fe/H]~$\sim$~$-$4~dex. Andrievsky et al. (2010) studied the same sample, 
and their results present small differences as compared with Takeda et al. (2009) 
because a direct fit of the profiles was
carried out instead of employing EWs. The spectral synthesis was also
used in the present work, so the NLTE corrections were adopted from
Andrievsky et al. (2010), and to estimate the effect $\Delta_{NLTE}$ on
our sample we used a regression line as a function of metallicities
based on their results, $\Delta_{NLTE}=-0.125\times\hbox{[Fe/H]}-0.6$.
This method is a first rough estimate, because it is known that the
temperature and gravity also play a role in the calculations.

\begin{figure}
\centering
\resizebox{80mm}{!}{\includegraphics[angle=0]{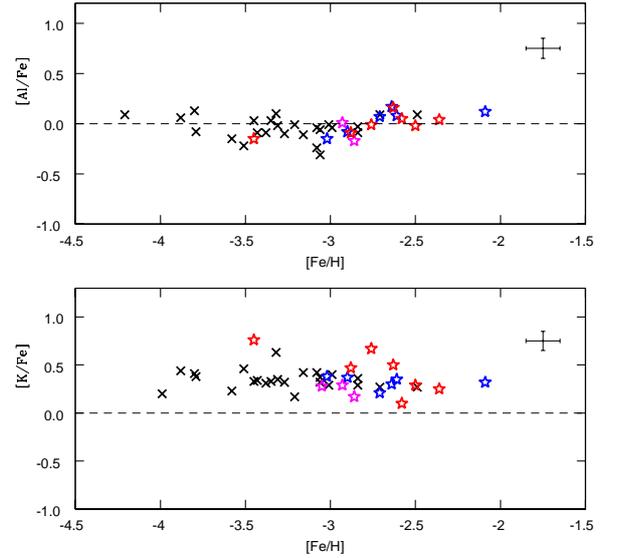}}
\caption{Comparison of the abundance ratios for [Al/Fe] and [K/Fe] obtained in this 
work with the results from Cayrel et al. (2004). NLTE corrections were taken into account. 
Symbols are the same as in Fig. \ref{Mg_Si_fig}.}
\label{oddLP_fig}
\end{figure}

Fig. \ref{oddLP_fig} compares the NLTE results in this work for the
abundance ratios [Al/Fe] and [K/Fe] with the literature (see text
above). The 3D corrections are listed in the text for reference only. 
An excellent agreement was obtained for Al, and again no significant
global difference was clearly found for the r-I stars relative to r-II
stars (or the normal stars). Within the r-I stars, the abundances of
S and Al are remarkably constant, while the K abundances are more widely
scattered. Indeed, some r-I stars analyzed in our sample seem to present
higher K abundances than the general trend. The difference might arise because of 
the difficulty in measuring this element, or because of the rough NLTE 
correction assumed above.

\begin{figure}
\centering
\resizebox{80mm}{!}{\includegraphics[angle=0]{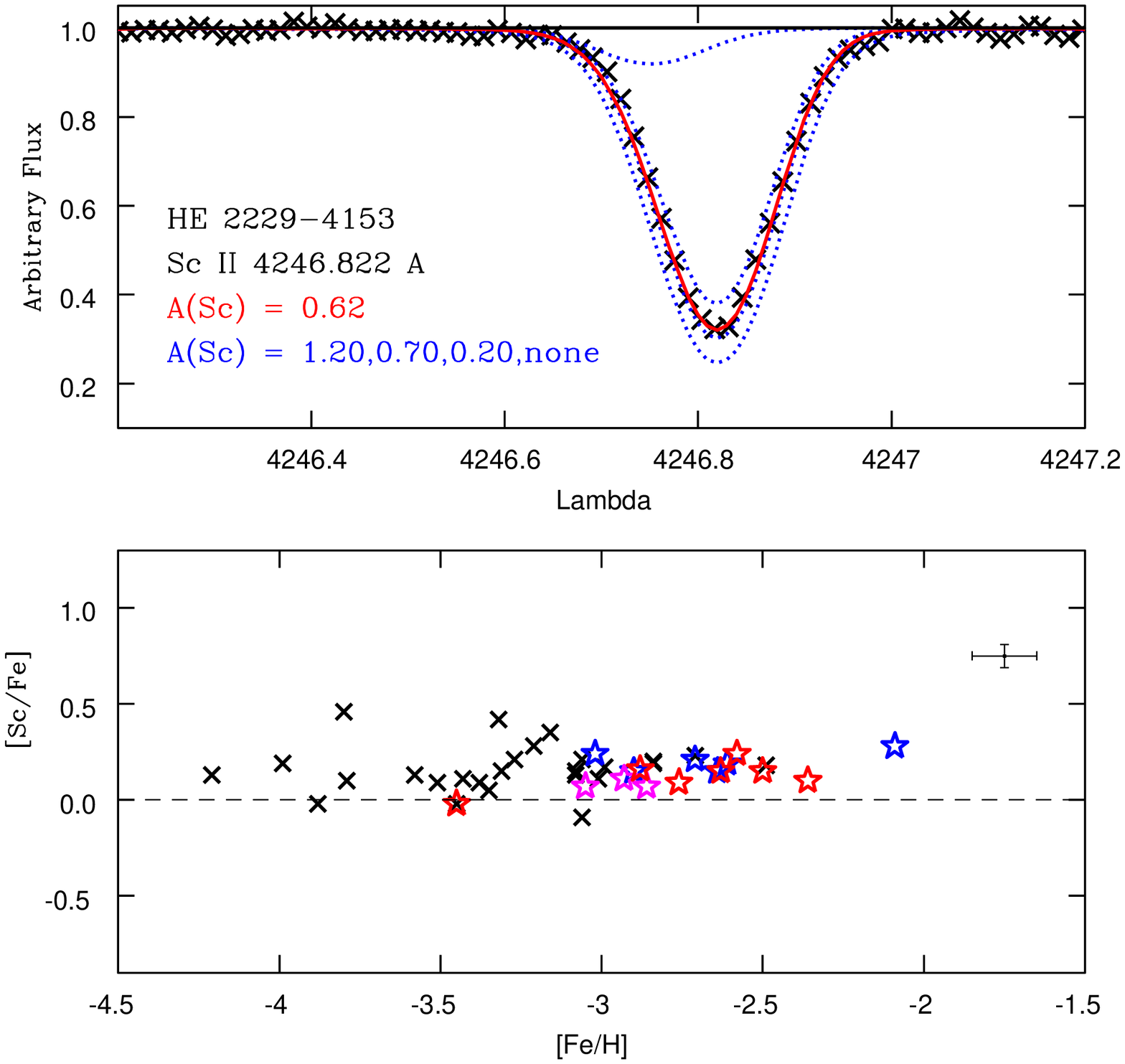}}
\caption{Upper panel: Sc abundances from the \ion{Sc}{II}~4246.82\,{\rm
\AA} line for HE~2229-4153. 
Symbols are the same as in Fig. \ref{Li_fig}. Lower panel: comparison of the abundance ratio [Sc/Fe] 
obtained in this work with the LTE results from Cayrel et al. (2004). 
Symbols are the same as in Fig. \ref{Mg_Si_fig}.}
\label{Sc_fig}
\end{figure}

The scandium abundances were derived from 9 \ion{Sc}{II} lines. Fig.
\ref{Sc_fig} shows the \ion{Sc}{II}~4246.82\,{\rm \AA} line for
HE~2229-4153 (upper panel). The results agree with the
abundances obtained in Cayrel et al. (2004), as shown in Fig.
\ref{Sc_fig} (lower panel). Compared with the other odd-Z elements
(Al and K), the scatter in Sc results is much lower. Based on the
\ion{Sc}{II} lines located at 4246.82~{\rm \AA}, 4320.73~{\rm \AA}, and
4415.56~{\rm \AA}, the 3D effects were studied with the approach
described above, and the estimated corrections obtained are
$\Delta_{3D}=+0.08$~dex and $\Delta_{3D}=+0.01$~dex, using
[Fe/H]~=~$-2$ and [Fe/H]~=~$-3$, respectively.

\subsubsection{Iron-peak elements: V, Cr, Mn, Fe, Co, and Ni}

The equivalent widths presented in Table \ref{EW_measurements} of about
40 lines of \ion{Fe}{I} and 5 lines of \ion{Fe}{II} were employed in the
analysis. For other elements spectrum synthesis was employed.

\begin{figure}
\centering
\resizebox{80mm}{!}{\includegraphics[angle=0]{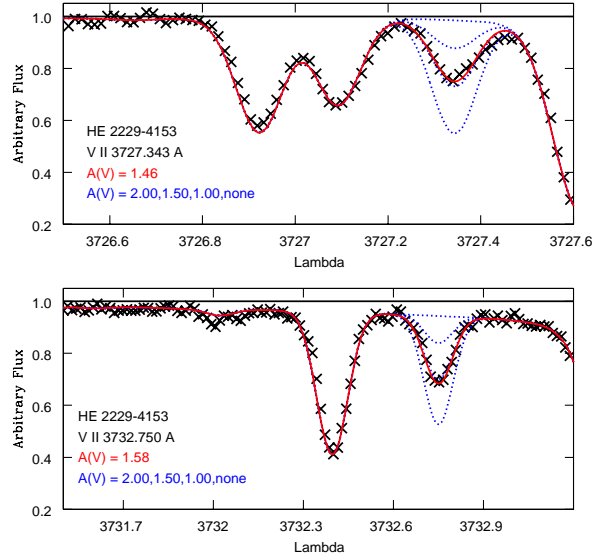}}
\caption{V abundances from the \ion{V}{II}~3727.34\,{\rm \AA} line (upper panel) and line 
\ion{V}{II}~3732.75\,{\rm \AA} line (lower panel) for HE 2229-4153.
Symbols are the same as in Fig. \ref{Li_fig}.}
\label{V_fig}
\end{figure}

For the vanadium abundance derivation, we checked 6 \ion{V}{I} and 11
\ion{V}{II} lines, but for half of our program stars only a few lines are strong
enough to be useful. Fig. \ref{V_fig} shows the line
\ion{V}{II}~3727.34~{\rm \AA} (upper panel) and \ion{V}{II}~3732.75~{\rm
\AA} (lower panel) for HE 2229-4153. All stars exhibit higher abundances
from the \ion{V}{II} lines than in the results from the \ion{V}{I}
lines. The average difference in the present sample is $+$0.23 dex. The
same discrepancy was also found by Siqueira-Mello et al. (2012) for
HD~140283, and by Johnson (2002), who analyzed the V
abundance in 23 metal-poor stars. Johnson suggested that the cause of
this discrepancy might be that hyperfine splitting and NLTE effects had
been neglected in the computations.

To evaluate the 3D effects on the V abundances, the 
\ion{V}{I}~4379.23~{\rm \AA} and \ion{V}{II}~3951.96~{\rm \AA} 
lines were studied with hydrodynamical models, following the 
method already described. For \ion{V}{I}, the results obtained 
are $\Delta_{3D}=-0.19$~dex and $\Delta_{3D}=-0.55$~dex, 
using [Fe/H]~=~$-$2 and [Fe/H]~=~$-$3, respectively. On the other hand, 
the estimated corrections for \ion{V}{II} are $\Delta_{3D}=+0.04$~dex 
and $\Delta_{3D}=+0.05$~dex. As a general trend, the 3D effects are 
stronger for \ion{V}{I} lines and make the differences in abundances 
obtained from neutral and ionized species bigger.

\begin{figure}
\centering
\resizebox{80mm}{!}{\includegraphics[angle=0]{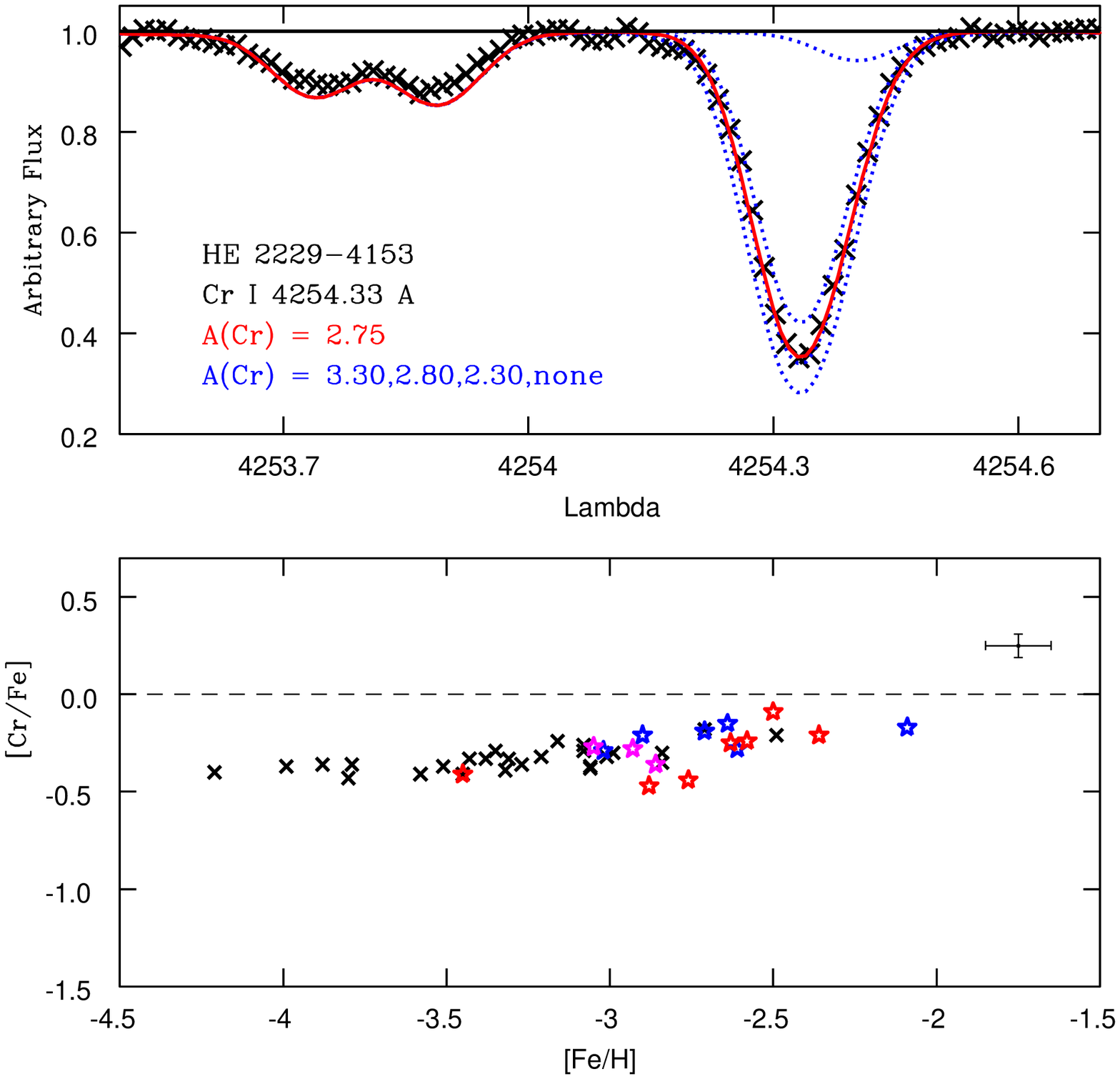}}
\caption{Upper panel: Cr abundances from the \ion{Cr}{I}~4254.33\,{\rm
\AA} line for HE~2229-4153. 
Symbols are the same as in Fig. \ref{Li_fig}. Lower panel: comparison of the abundance ratio 
[Cr/Fe] obtained in this work with the LTE results from Cayrel et al. (2004). 
Symbols are the same as in Fig. \ref{Mg_Si_fig}.}
\label{Cr_fig}
\end{figure}

The abundance of chromium was measured using three lines:
\ion{Cr}{I}~4254.33\,{\rm \AA}, \ion{Cr}{I}~4274.80\,{\rm \AA}, and
\ion{Cr}{I}~4289.72\,{\rm \AA}. Fig. \ref{Cr_fig} shows the results for
\ion{Cr}{I}~4254.33\,{\rm \AA} for HE~2229-4153 (upper panel), and the
abundance ratios [Cr/Fe] obtained in the present work, overlapping the
results of Cayrel et al. (2004) (lower panel). The lower abundance of Cr
with respect to Fe observed in the metal-poor stars was also found in
the present results. The decrease of [Cr/Fe] with [Fe/H] is confirmed by
the present result. However, Bonifacio et al. (2009) were also able to
observe a
\ion{Cr}{II} line, and have shown (see their Fig. 8) an increasing
discrepancy between the Cr abundance deduced from the neutral and the
ionized species when [Fe/H] decreases (see also Lai et al. 2008). As a
consequence, the trend of [Cr/Fe] vs [Fe/H] is very probably due to a
NLTE effect.

\begin{figure}
\centering
\resizebox{80mm}{!}{\includegraphics[angle=0]{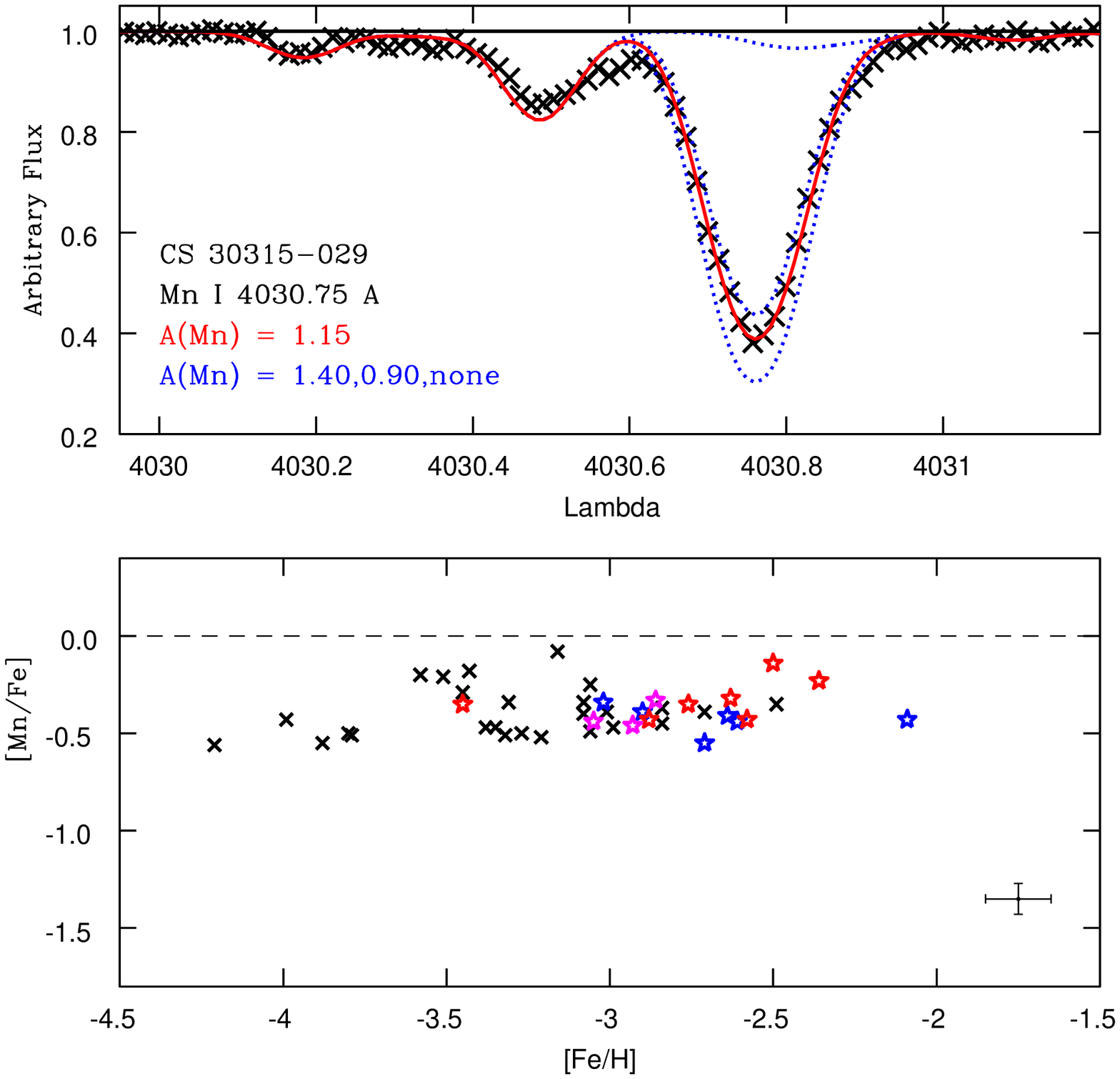}}
\caption{Upper panel: Mn abundance from the \ion{Mn}{I}\,4030.75~{\rm
\AA} line for CS~30315-029. 
Symbols are the same as in Fig. \ref{Li_fig}. Lower panel: comparison of the abundance ratio [Mn/Fe] 
obtained in this work with the LTE results from Cayrel et al. (2004).  
Symbols are the same as in Fig. \ref{Mg_Si_fig}.}
\label{Mn_fig}
\end{figure}

Five lines of Mn I were analyzed to measure the manganese abundance, and
the hyperfine structure was properly taken into account (Kurucz \& Bell
1995, Ivans et al. 2006). However, the \ion{Mn}{I}~4082.94~{\rm \AA}
line is weaker than the others and was measured only in
HE~0524-2055. Moreover, three lines belong to the resonance triplet:
\ion{Mn}{I}~4030.75~{\rm \AA}, \ion{Mn}{I}~4033.06~{\rm \AA}, and
\ion{Mn}{I}~4034.48~{\rm \AA}. Fig. \ref{Mn_fig} shows the line
\ion{Mn}{I}~4030.75~{\rm \AA} in CS~30315-029 (upper panel) as an
example. It is well-known that the abundances derived from this triplet
are systematically lower than the results from subordinate lines in very
metal-poor giant stars. Cayrel et al. (2004) reported an average
difference of 0.4~dex among the stars of the ESO LP ``First Stars,'' and
the present results confirm their value.

We adopted the abundances of Mn deduced from the
subordinate lines and not the resonance lines as the final results,
because the resonance lines are more susceptible to NLTE effects
(Bonifacio et al. 2009). Fig. \ref{Mn_fig} compares our results with
those obtained by Cayrel et al. (2004), which also gave preference to
the abundances deduced from the subordinate lines (lower panel). The
underabundances of Mn with respect to Fe derived for our sample stars agree 
with results from the literature.

The Mn abundances obtained from the resonance triplet are also affected 
more strongly by the 3D modeling. Our estimated computations 
with hydrodynamical models applied to 
\ion{Mn}{I}~4030.75~{\rm \AA}, \ion{Mn}{I}~4033.06~{\rm \AA}, and
\ion{Mn}{I}~4034.48~{\rm \AA} lines obtained the corrections 
$\Delta_{3D}=-0.18$~dex and $\Delta_{3D}=-0.57$~dex, with 
[Fe/H]~=~$-$2 and [Fe/H]~=~$-$3, respectively. The same procedure 
applied to the subordinate line located at 4041.35~{\rm \AA} 
results in $\Delta_{3D}=+0.02$~dex and $\Delta_{3D}=-0.18$~dex.

\begin{figure}
\centering
\resizebox{80mm}{!}{\includegraphics[angle=0]{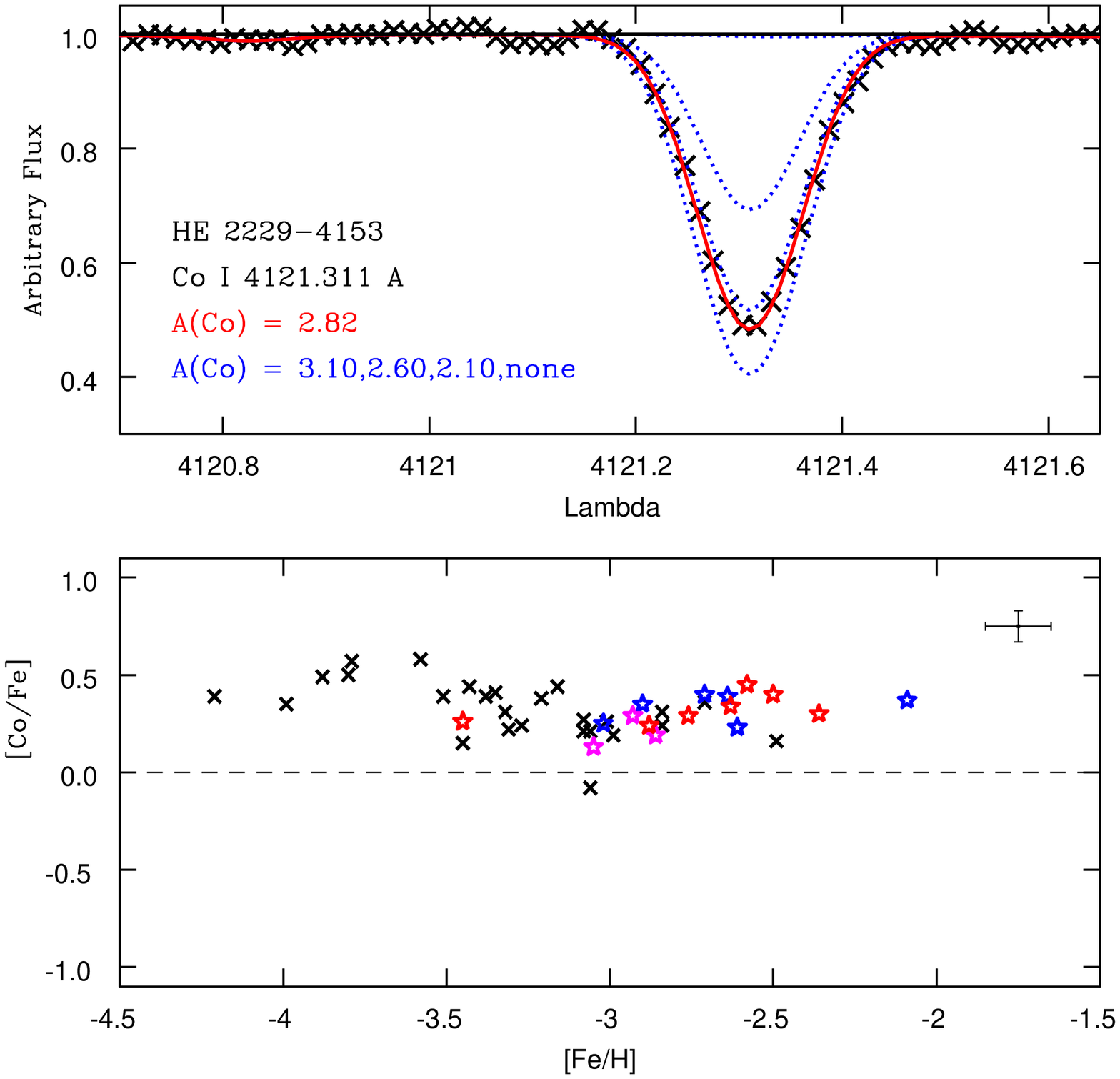}}
\caption{Upper panel: Co abundance from the \ion{Co}{I}~4121.31\,{\rm
\AA} line for HE~2229-4153. 
Symbols are the same as in Fig. \ref{Li_fig}. Lower panel: comparison of the abundance ratio [Co/Fe] 
obtained in this work with the LTE results from Cayrel et al. (2004).  
Symbols are the same as in Fig. \ref{Mg_Si_fig}.}
\label{Co_fig}
\end{figure}

The same good agreement is obtained for cobalt abundances, which were
derived based on four \ion{Co}{I} lines, and Fig. \ref{Co_fig} shows (in the
upper panel) the \ion{Co}{I}~4121.31\,{\rm \AA} line for HE~2229-4153 as an
example. The results for all the stars are compared with the abundances
obtained in Cayrel et al. (2004) (lower panel); in this case there
is an overabundance of Co with respect to Fe.

\begin{figure}
\centering
\resizebox{80mm}{!}{\includegraphics[angle=0]{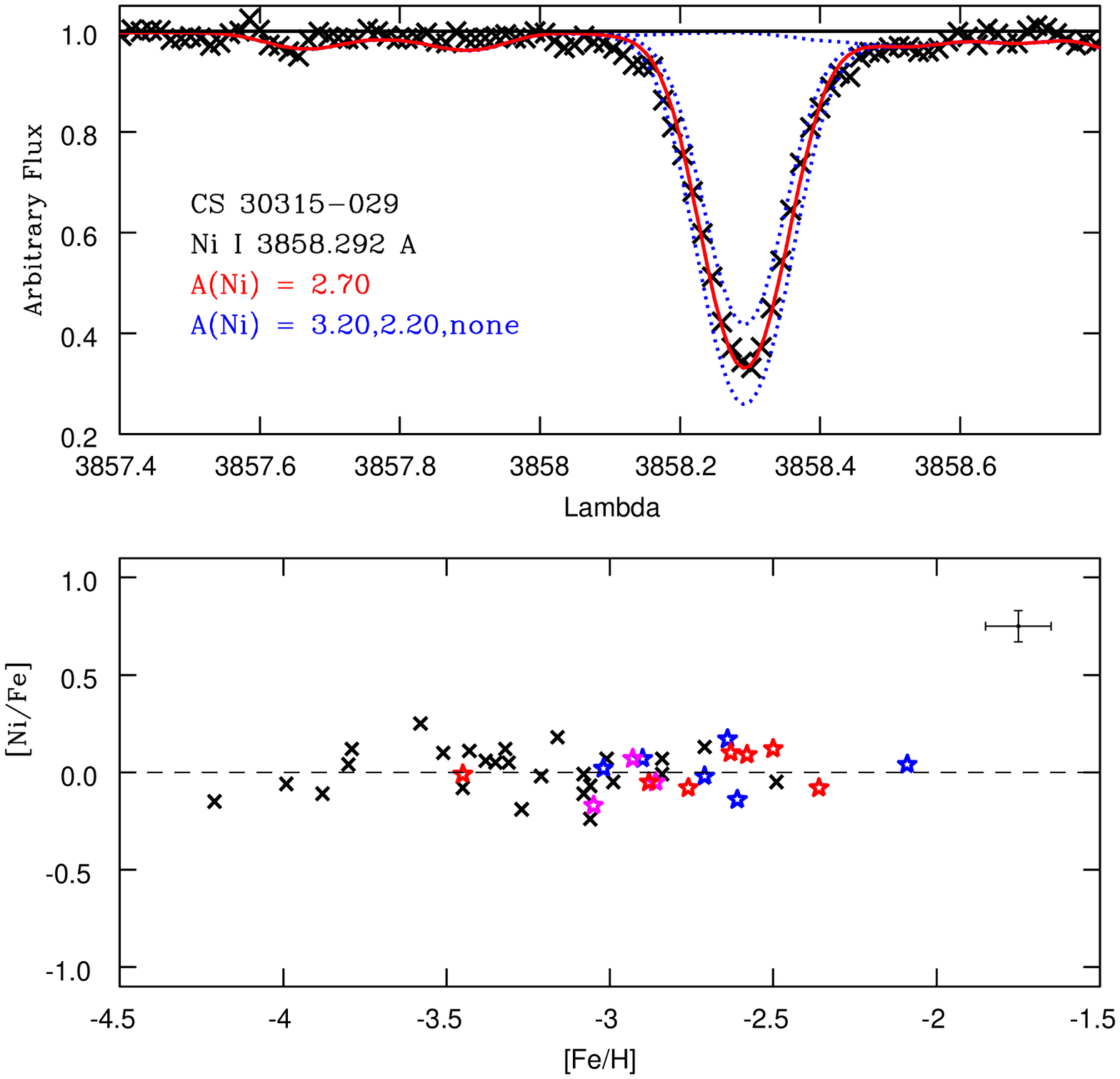}}
\caption{Upper panel: Ni abundance from the \ion{Ni}{I}~3858.29\,{\rm
\AA} line for CS~30315-029. 
Symbols are the same as in Fig. \ref{Li_fig}. Lower panel: comparison of the abundance ratio [Ni/Fe] 
obtained in this work with the LTE results from Cayrel et al. (2004).  
Symbols are the same as in Fig. \ref{Mg_Si_fig}.}
\label{Ni_fig}
\end{figure}

Nickel was the heaviest iron-peak element derived in the present work,
based on 2 \ion{Ni}{I} lines: 3807.14\,{\rm \AA} and 3858.29\,{\rm \AA}.
Fig. \ref{Ni_fig} presents (in the upper panel) the fitting used for
\ion{Ni}{I}~3858.29\,{\rm \AA} for CS~30315-029 as an example, and
compares the results with the abundances from the ESO LP ``First Stars''
in the lower panel. As for other iron-peak elements, there is no
apparent difference in the Ni abundances among the r-I and r-II stars
(or the normal stars).

Zinc is an interesting element because it represents 
(together with other elements) the transition between the iron peak 
and the heavy elements region. However, the Zn abundance is mainly 
measured based on the \ion{Zn}{I}~4722.15\,{\rm \AA} and 
\ion{Zn}{I}~4810.53\,{\rm \AA} lines, not covered by the wavelength ranges 
of the present spectra.

\subsection{Neutron-capture elements}

The line list retained to derive the abundances of the heavy elements
was established carefully, by verifying the intensity of different lines
of a same element, the presence of blends, and the signal-to-noise ratio
(S/N) of the observed spectra. These verifications are particularly
important in the blue regions. The final line list with individual
abundances is listed in Table \ref{linelist}.

To organize the discussion in terms of the r-process, the 
trans-iron elements are grouped into the first peak (from gallium to cadmium), 
the second peak (from barium to tantalum), the third peak 
(from tungsten to bismuth), and the actinides region (thorium and uranium).

\subsubsection{First-peak region elements}

\begin{figure}
\centering
\resizebox{80mm}{!}{\includegraphics[angle=0]{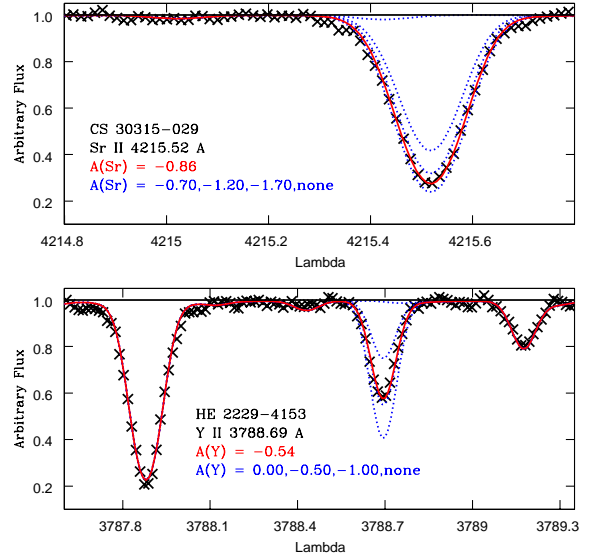}}
\caption{Upper panel: Sr abundance from the 
\ion{Sr}{II}~4215.52\,{\rm \AA} line for CS~30351-029. 
Lower panel: Y abundance from the line \ion{Y}{II}~3788.69\,{\rm \AA}
line for HE~2229-4153. 
Symbols are the same as in Fig. \ref{Li_fig}.}
\label{Sr_Y_fig}
\end{figure}

The strontium abundances were derived from the \ion{Sr}{II}~4077.72\,{\rm
\AA} and \ion{Sr}{II}~4215.52\,{\rm \AA} lines. A third line,
\ion{Sr}{II}~4161.79\,{\rm \AA}, was also checked, but it was too weak in 
most stars and was excluded from the final results. Fig.
\ref{Sr_Y_fig} shows (in the upper panel) the fit of the
\ion{Sr}{II}~4215.52\,{\rm \AA} line for CS~30351-029. 
Using NLTE computations, Andrievsky et al. (2011) studied the Sr 
abundances in the sample observed during the LP ``First Stars'', 
and they obtained the average correction $\Delta_{NLTE}=+0.09$~dex.

\begin{figure}
\centering
\resizebox{80mm}{!}{\includegraphics[angle=0]{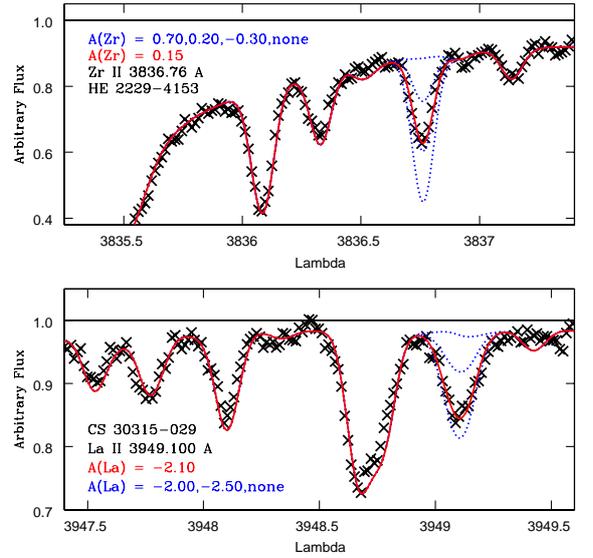}}
\caption{Upper panel: Zr abundance from the \ion{Zr}{II}~3836.76\,{\rm
\AA} line for HE~2229-4153. 
Lower panel: La abundance from the line \ion{La}{II}~3949.10\,{\rm \AA}
line for CS~30351-029. 
Symbols are the same as in Fig. \ref{Li_fig}.}
\label{Zr_La_fig}
\end{figure}

The yttrium abundances were obtained after checking five \ion{Y}{II} lines:
3774.33\,{\rm \AA}, 3788.69\,{\rm \AA}, 3818.34\,{\rm \AA}, 3950.35\,
{\rm \AA}, and 4398.01\,{\rm \AA}. In some stars, one or more lines are
too weak and were not used. Fig. \ref{Sr_Y_fig} presents the result for
the \ion{Y}{II}~3788.69\,{\rm \AA} line for HE~2229-4153. For zirconium,
the abundances were derived from four \ion{Zr}{II} lines:
\ion{Zr}{II}~3836.76\,{\rm \AA}, \ion{Zr}{II}~4161.21\,{\rm \AA}, 
\ion{Zr}{II}~4208.98\,{\rm \AA}, and \ion{Zr}{II}~4317.30\,{\rm \AA}.
Fig. \ref{Zr_La_fig} shows (in the upper panel) the result obtained for
the \ion{Zr}{II}~3836.76\,{\rm \AA} for HE~2229-4153. 

\begin{figure}
\centering
\resizebox{90mm}{!}{\includegraphics[angle=0]{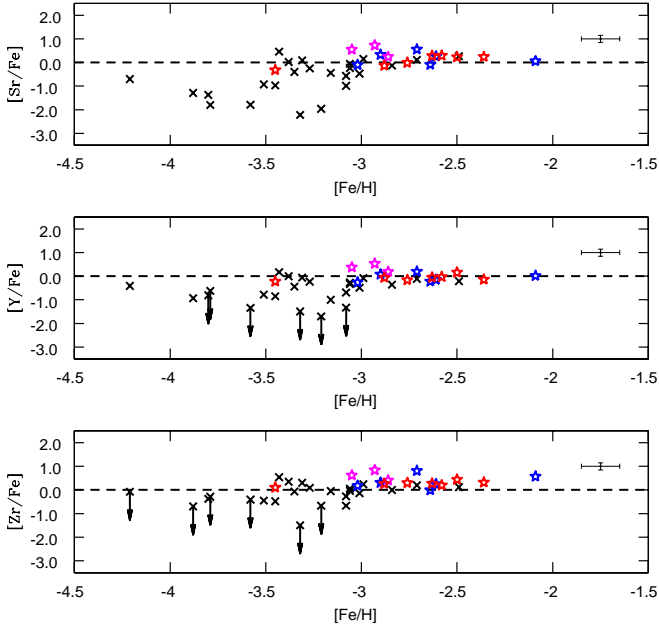}}
\caption{[Sr, Y, Zr/Fe] vs. [Fe/H] for the present sample compared with results for metal-poor stars 
by Fran\c{c}ois et al. (2007). Symbols are the same as in Fig. \ref{Mg_Si_fig}. Upper limits from Fran\c{c}ois et al. 
are indicated as black arrows.}
\label{first_peak}
\end{figure}

Fig. \ref{first_peak} shows the abundances ratios [Sr/Fe], [Y/Fe], and
[Zr/Fe], as a function of metallicity, [Fe/H], for the present
sample. These ratios are also compared with the results obtained by Fran\c{c}ois
et al. (2007), showing excellent agreement. It is interesting to note
that the r-II stars are located around [Fe/H]~$\sim$~$-$3, and most of
r-I stars appear at higher metallicities. The abundance ratios are
stable, but the scatter seems to become important for [Fe/H]~$<$~$-$3 
in normal stars. For the r-I and r-II stars, a remarkably consistent 
first-peak abundance is observed across the entire metallicity range.

The best lines to derive the molybdenum abundances are located in the
ultraviolet region, but it was possible to derive upper limits for two sample
stars, CS~30315-029 and HE~0524-2055, using the 
\ion{Mo}{I} line located at 3864.10\,{\rm \AA}, which is rather weak. 
In the case of ruthenium, the abundance was derived for three stars:
CS~30315-029, HE~0057-4541, and HE~0524-2055. The \ion{Ru}{I} lines used 
were 3498.94\,{\rm \AA}, 3728.02\,{\rm \AA}, and 3799.35\,{\rm \AA}, which
also allowed us to derive upper limits for HE~0240-0807 and HE~2229-4153.
However, the difficulty in placing the continuum, as well as the blends
with strong lines, indicate that the reported Ru abundances should be taken with
caution.

The same difficulty due to the weakness of the lines was found for
palladium. Based on two \ion{Pd}{I} lines, located at 3404.58\,{\rm \AA}
and 3516.94\,{\rm \AA}, only upper limits were derived for CS~30315-029
and HE~2229-4153.
 
\subsubsection{Second-peak region elements}

The barium abundance was derived from the \ion{Ba}{II} 4554.02\,{\rm
\AA} line. This wavelength is not covered by our new high-resolution
spectra, thus the HERES spectra were used. The exception was
CS~30315-029, for which a high-resolution spectrum obtained in the
framework of the LP ``First Stars'' -- and not analyzed previously --
was used. The hyperfine structure was taken into account based on
McWilliam (1998). The \ion{Ba}{II} 3891.78\,{\rm \AA} and \ion{Ba}{II}
4130.64\,{\rm \AA} lines, which are in the wavelength range covered by
the new data, were also checked, but they are too weak to yield reliable
abundances. The barium abundances derived are similar to those reported
by B05 (Table \ref{atm_barklem}). Andrievsky et al. (2009) used NLTE
computations to obtain the Ba abundances in the sample of the LP ``First
Stars,'' reporting an average correction of $\Delta_{NLTE}=+0.17$~dex.
However, it is important to note that this correction strongly depends 
on the effective temperature and metallicity (see Fig. 4 in their
paper).

For lanthanum abundances, we checked 14 \ion{La}{II} lines using
experimental oscillator strengths from Lawler et at. (2001a) and
hyperfine structures from Ivans et al. (2006). Fig. \ref{Zr_La_fig}
shows the fit used for the \ion{La}{II}~3949.10\,{\rm \AA} line for
CS~30351-029 (lower panel).

\begin{figure}
\centering
\resizebox{80mm}{!}{\includegraphics[angle=0]{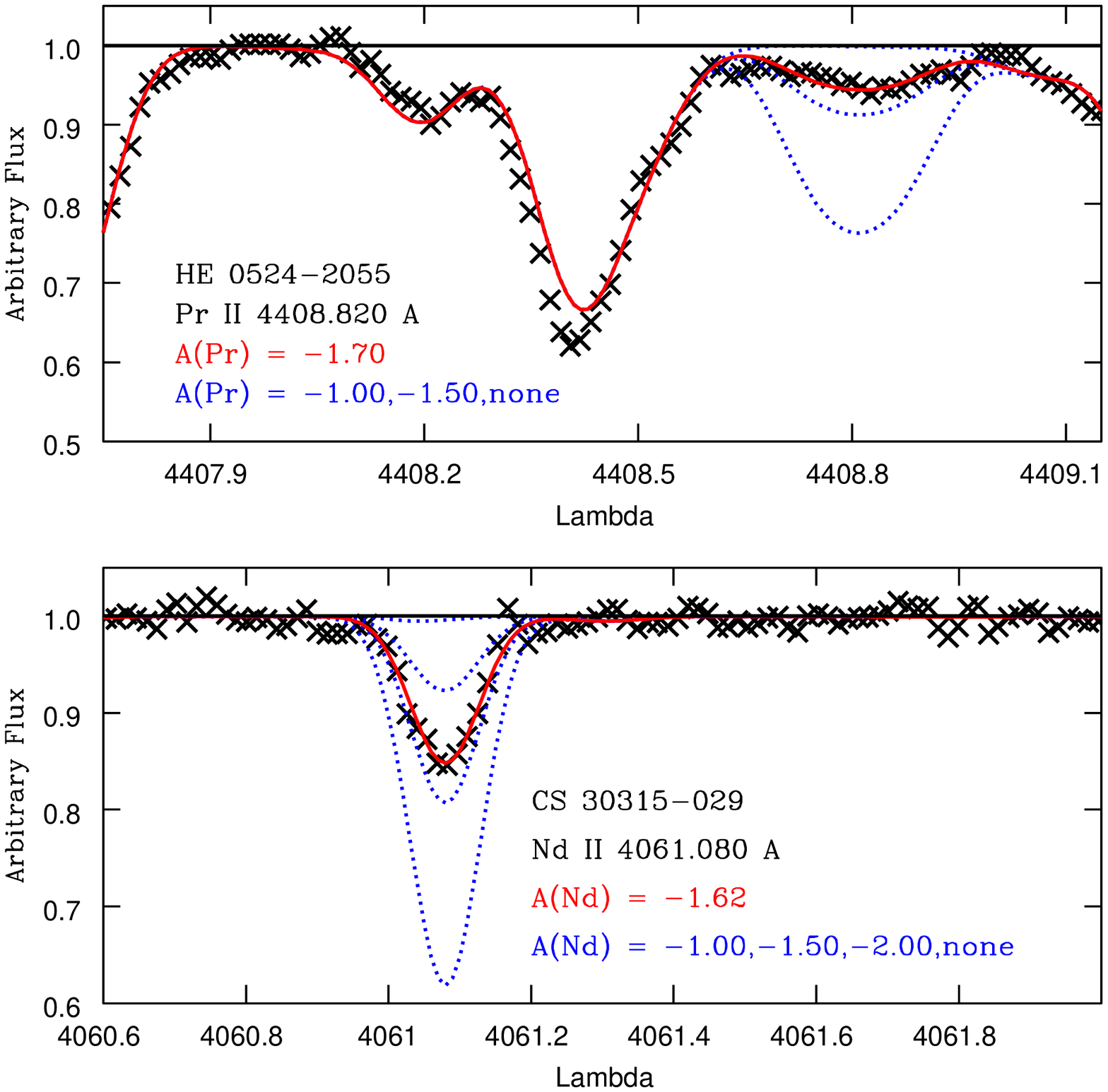}}
\caption{Upper panel: Pr abundance from the \ion{Pr}{II}~4408.82\,{\rm
\AA} line for HE~0524-2055. 
Lower panel: Nd abundance from the \ion{Nd}{II}~4061.08\,{\rm \AA} line
for CS~30351-029. 
Symbols are the same as in Fig. \ref{Li_fig}.}
\label{Pr_Nd_fig}
\end{figure}

Based on consideration of 15 \ion{Ce}{II} lines, the abundance of cerium
was derived for six of our program stars. For HE~0105-6141 only an upper limit was obtained. Improved
laboratory transition probabilities were adopted from Lawler et al.
(2009). The praseodymium abundance was derived only for CS~30315-029,
HE~0240-0807, HE~0524-2055, and HE~2229-4153, after checking 11
\ion{Pr}{II} lines. Upper limits were obtained for the other stars. New
oscillator strengths were obtained from Li et al. (2007) or Ivarsson et
al. (2001), and the hyperfine structures were calculated based on Sneden
et al. (2009). Fig. \ref{Pr_Nd_fig} shows in the upper panel the result
for the \ion{Pr}{II}~4408.82\,{\rm \AA} line for HE~0524-2055.

For neodymium, the abundance derivation was robust for all of our
program stars, based on 16 \ion{Nd}{II} lines, with experimental
oscillator strengths from Den Hartog et al. (2003). Fig. \ref{Pr_Nd_fig}
presents (in the lower panel) the \ion{Nd}{II}~4061.08\,{\rm \AA} line
fit for CS~30351-029, as an example. The samarium abundance was derived
after checking ten \ion{Sm}{II} lines, but for HE~0516-3820 only an upper
limit was obtained. Note that the odd-Z elements have lower abundances
than the even-Z elements.

\begin{figure}
\centering
\resizebox{80mm}{!}{\includegraphics[angle=0]{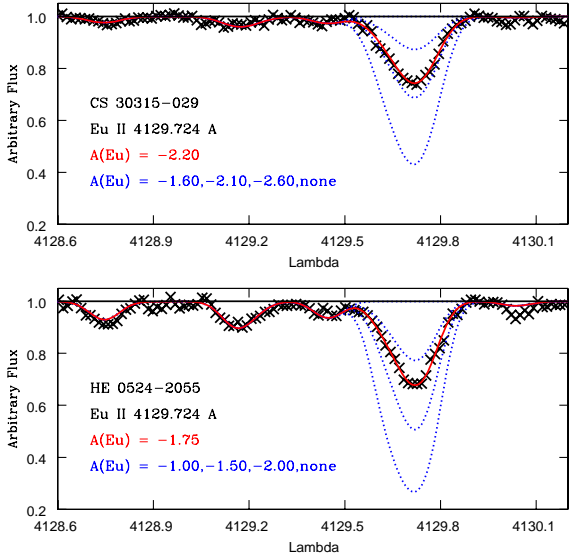}}
\caption{Europium abundance from the line \ion{Eu}{II}~4129.72~{\rm \AA}
line for CS~30351-029 
(upper panel) and for HE~0524-2055 (lower panel). 
Symbols are the same as in Fig. \ref{Li_fig}.}
\label{Eu_fig}
\end{figure}

Europium was derived for all of our program stars, using five \ion{Eu}{II} lines in
the best cases, and the hyperfine structures for the transitions were
obtained from Kurucz data
base\footnote{http://kurucz.harvard.edu/atoms/6301/}. Europium presents
two stable isotopes, $^{151}$Eu and $^{153}$Eu, and the solar system 
isotopic ratio of 0.48:0.52 (Arlandini et al. 1999) for Eu 151:153 was
chosen for the calculation. Fig. \ref{Eu_fig} shows the results for
CS~30351-029 (upper panel) and HE~0524-2055 (lower panel), using
\ion{Eu}{II}~4129.72\,{\rm \AA} line. 
Mashonkina et al. (2012) present NLT corrections for Eu abundances 
in cool stars, but the estimated values are lower than $\Delta_{NLTE}=+0.10$~dex 
for our sample stars. 

\begin{figure}
\centering
\resizebox{90mm}{!}{\includegraphics[angle=0]{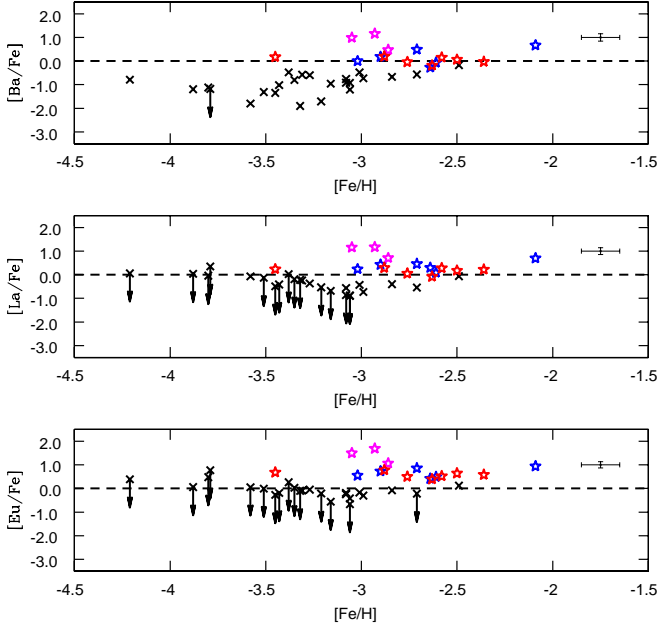}}
\caption{[Ba, La, Eu/Fe] vs. [Fe/H] for the present sample compared with results for metal-poor stars 
by Fran\c{c}ois et al. (2007). Symbols are the same as in Fig. \ref{Mg_Si_fig}. Upper limits from Fran\c{c}ois et al. 
are indicated as black arrows.}
\label{second_peak}
\end{figure}

Fig. \ref{second_peak} shows the abundances ratios [Ba/Fe], [La/Fe], and
[Eu/Fe], as a function of metallicity, [Fe/H], for the present sample. 
These ratios are also compared with the results obtained by 
Fran\c{c}ois et al. (2007).

\begin{figure}
\centering
\resizebox{80mm}{!}{\includegraphics[angle=0]{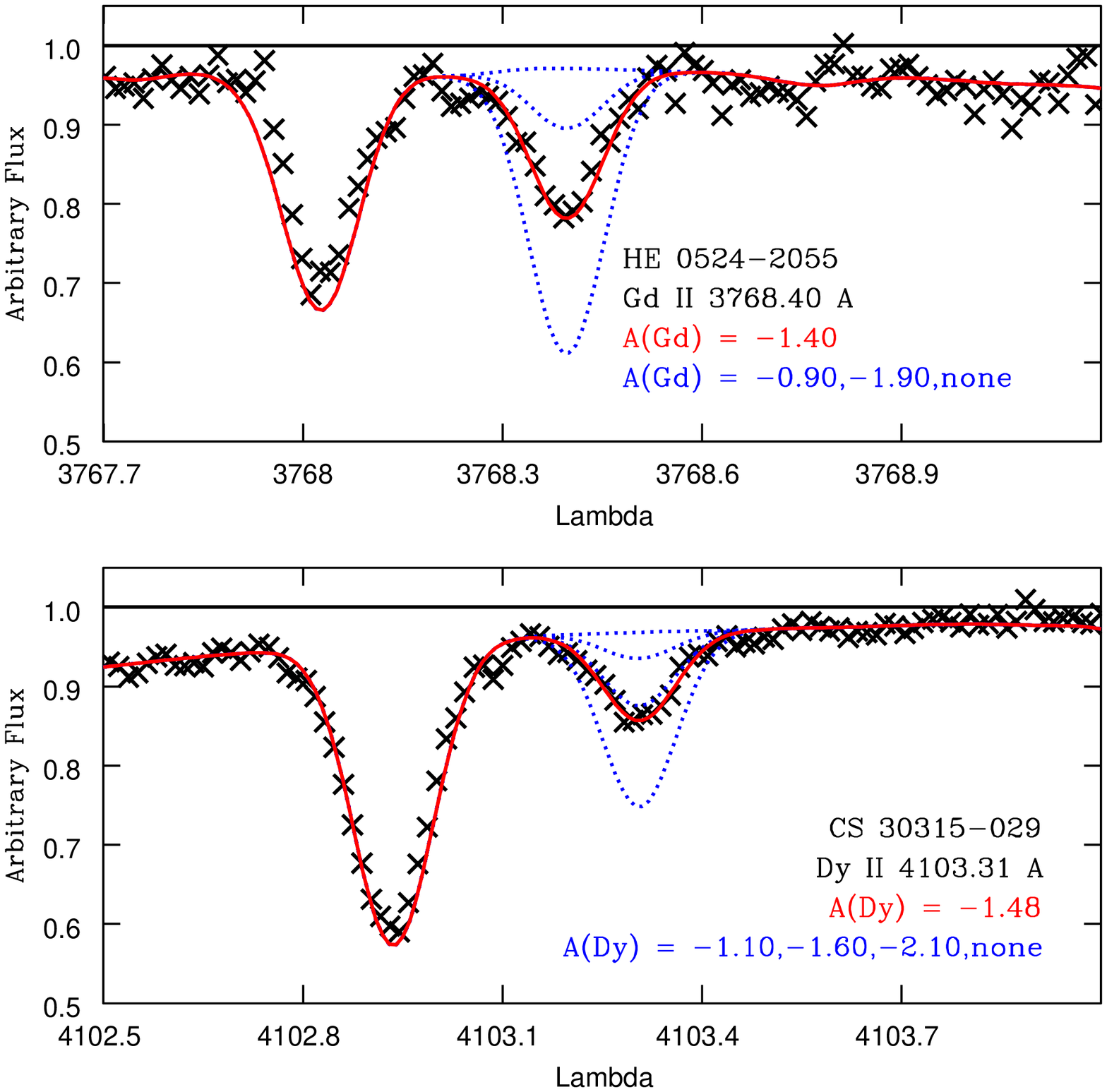}}
\caption{Upper panel: Gd abundance from the \ion{Gd}{II}~3768.40\,{\rm
\AA} line for HE~0524-2055. 
Lower panel: Dy abundance from the \ion{Dy}{II}~4103.31\,{\rm \AA}
line for CS~30351-029. 
Symbols are the same as in Fig. \ref{Li_fig}.}
\label{Gd_Dy_fig}
\end{figure}

After checking seven \ion{Gd}{II} lines, the gadolinium abundance was
derived for six of our program stars. For HE~0105-6141, only an upper
limit was obtained. Fig. \ref{Gd_Dy_fig} shows (in the upper panel) the
result for the \ion{Gd}{II}~3768.40\,{\rm \AA} line for HE~0524-2055. The
terbium abundance was derived only for CS~30315-029, and an upper limit
was obtained for HE~0240-0807 and HE~0524-2055, based on three weak
\ion{Tb}{II} lines with new atomic transition probabilities from Lawler
et al. (2001b), and hyperfine structure from Lawler et al. (2001c). 

The dysprosium abundance was derived from nine \ion{Dy}{II} lines in the
best case. Fig. \ref{Gd_Dy_fig} presents the  
\ion{Dy}{II}~4103.31\,{\rm \AA} line for CS~30351-029 as an example of
the fitting procedure. For the holmium abundance, the
\ion{Ho}{II}~3796.74\,{\rm \AA} and \ion{Ho}{II}~3810.73\,{\rm \AA} 
lines were checked using the hyperfine structure from Worm et al. (1990), as reported 
and adopted by Lawler et al. (2004), who provided \loggf's for these lines. 
An abundance estimate was not possible to obtain for HE~0105-6141, and for HE~0516-3820 
an upper limit was obtained. 

We also checked five \ion{Er}{II} lines, three \ion{Tm}{II} lines, and one
\ion{Yb}{II} line. The erbium abundance was derived for all of our
program stars except HE~0105-6141, and for HE~0516-3820 only an upper limit was
obtained. A robust result for the thulium abundance was derived only for
CS~30315-029, and upper limits only were obtained for HE~0057-4541, HE~0240-0807, and
HE~2229-4153. From the \ion{Yb}{II}~3694.20\,{\rm \AA} line, the
ytterbium abundance was derived for all of our program stars. The heavier
elements from the second-peak region were explored, but no useful lines were
found.

\subsubsection{Third-peak region elements and the actinides}

Upper limits on the osmium abundance were derived for CS~30315-029 and
HE~0240-0807, based on the lines \ion{Os}{I}~4260.85\,{\rm \AA} and
\ion{Os}{I}~4420.47\,{\rm \AA}. For the iridium abundance, the lines 
\ion{Ir}{I}~3513.65\,{\rm \AA} and \ion{Ir}{I}~3800.12\,{\rm \AA} were
checked, and upper limits were obtained for CS~30315-029, HE~057-4541,
HE~0240-0807, and HE~0524-2055.

The \ion{Th}{II}~4019.13\,{\rm \AA} line was used to derive the thorium
abundance. A robust result was obtained for CS~30315-029, and upper 
limits were obtained for HE~057-4541, HE~0240-0807, and HE~0524-2055. 
According to Mashonkina et al. (2012), an NLTE correction of 
$\Delta_{NLTE}=+0.12$~dex should be applied for the Th abundance 
in a star with \Teff~=~4500~K, \logg~=~1.0~[cgs], and [Fe/H]~=~$-3$, 
similar with the atmospheric parameters adopted for CS~30315-029.

\section{Discussion}

\begin{figure}
\centering
\resizebox{90mm}{!}{\includegraphics[angle=0]{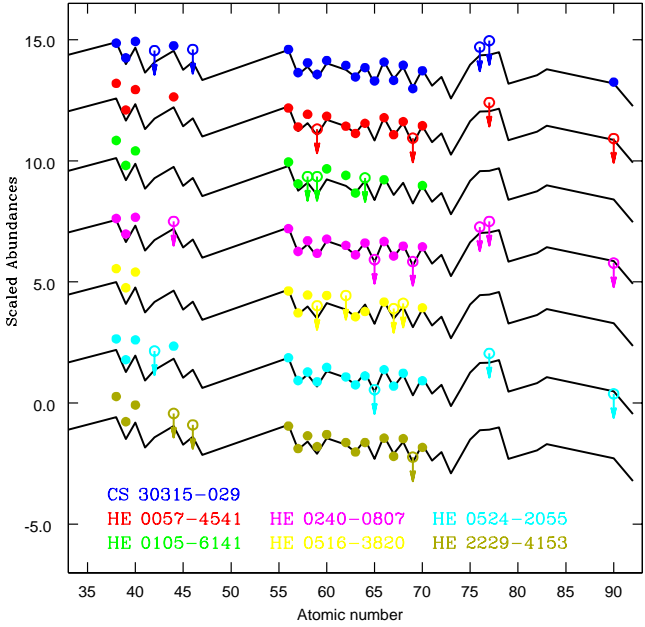}}
\caption{Comparison of the element abundances obtained in this work with
the abundance pattern of the extremely metal-poor r-II 
uranium-rich star CS~31082-001 (black line). An arbitrary offset was
applied to distinguish the individual program stars. The error bars are 
smaller than the dot sizes.}
\label{compara}
\end{figure}

Fig. \ref{compara} compares for each of our program stars the heavy-element abundances
obtained in this work with the abundance pattern of the extremely metal-poor r-II
uranium-rich star CS~31082-001, for which spectra from UVES and
STIS/Hubble Space Telescope were used to derive accurate abundances
(Barbuy et al. 2011; Siqueira-Mello et al. 2013). This star is taken 
to be representative of r-II stars, whose abundance patterns are 
almost identical, except for those beyond the third-peak region (Pb and Th). 
The abundances for CS~31082-001 were rescaled to match the europium 
abundance for each r-I star, and the results were shifted vertically to
aid readability. This comparison shows a remarkable agreement of the
abundance pattern of the second-peak region elements, suggesting a common
origin for these elements in these two classes of stars (r-I and r-II).
In other words, in terms of chemical enrichment of the second-peak region 
neutron-capture elements, there seems to be little or no difference
between r-I and r-II stars. Usually, the main r-process is invoked to
explain the origin of this pattern. 

\begin{figure}
\centering
\resizebox{90mm}{!}{\includegraphics[angle=0]{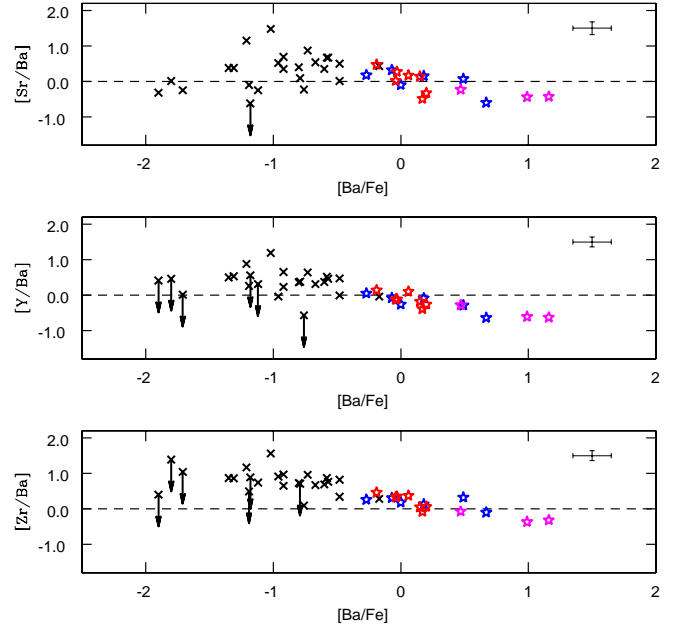}}
\caption{[Sr, Y, Zr/Ba] as functions of [Ba/Fe] obtained in this work 
compared with results for metal-poor stars by Fran\c{c}ois et al.
(2007). Symbols are the same as in Fig. \ref{Mg_Si_fig}.}
\label{first_second}
\end{figure}

By way of comparison, Fig. \ref{compara} also shows the enhancement of
first peak region elements for r-I stars with respect to r-II stars. The
abundance ratios [Sr/Fe], [Y/Fe], and [Zr/Fe] agree well with 
the data of Fran\c{c}ois et al. (2007), as shown in Fig.
\ref{first_peak}, and our data confirm that the dispersion of these
ratios is remarkably low for stars with [Fe/H]~$>$~$-$3. 
To evaluate the overabundance of the first-peak elements compared with 
the general abundance level of the main r-process,
Fig. \ref{first_second} shows the abundance ratios [Sr/Ba], [Y/Ba], 
and [Zr/Ba], as functions of [Ba/Fe]. These comparisons, already done 
in the literature (e.g., B05; Fran\c{c}ois et al. 2007), represent the ratio
[first-peak/second-peak], as a function of r-process enrichment. 
The [first-peak/second-peak] ratio increases as [Ba/Fe] values
decreases, until $\sim-1.5$, thus r-I stars have intermediate
values of [Sr, Y, Zr/Ba] between r-II stars and the stars without r-process
enhancement (with low values of [Ba/Fe]). In addition, a break in the
abundance trends may be present at the lowest [Ba/Fe] ratios. The 
results point toward a production of first-peak elements that is, at
least in part, independent of the second-peak element production.

\begin{figure}
\centering
\resizebox{90mm}{!}{\includegraphics[angle=0]{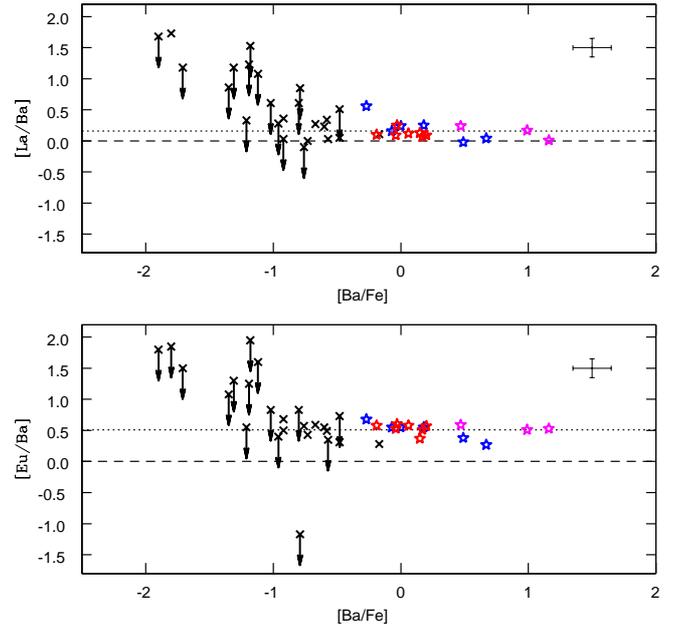}}
\caption{[La/Ba] (upper panel) and [Eu/Ba] (lower panel) as functions
of [Ba/Fe] obtained in this work compared with results for 
metal-poor stars by Fran\c{c}ois et al. (2007). Symbols 
are the same as in Fig. \ref{Mg_Si_fig}. The dotted lines 
represent [La/Ba]~=~$+$0.16 and [Eu/Ba]~=~$+$0.51.}
\label{la_eu}
\end{figure}

Among the elements from the second-peak region, the abundance ratios as a
function of r-process enrichment present constant values. Fig.
\ref{la_eu} shows the [La/Ba] and [Eu/Ba] abundance ratios, as a function
of [Ba/Fe]. Setting aside upper limits, these ratios give the average values 
[La/Ba]~=~$+$0.16 and [Eu/Ba]~=~$+$0.51 (dotted lines in the figure). 
As comparison, the solar system r-process abundances, according to Simmerer et al. (2004), 
give the abundance ratios [La/Ba]$_{\rm r}$~=~$+$0.17 and [Eu/Ba]$_{\rm r}$~=~$+$0.70. 
On the other hand, the abundance pattern obtained in the atmosphere of CS~31082-001, a template of main 
r-process (r-II) star, give the ratios [La/Ba]$_{\rm r-II}$~=~$+$0.01 and [Eu/Ba]$_{\rm r-II}$~=~$+$0.53. 
The flatness of the second-peak elemental abundances for both r-I and r-II stars, as can be 
seen in Fig. \ref{la_eu}, is evidence that the main r-process similarly affects the 
elements in r-I and r-II stars, because the dispersion around these mean lines can account for the errors.

\begin{figure}
\centering
\resizebox{90mm}{!}{\includegraphics[angle=0]{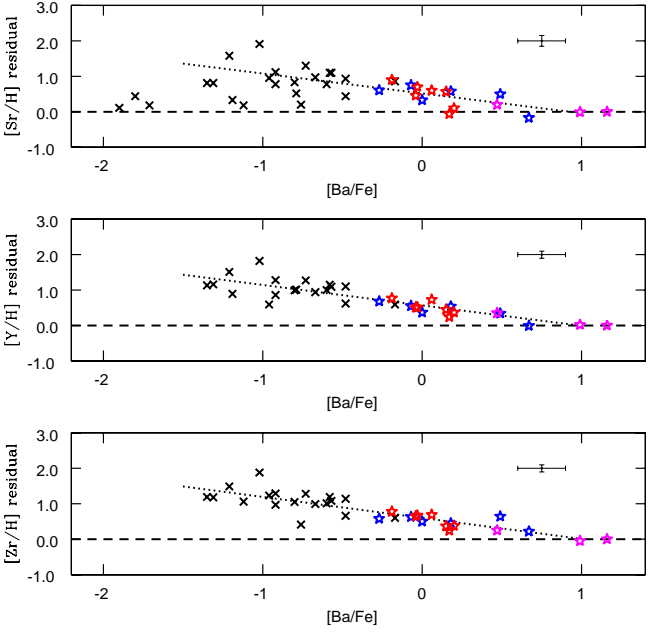}}
\caption{Residual [Sr, Y, Zr/H] abundance ratios in comparison with the r-II star CS~31082-001 
as a function of [Ba/Fe] for our program stars compared with results for metal-poor 
stars by Fran\c{c}ois et al. (2007). Symbols are the same as in Fig. \ref{Mg_Si_fig}.}
\label{residual}
\end{figure}

Indeed, the r-II stars are considered to be enriched in heavy elements
by a main r-process component, and the difference observed between
the abundance enhancement of first-peak elements in r-I relative to r-II
stars, instead, should be due to a so-called weak component 
(e.g., Wanajo \& Ishimaru 2006; Montes et al. 2007). We computed the residual abundances 
of Sr, Y, and Zr in our sample stars and in the sample from Fran\c{c}ois et al. (2007), 
with respect to CS~31082-001, which was taken to be representative of r-II stars. 
Fig. \ref{residual} shows the results, and the dotted lines represent the average trend 
observed in the residual abundances. For [Sr/H], the stars with the lowest 
[Ba/Fe] values were not taken into account because of the break in the trend that 
might be present. The results present a continuum behavior for the enhancement 
in weak r-process, from the r-II stars to the normal metal-poor stars, with 
an intermediate value for the r-I stars, therefore showing that the classification 
in r-I and r-II, as established originally by Beers \& Christlieb (2005), could be 
seen as \textit{ad hoc}. This pattern should appear because of differences in the level 
of main r-process, which 
hide the weak r-process products with different efficiencies. The discussion suggests 
that the r-I stars therefore appear to exhibit a clear fraction of weak r-process 
in their abundance patterns, but in the r-II stars this weak component should be 
completely hidden.

The slopes $a$ (and coefficients of determination R$^{2}$) of
the regression lines are $a$=$-$0.5588 (R$^{2}$=0.5498), $a$=$-$0.5730
(R$^{2}$=0.7382), and $a$=$-$0.5910 (R$^{2}$=0.7448), for [Sr/H], [Y/H],
and [Zr/H], respectively. These results show that the contributions from
the weak r-process to the abundances of Sr, Y, and Zr are very
similar. For the other first peak region elements, it was not possible to obtain
a clear picture because of the low number of stars with derived abundances. 
Considering [Ba/Fe]$\sim-1.5$ as the minimum limit for the presence of 
weak r-process, according to the break observed in [Sr/H], it is possible 
to estimate the maximum contribution from weak r-process expected for these 
elements by comparing with the pattern observed in r-II stars. 
This exercise gives the absolute abundances A(Sr)$_{max}=4.28$, 
A(Y)$_{max}=3.64$, and A(Zr)$_{max}=4.07$, which should represent the maximum 
abundances that can be produced by the weak process in the chemical content of 
metal-poor stars. It is important to note that the estimates were made assuming that 
there is no weak r-process in stars with [Ba/Fe]~$<$~$-1.5$.

The classification of r-I and r-II stars used in this work 
is strictly based on the criteria of Beers \& Christlieb (2005). 
The three r-II stars from the LP ``First Stars'' are CS~22892-052
([Eu/Fe]~=~$+$1.49, [Ba/Eu]~=~$-$0.48), CS~22953-003
([Eu/Fe]~=~$+$1.05, [Ba/Eu]~=~$-$0.56), and CS~31082-001
([Eu/Fe]~=~$+$1.69, [Ba/Eu]~=~$-$0.53). However, as shown in Fig.
\ref{first_second}, the ratios [first-peak/second-peak] exhibit the
lowest values in CS~22892-052 and CS~31082-001, which can be assumed 
as the ratio that arises only from the main r-process, whereas 
CS~22953-003 appears to exhibit an excess of first-peak elements, 
typical of r-I stars. If [first-peak/second-peak] ratio could be 
used as an additional criterion to classify stars in r-I or r-II, 
CS~22953-003 should be re-classified as an r-I star. On the other hand, 
the reverse is also true: BD+17:3248 (Fran\c{c}ois et al. 2007) is an 
r-I star ([Eu/Fe]~=~$+$0.93, [Ba/Eu]~=~$-$0.24) showing no evidences 
of weak r-process component in the abundance pattern 
(see Fig. \ref{first_second} and Fig. \ref{residual}), 
and should be re-classified as an r-II star with the additional 
criterion. It is important to note that the cases of borderline stars 
can be now discussed owing to the good quality of the data. In this sense, 
it is possible to find a real difference between the two 
groups of stars: the fraction of weak r-process is not relevant in 
the abundance pattern of r-II stars, but becomes evident in r-I stars. 
In other words, r-I and r-II stars are now confirmed to indicate different 
chemical enrichment histories.

\begin{figure}
\centering
\resizebox{90mm}{!}{\includegraphics[angle=0]{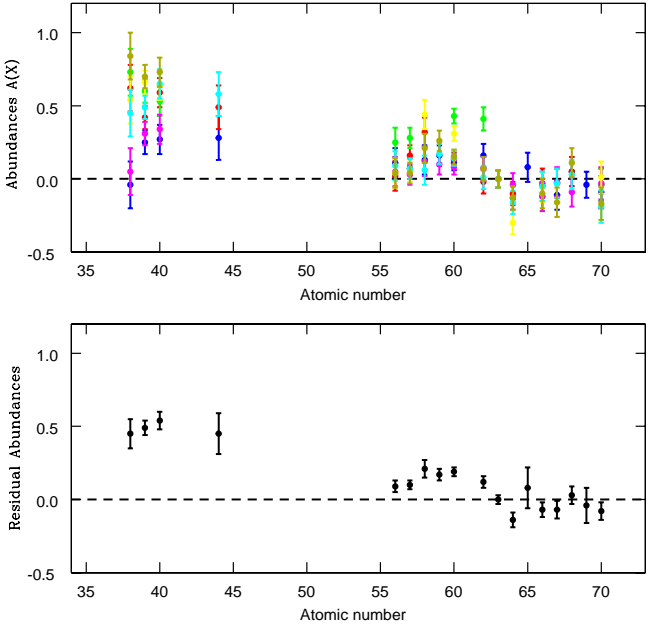}}
\caption{Upper panel: the residual abundances with respect to an average 
pattern based on the abundances of CS~31082-001 and CS~22892-052 in the present sample. 
Lower panel: the average residual abundance. Symbols are the same as in Fig. \ref{compara}.}
\label{residual_weak}
\end{figure}

Fig. \ref{residual_weak} shows the residual abundances in the sample stars 
with respect to an average pattern based on the abundances of CS~31082-001 and 
CS~22892-052 (upper panel), as well as an average of the results (lower panel). 
The dashed lines represent the null value to indicate the agreement between 
the abundance level in the r-I stars and in the r-II star. The abundance pattern shown 
in the lower panel of Fig. \ref{residual_weak} should represent the products from 
this incomplete r-process mechanism, and from the present results, a contribution 
to the abundances of some second-peak region elements from the weak r-process 
may be also present.

As a general conclusion, we suggest that to understand the
nucleosynthesis of the r-elements, it is important to study r-I and r-II stars together.

\subsection{Radioactive chronometry for CS~30315-029}

CS~30315-029 is the only star in the sample in which Th was detected. 
In Hill et al. (2002) and Barbuy et al. (2011), the r-II star CS~31082-001 was 
shown to present an actinide boost disturbing the
radioactive chronometers, with the consequence that chronometer
indicators between a radioactive element and Eu would result in negative
ages. The U/Th ratio appears to be the most reliable chronometer in such
cases. Fig. \ref{compara} shows that the thorium abundance in the r-I
star CS~30315-029 seems to be at the same level relative to the second 
r-process peak elements as the one observed for CS~31082-001, which indicates 
the likely presence of an actinide boost. Indeed, for CS~31082-001 we obtain 
log(Th/Eu)~$=-0.26\pm0.14$ and for CS~30315-029 this value is log(Th/Eu)~$=-0.21\pm0.16$.

In the present sample, \ion{U}{II} lines were not measurable, excluding
the ratio U/Th as a possible age indicator. Therefore, we applied age
derivation using the radioactive element Th alone. The ages can be
derived from the ratios, as given for example in Cayrel et al. (2001):

\smallskip
$$\Delta t(\rm{ Gyr)} = 46.7 [log(\rm{Th/r)}_{\rm init}-log(\rm{Th/r)}_{\rm
now}]. $$
\smallskip

We checked several production ratios (PR) from
theoretical work. These consist of the initial abundance ratios
(zero-decay) expected to have occurred in the early Galaxy, obtained
from r-process models, which are compared with the currently derived
abundances. All the results are obtained from site-independent models or 
those involving high-entropy neutrino-driven winds of neutron-rich matter in
core-collapse supernova, the classic astrophysical site of the main r-process. 
It is important to note that the chronometric ages are found to be 
relatively insensitive to astrophysical modeling (if reasonable parameters are 
chosen; Wanajo et al. 2002, 2003), but are sensitive to the nuclear data adopted 
(Goriely et al. 2001; Schatz et al. 2002). From Cowan et al. (1999), 
we used their best result for
Th/Eu based on the extended Thomas-Fermi model with quenched shell
effects far from stability ETFSI-Q (least-squares). The same model was
used by Cowan et al. (2002) for Th/Ir, but an average value based on the
ETFSI-Q, ETFSI-Q (least-squares) and finite-range droplet model with
microscopic shell corrections (FRDM+HFB) was adopted for Th/Eu in
this case. Schatz et al. (2002) have conducted an investigation of the
influence of $\beta$-decay rates, mass models, fission processes, and
model uncertainties on the age determination, presenting their best
predictions of the zero-age abundance for several ratios. Different
initial electron fractions ($Y_\mathrm{e}$) were used by Wanajo et al.
(2002) to derive the consistent age of CS~31082-001 for Th/Eu and 
U/Th, and we used the results for Th/Eu obtained with $Y_\mathrm{e}$~=~0.40. 
Wanajo (2007) used a similar model, but adopting hot and cold r-process conditions 
(see details in his paper).
From Kratz et al. (2007), we adopted the Th results based on their
new r-process calculations using Fe-peak seeds, which yield the best overall
fit to the stable abundance data for masses $A>$~125, with the
present-day observed solar system elemental abundances for the stable
elements. 

The ages derived are very sensitive to the adopted production
ratios. In general, the radioactive chronometers involving second-peak 
region stable elements give ages that are too young, or even negative ages,
stressing the occurrence of an actinide boost in CS~30315-029. Table
\ref{ages} presents the results, and the best values (in boldface) were selected 
based on the ages obtained for CS~31082-001. Hill et al. (2002) reported an age of
14.0~$\pm$~2.4~Gyr for CS~31082-001. The radioactive pairs used for
CS~30315-029 are those that give results for CS~31082-001 consistent
with the age from Hill et al. (2002). Only the production ratios from
Wanajo et al. (2002) and Wanajo (2007) give reasonable ages, as a result
of their models adjusted to the abundance pattern of CS~31082-001. The
average age from the best production ratios in Table \ref{ages} obtained for
CS~30315-029 is 13.5~$\pm$~3.4~Gyr, with the standard deviation of the
values taken as the uncertainty. 

\begin{table}
\caption{Computed production ratios (PR) of Th vs.some stable {\it r}-process elements, observed abundance ratios, 
and age estimates.}             
\label{ages}      
\scalefont{0.65}
\centering                          
\begin{tabular}{ccccrcr}        
\hline\hline                 
\noalign{\smallskip}
\hbox{Chrono.} & \hbox{log (PR)} & \hbox{Ref.} & \hbox{CS~31082-001} & \hbox{Age}  &  \hbox{CS~30315-029} & \hbox{Age}  \\
\noalign{\smallskip}
\hline                        
\noalign{\smallskip}
\hbox{Pair} &   &  & \hbox{log (Th/X)} & \hbox{(Gyr)} & \hbox{log (Th/X)} & \hbox{(Gyr)} \\
\noalign{\smallskip}
\hline                        
\noalign{\smallskip}
\hbox{Th/La}    & $-$0.60  & S02  & $-$0.36 & $-$11.21 & $-$0.39 & $-$10.01 \\
                & \textbf{$-$0.01}  & \textbf{W07a} & \textbf{$-$0.36} & \textbf{$+$16.35} & \textbf{$-$0.39} & \textbf{$+$17.55} \\
                & $+$0.12  & W07b & $-$0.36 & $+$22.42 & $-$0.39 & $+$23.62 \\
\hbox{Th/Ce}    & $-$0.79  & S02  & $-$0.69 & $-$4.67 & $-$0.80 &  $+$0.25 \\
                & $+$0.06  & W07a & $-$0.69 & $+$35.03 & $-$0.80 & $+$39.94 \\
                & $-$0.25  & W07b & $-$0.69 & $+$20.55 & $-$0.80 & $+$25.47 \\
\hbox{Th/Pr}    & $-$0.30  & S02  & $-$0.19 &  $-$5.14 & $-$0.32 &  $+$1.03 \\
                & $+$0.60  & W07a & $-$0.19 & $+$36.89 & $-$0.32 & $+$43.06 \\
                & $+$0.25  & W07b & $-$0.19 & $+$20.55 & $-$0.32 & $+$26.71 \\
\hbox{Th/Nd}    & $-$0.91  & S02  & $-$0.83 &  $-$3.74 & $-$0.89 &  $-$0.86 \\
                & $-$0.26  & W07a & $-$0.83 & $+$26.62 & $-$0.89 & $+$29.50 \\
                & $-$0.68  & W07b & $-$0.83 &  $+$7.01 & $-$0.89 &  $+$9.88 \\
\hbox{Th/Sm}    & $-$0.61  & S02  & $-$0.56 &  $-$1.87 & $-$0.69 &  $+$4.20 \\                
                & $+$0.01  & W07a & $-$0.56 & $+$26.62 & $-$0.69 & $+$32.69 \\                
                & \textbf{$-$0.27}  & \textbf{W07b} & \textbf{$-$0.56} & \textbf{$+$13.54} & \textbf{$-$0.69} & \textbf{$+$19.61} \\                
\hbox{Th/Eu}    & $-$0.32  & C99  & $-$0.26 &  $-$2.80 & $-$0.21 &  $-$5.32 \\ 
                & $-$0.29  & C02  & $-$0.26 &  $-$1.40 & $-$0.21 &  $-$3.92 \\
                & $-$0.33  & S02  & $-$0.26 &  $-$3.27 & $-$0.21 &  $-$5.79 \\ 
                & \textbf{$+$0.05}  & \textbf{W02}  & \textbf{$-$0.26} & \textbf{$+$14.48} & \textbf{$-$0.21} & \textbf{$+$11.96} \\
                & $-$0.37  & K07  & $-$0.26 &  $-$5.14 & $-$0.21 &  $-$7.66 \\
                & $+$0.49  & W07a & $-$0.26 & $+$35.03 & $-$0.21 & $+$32.50 \\ 
                & \textbf{$+$0.07}  & \textbf{W07b} & \textbf{$-$0.26} & \textbf{$+$15.41} & \textbf{$-$0.21} & \textbf{$+$12.89} \\                   
\hbox{Th/Gd}    & $-$0.81  & S02  & $-$0.77 &  $-$1.87 & $-$0.60 &  $-$9.81 \\ 
                & $+$0.42  & W07a & $-$0.77 & $+$55.57 & $-$0.60 & $+$47.63 \\ 
                & \textbf{$-$0.37}  & \textbf{W07b} & \textbf{$-$0.77} & \textbf{$+$18.68} & \textbf{$-$0.60} & \textbf{$+$10.74} \\ 
\hbox{Th/Tb}    & $-$0.12  & S02  & $+$0.03 &  $-$7.01 & $-$0.05 &  $-$3.27 \\ 
                & $+$0.73  & W07a & $+$0.03 & $+$32.69 & $-$0.05 & $+$36.43 \\
                & $+$0.48  & W07b & $+$0.03 & $+$21.02 & $-$0.05 & $+$24.75 \\
\hbox{Th/Dy}    & $-$0.89  & S02  & $-$0.91 &  $+$0.93 & $-$0.83 &  $-$2.80 \\ 
                & $-$0.07  & W07a & $-$0.91 & $+$39.23 & $-$0.83 & $+$35.49 \\
                & $-$0.13  & W07b & $-$0.91 & $+$36.43 & $-$0.83 & $+$32.69 \\
\hbox{Th/Er}    & $-$0.68  & S02  & $-$0.68 &     0.00 & $-$0.70 &  $+$0.93 \\ 
                & $-$0.02  & W07a & $-$0.68 & $+$30.82 & $-$0.70 & $+$31.76 \\
                & $-$0.28  & W07b & $-$0.68 & $+$18.68 & $-$0.70 & $+$19.61 \\
\hbox{Th/Tm}    & $+$0.12  & S02  & $+$0.17 &  $-$2.34 & $+$0.27 & $-$6.77 \\
                & $+$0.65  & W07a & $+$0.17 & $+$22.42 & $+$0.27 & $+$17.98 \\
                & \textbf{$+$0.44}  & \textbf{W07b} & \textbf{$+$0.17} & \textbf{$+$12.61} & \textbf{$+$0.27} & \textbf{$+$8.17} \\
\noalign{\smallskip}
\hline
\end{tabular}
\tablebib{S02: Schatz et al. (2002); W07: Wanajo (2007), (a) hot and (b) cold model; 
C99: Cowan et al. (1999); C02: Cowan et al. (2002); W02: Wanajo et al. (2002), with 
initial electron fraction $Y_\mathrm{e}$~=~0.40; K07: Kratz et al. (2007).}
\end{table}

\section{Conclusions}

We analyzed seven r-I stars and derived the chemical abundances based 
on high-quality, high-resolution spectra. 

The results obtained for the lighter element Li show that the stars have
undergone the first dredge-up, and for the objects with no Li detectable
an extra-mixing event may have occurred. The C abundances obtained confirm
that the sample is not carbon-enhanced, per our original selection
criterion. For nitrogen, the abundances were derived from the CN and NH
lines, and the present values confirm the discrepancy found in the
literature between these two abundance indicators.

For the $\alpha$-elements Mg, Si, S, Ca, and Ti, the odd-Z elements Al,
K, and Sc, and the iron-peak elements V, Cr, Mn, Fe, Co, and Ni, the
results obtained in our present sample agree excellently with 
the values from the literature. There are no differences in the chemical
content between r-I and r-II stars (or even the normal stars), in
terms of these elements. Hydrodynamics models were also computed 
for some elements to explore the 3D corrections.

The same excellent agreement is obtained among the abundance patterns of
the second-peak region elements Ba, La, Ce, Pr, Nd, Sm, Eu, Gd, Tb, Dy, Ho, Er,
Tm, and Yb. The results of the r-I stars analyzed reproduce the pattern
observed in the r-II star CS~31082-001. The origin of this region is
associated with the main r-process, which must operate in the same
way to build the chemical content observed in the atmospheres of r-I and
r-II stars. The upper limits derived for the abundances of the
third-peak elements also agree with the expected values from
the r-II abundance pattern. 

On the other hand, the derived abundances for the first peak region 
elements Sr, Y, Zr, Mo, Ru, and Pd, some of them as upper limits only, are
enhanced with respect to the level observed in the r-II stars. In
addition, this overabundance is higher in stars with lower [Ba/Fe]
ratios. In other words, the abundance level obtained in r-I stars is
between the lowest value derived in r-II stars and the highest one
observed in the normal stars. It is important to note that the
behavior of [first-peak/Fe] ratios as a function of metallicity [Fe/H]
seems to be the same among all the stars (r-I, r-II, and normal stars). 
Indeed, a constant ratio appears for [Fe/H]~$>$~$-$3, which is
the metallicity region of the r-process-element enriched stars,
indicating a co-production of iron and the first-peak elements. A weak 
r-process is claimed to explain the origin of these elements, and
several models are available in the literature (e.g., Montes et al. 2007; 
Wanajo 2013). The comparison between the calculated patterns and the observational 
evidence can perhaps shed some light on this problem. 

The thorium abundance derived in CS~30315-029 shows that an actinide
boost most likely exists for this star, which currently is the most metal-deficient 
object with r-process enhancement. Several r-process models were applied to calculate
the age of the star based on radioactive chronometry, but the lack of
uranium abundance in CS~30315-029 does not permit us to use the ratio U/Th,
the only robust radioactive pair in stars with actinide boost.

\begin{acknowledgements}
CS and BB acknowledge grants from CAPES, CNPq and FAPESP. 
MS and FS acknowledge the support of CNRS (PNCG and PNPS).
TCB and HS acknowledge support from grant PHY 08-22648:
Physics Frontiers Center/Joint Institute for Nuclear Astrophysics
(JINA), awarded by the U.S. National Science Foundation. 
HS acknowledges support from NSF grant PHY1102511. 
EC is grateful to the FONDATION MERAC for funding her fellowship. 
SW acknowledges support from the JSPS Grants-in-Aid for Scientific Research (23224004). 
This work has made use of the ADS Service (SAO/NASA), the SIMBAD database 
(Centre de donn\'ees de Strasbourg), and of the arXiv Server (Cornell University).
\end{acknowledgements}


\begin{appendix} 
\section{Line lists}

\scalefont{0.7}
\begin{longtable}{cccrrrrrrrr}
\caption{\label{EW_measurements} Equivalent widths (EW) measured and used to derive new atmospheric parameters 
and abundances of titanium and iron.}\\
\hline\hline
\hbox{Species} &  \hbox{$\lambda$({\rm\AA})} & \hbox{$\chi$$_{ex}$(eV)}
& \hbox{log $gf$} & \hbox{EW(m\AA)} & \hbox{EW(m\AA)} & \hbox{EW(m\AA)} & \hbox{EW(m\AA)} & \hbox{EW(m\AA)} & \hbox{EW(m\AA)} & \hbox{EW(m\AA)}\\
\hbox{} & \hbox{} & \hbox{} & \hbox{} & \hbox{CS~30315-029} & \hbox{HE~0057-4541} & \hbox{HE~0105-6141} & \hbox{HE~0240-0807} & \hbox{HE~0516-3820} & \hbox{HE~0524-2055} & \hbox{HE~2229-4153}\\
\hline
\endfirsthead
\caption{continued.}\\
\hline\hline
\hbox{Species} & \hbox{$\lambda$({\rm\AA})} &  \hbox{$\chi$$_{ex}$(eV)} & \hbox{log$gf$} & \hbox{EW(m\AA)} & \hbox{EW(m\AA)} & \hbox{EW(m\AA)} & \hbox{EW(m\AA)} & \hbox{EW(m\AA)} & \hbox{EW(m\AA)} & \hbox{EW(m\AA)} \\
\hbox{} & \hbox{} & \hbox{} & \hbox{} & \hbox{CS~30315-029} & \hbox{HE~0057-4541} & \hbox{HE~0105-6141} & \hbox{HE~0240-0807} & \hbox{HE~0516-3820} & \hbox{HE~0524-2055} & \hbox{HE~2229-4153}\\
\hline
\endhead
\hline
\endfoot
\hbox{Ti I}  & 3729.807 & 0.000 & $-$0.351 & 33.87  & 58.40  & 45.56  & 60.48  & 40.79  & 65.04  & ------ \\
\hbox{Ti I}  & 3741.059 & 0.021 & $-$0.213 & 36.33  & 60.51  & 42.88  & 55.30  & 42.77  & 60.52  & ------ \\
\hbox{Ti I}  & 3752.859 & 0.048 & $-$0.019 & 45.86  & 56.77  & 49.56  & 63.77  & 51.83  & 64.67  & 47.54 \\
\hbox{Ti I}  & 3924.527 & 0.021 & $-$0.937 & 15.83  & 31.36  & 18.35  & 34.07  & 15.00  & 34.23  & 19.56 \\
\hbox{Ti I}  & 3929.874 & 0.000 & $-$1.060 & ------ & ------ & ------ & ------ & ------ & ------ & 10.79 \\
\hbox{Ti I}  & 3962.851 & 0.000 & $-$1.167 & ------ & ------ & 12.34  & ------ & ------ & 23.86  & 10.78 \\
\hbox{Ti I}  & 3964.269 & 0.021 & $-$1.184 & ------ & ------ & ------ & ------ & ------ & ------ &  8.79 \\
\hbox{Ti I}  & 3998.636 & 0.048 & $-$0.056 & 44.95  & ------ & 53.57  & 68.53  & 53.37  & 71.27  & 55.75 \\
\hbox{Ti I}  & 4112.709 & 0.048 & $-$1.758 & ------ &  7.68  & ------ & ------ & ------ &  8.58  & ------ \\
\hbox{Ti I}  & 4287.403 & 0.836 & $-$0.442 &  5.56  & ------ & 13.14  & 15.12  & ------ & 21.39  & ------ \\
\hbox{Ti I}  & 4453.312 & 1.430 & $-$0.051 & ------ &  9.81  & ------ & 11.50  &  4.69  &  9.14  &  6.30 \\
\hbox{Ti I}  & 4512.734 & 0.836 & $-$0.480 &  6.30  & 15.00  &  9.10  & 15.66  &  6.28  & 14.18  &  8.68 \\
\hbox{Ti I}  & 4518.022 & 0.826 & $-$0.325 &  9.20  & ------ & ------ & 14.47  & ------ & ------ & 11.79 \\
\hbox{Ti II} & 3913.461 & 1.116 & $-$0.420 & ------ & ------ & 94.70  & ------ & 91.38  & ------ & 95.43 \\
\hbox{Ti II} & 4012.383 & 0.574 & $-$1.840 & 87.96  & 91.13  & 73.63  & ------ & 72.16  & ------ & 73.64 \\
\hbox{Ti II} & 4028.338 & 1.892 & $-$0.960 & 30.45  & 59.99  & 34.18  & 56.68  & 33.69  & 59.15  & 38.31 \\
\hbox{Ti II} & 4290.215 & 1.165 & $-$0.850 & 85.88  & ------ & 78.65  & ------ & 82.98  & ------ & 85.41 \\
\hbox{Ti II} & 4337.914 & 1.080 & $-$0.960 & 84.48  & ------ & 69.47  & 97.88  & 66.18  & ------ & 73.88 \\
\hbox{Ti II} & 4394.051 & 1.221 & $-$1.780 & 39.89  & 56.80  & 41.36  & 63.31  & 36.88  & 67.32  & 43.75 \\
\hbox{Ti II} & 4395.031 & 1.084 & $-$0.540 & ------ & ------ & 97.48  & ------ & ------ & ------ & ------ \\
\hbox{Ti II} & 4399.765 & 1.237 & $-$1.190 & 66.24  & 89.52  & 66.97  & 93.41  & 62.66  & 96.36  & 73.96 \\
\hbox{Ti II} & 4417.714 & 1.165 & $-$1.190 & 75.73  & 81.96  & 66.35  & 97.18  & 65.75  & 97.47  & 70.73 \\
\hbox{Ti II} & 4418.330 & 1.237 & $-$1.970 & 29.46  & 42.92  & 23.50  & 47.88  & 20.14  & 53.55  & 28.64 \\
\hbox{Ti II} & 4443.794 & 1.080 & $-$0.720 & 98.70  & ------ & 88.00  & ------ & 84.90  & ------ & 90.03 \\
\hbox{Ti II} & 4444.555 & 1.116 & $-$2.240 & 24.56  & 38.28  & 22.34  & 48.98  & 16.78  & 50.11  & 23.55 \\
\hbox{Ti II} & 4450.482 & 1.084 & $-$1.520 & 62.47  & 72.85  & 56.72  & 91.99  & 53.17  & 89.45  & 59.69 \\
\hbox{Ti II} & 4464.449 & 1.161 & $-$1.810 & 43.61  & 54.46  & 37.22  & 68.11  & 34.86  & 70.86  & 40.69 \\
\hbox{Ti II} & 4468.507 & 1.131 & $-$0.600 & ------ & ------ & 90.22  & ------ & 88.35  & ------ & 94.99 \\
\hbox{Ti II} & 4470.853 & 1.165 & $-$2.020 & 27.84  & 35.31  & 19.01  & 44.12  & 18.75  & 52.32  & 25.58 \\
\hbox{Ti II} & 4501.270 & 1.116 & $-$0.770 & 95.46  & 99.82  & 90.15  & ------ & 84.53  & ------ & 86.46 \\
\hbox{Fe I}  & 3753.611 & 2.176 & $-$0.890 & 55.33  & 82.77  & 58.07  & 83.46  & 59.91  & 87.73  & 59.34 \\
\hbox{Fe I}  & 3765.539 & 3.237 &    0.482 & 50.72  & 82.44  & 63.82  & 74.11  & 67.17  & 83.57  & 62.68 \\
\hbox{Fe I}  & 3805.342 & 3.301 &    0.312 & 46.07  & 67.63  & 63.82  & 75.49  & 56.50  & 71.32  & 59.57 \\
\hbox{Fe I}  & 3997.392 & 2.727 & $-$0.479 & 43.99  & 71.10  & 59.48  & 68.81  & 59.59  & 73.63  & 57.97 \\
\hbox{Fe I}  & 4005.242 & 1.557 & $-$0.610 & ------ & ------ & 93.89  & ------ & ------ & ------ & ------ \\
\hbox{Fe I}  & 4021.866 & 2.758 & $-$0.729 & ------ & 61.16  & 40.77  & 58.43  & 50.06  & 59.52  & 44.45 \\
\hbox{Fe I}  & 4032.627 & 1.485 & $-$2.377 & 26.88  & 44.21  & 29.80  & 49.88  & 27.55  & 52.40  & 30.93 \\
\hbox{Fe I}  & 4062.441 & 2.845 & $-$0.862 & 19.18  & 46.89  & 30.52  & 38.17  & 37.01  & 44.81  & 31.38 \\
\hbox{Fe I}  & 4067.978 & 3.211 & $-$0.472 & 16.44  & 47.51  & 36.75  & 42.40  & 38.83  & 45.99  & 34.37 \\
\hbox{Fe I}  & 4076.629 & 3.211 & $-$0.529 & ------ & ------ & ------ & ------ & 37.71  & ------ & ------ \\
\hbox{Fe I}  & 4107.488 & 2.831 & $-$0.879 & 25.17  & 56.41  & 37.87  & 53.20  & 40.81  & 57.86  & 42.82 \\
\hbox{Fe I}  & 4114.445 & 2.831 & $-$1.303 & 11.34  & 33.14  & 16.63  & 24.37  & 18.95  & 29.26  & 18.74 \\
\hbox{Fe I}  & 4134.677 & 2.831 & $-$0.649 & 31.73  & 61.35  & 44.41  & 57.66  & 49.15  & 63.34  & 47.63 \\
\hbox{Fe I}  & 4136.998 & 3.415 & $-$0.453 &  8.98  & 35.31  & 19.09  & 24.41  & 23.61  & 29.33  & 21.59 \\
\hbox{Fe I}  & 4147.669 & 1.485 & $-$2.104 & 50.25  & 68.46  & 45.53  & 78.33  & 53.53  & 76.80  & 53.24 \\
\hbox{Fe I}  & 4153.900 & 3.396 & $-$0.321 & 17.01  & 48.52  & 31.17  & 38.80  & 36.73  & 50.09  & 32.78 \\
\hbox{Fe I}  & 4154.499 & 2.831 & $-$0.688 & 25.71  & ------ & 43.63  & 49.27  & 49.67  & 59.04  & 45.02 \\
\hbox{Fe I}  & 4154.805 & 3.368 & $-$0.400 & 15.83  & 45.51  & 29.84  & 34.83  & 38.42  & 45.38  & 33.43 \\
\hbox{Fe I}  & 4156.799 & 2.831 & $-$0.809 & 24.92  & 58.68  & 44.02  & 51.50  & 42.34  & 61.72  & 44.39 \\
\hbox{Fe I}  & 4157.780 & 3.417 & $-$0.403 & 12.87  & 44.75  & 30.18  & 33.58  & 32.48  & 38.88  & 31.79 \\
\hbox{Fe I}  & 4175.636 & 2.845 & $-$0.827 & 28.18  & 54.87  & 40.22  & 52.28  & 43.04  & 55.38  & 38.86 \\
\hbox{Fe I}  & 4176.566 & 3.368 & $-$0.805 & 12.43  & 36.47  & 20.32  & 25.93  & 25.63  & 31.03  & 23.19 \\
\hbox{Fe I}  & 4181.755 & 2.831 & $-$0.371 & 42.93  & 80.11  & 57.29  & 69.04  & 62.77  & 76.12  & 61.31 \\
\hbox{Fe I}  & 4182.383 & 3.017 & $-$1.180 &  6.88  & 28.61  & 16.67  & 18.49  & 17.39  & 24.40  & 15.01 \\
\hbox{Fe I}  & 4184.891 & 2.831 & $-$0.869 & 18.19  & 46.34  & 31.31  & 40.77  & 35.51  & 45.86  & 33.15 \\
\hbox{Fe I}  & 4187.039 & 2.449 & $-$0.548 & 62.34  & 81.58  & 66.77  & 85.12  & 69.22  & 89.05  & 71.25 \\
\hbox{Fe I}  & 4187.795 & 2.425 & $-$0.554 & 67.04  & 91.71  & 70.51  & 89.85  & 74.55  & 93.16  & 76.14 \\
\hbox{Fe I}  & 4191.430 & 2.469 & $-$0.666 & 55.47  & 75.82  & 63.40  & 82.67  & 65.01  & 85.41  & 63.29 \\
\hbox{Fe I}  & 4195.329 & 3.332 & $-$0.492 & 19.55  & 56.67  & 34.74  & 44.07  & 36.00  & 48.54  & 38.47 \\
\hbox{Fe I}  & 4199.095 & 3.047 &    0.155 & 52.25  & 83.34  & 64.78  & 81.07  & 65.87  & 85.00  & 69.21 \\
\hbox{Fe I}  & 4202.029 & 1.485 & $-$0.708 & ------ & ------ & 98.63  & ------ & ------ & ------ & ------ \\
\hbox{Fe I}  & 4213.647 & 2.845 & $-$1.290 &  5.85  & 27.94  & ------ & 22.84  & ------ & 26.79  & 17.86 \\
\hbox{Fe I}  & 4222.213 & 2.449 & $-$0.967 & 42.42  & 70.07  & 51.65  & 67.29  & 55.23  & 70.27  & 54.11 \\
\hbox{Fe I}  & 4233.602 & 2.482 & $-$0.604 & 58.40  & 80.63  & 64.47  & 78.95  & 67.24  & 86.66  & 68.23 \\
\hbox{Fe I}  & 4238.810 & 3.396 & $-$0.233 & 19.49  & 52.32  & 36.40  & 41.37  & 41.81  & 43.91  & 35.08 \\
\hbox{Fe I}  & 4250.119 & 2.469 & $-$0.405 & 67.91  & 90.42  & 71.71  & 91.67  & 72.36  & 95.17  & 76.36 \\
\hbox{Fe I}  & 4260.474 & 2.399 &    0.109 & 94.66  & ------ & 92.46  & ------ & 99.34  & ------ & 96.87 \\
\hbox{Fe I}  & 4271.153 & 2.449 & $-$0.349 & 76.56  & ------ & 87.23  & ------ & 96.48  & ------ & 92.10 \\
\hbox{Fe I}  & 4282.403 & 2.176 & $-$0.779 & 67.72  & ------ & 71.45  & ------ & 72.65  & 95.81  & 72.35 \\
\hbox{Fe I}  & 4337.046 & 1.557 & $-$1.695 & 66.34  & ------ & 60.76  & 85.76  & 68.69  & 89.26  & 63.66 \\
\hbox{Fe I}  & 4430.614 & 2.223 & $-$1.659 & 25.46  & 50.13  & 32.50  & 50.76  & 35.92  & 54.03  & 37.02 \\
\hbox{Fe I}  & 4442.339 & 2.198 & $-$1.255 & 46.21  & 79.59  & 53.66  & 75.54  & 58.65  & 80.42  & 58.74 \\
\hbox{Fe I}  & 4443.194 & 2.858 & $-$1.043 & 13.31  & 42.59  & 25.33  & 33.85  & 24.61  & 41.05  & 27.13 \\
\hbox{Fe I}  & 4447.717 & 2.223 & $-$1.342 & 37.50  & 69.72  & 48.78  & 68.01  & 49.06  & 71.68  & 50.40 \\
\hbox{Fe I}  & 4459.117 & 2.176 & $-$1.279 & 54.47  & 86.27  & 61.68  & 83.70  & 64.49  & 94.11  & 65.92 \\
\hbox{Fe I}  & 4466.551 & 2.831 & $-$0.600 & 52.03  & 74.98  & 54.73  & 77.26  & 56.46  & 85.01  & 57.88 \\
\hbox{Fe I}  & 4494.563 & 2.198 & $-$1.136 & 54.83  & 73.91  & 58.48  & 77.78  & 62.39  & 80.25  & 62.02 \\
\hbox{Fe II} & 4128.748 & 2.583 & $-$3.578 & ------ & 10.39  & ------ & ------ & ------ &  9.14  &  6.15 \\
\hbox{Fe II} & 4178.862 & 2.583 & $-$2.535 & 30.02  & 48.20  & 32.91  & 49.99  & 33.93  & 55.36  & 38.71 \\
\hbox{Fe II} & 4233.172 & 2.583 & $-$1.947 & 64.62  & ------ & 55.78  & 87.66  & 56.99  & ------ & 63.87 \\
\hbox{Fe II} & 4416.830 & 2.778 & $-$2.602 & 20.96  & 42.76  & 24.58  & 41.41  & 21.27  & 46.15  & 29.42 \\
\hbox{Fe II} & 4491.405 & 2.856 & $-$2.756 & 12.83  & 29.31  & 14.43  & 25.39  & 18.70  & 32.05  & 17.97 \\
\hbox{Fe II} & 4508.288 & 2.856 & $-$2.349 & 26.62  & 46.33  & 27.16  & 47.68  & 31.21  & 53.09  & 31.91 \\
\hbox{Fe II} & 4515.339 & 2.844 & $-$2.540 & 21.71  & 37.70  & 22.43  & 38.41  & 24.73  & ------ & 25.50 \\
\end{longtable}

\newpage

\scalefont{1.0}
\begin{longtable}{cccrrrrrrrr}
\caption{\label{linelist} List of lines used in the present analysis, with the individual abundances.}\\
\hline\hline
\hbox{Species} &  \hbox{$\lambda$({\rm\AA})} & \hbox{$\chi$$_{ex}$(eV)}&
\hbox{log $gf$} & \hbox{A(X)} & \hbox{A(X)} & \hbox{A(X)} & \hbox{A(X)} & \hbox{A(X)} & \hbox{A(X)} & \hbox{A(X)} \\
\hbox{} & \hbox{} & \hbox{} & \hbox{} & \hbox{CS~30315-029} & \hbox{HE~0057-4541} & \hbox{HE~0105-6141} & \hbox{HE~0240-0807} & \hbox{HE~0516-3820} & \hbox{HE~0524-2055} & \hbox{HE~2229-4153}\\
\hline
\endfirsthead
\caption{continued.}\\
\hline\hline
\hbox{Species} & \hbox{$\lambda$({\rm\AA})} &  \hbox{$\chi$$_{ex}$(eV)}  & \hbox{log$gf$} & \hbox{A(X)} & \hbox{A(X)} & \hbox{A(X)} & \hbox{A(X)} & \hbox{A(X)} & \hbox{A(X)} & \hbox{A(X)} \\
\hbox{} & \hbox{} & \hbox{} & \hbox{} & \hbox{CS~30315-029} & \hbox{HE~0057-4541} & \hbox{HE~0105-6141} & \hbox{HE~0240-0807} & \hbox{HE~0516-3820} & \hbox{HE~0524-2055} & \hbox{HE~2229-4153}\\
\hline
\endhead
\hline
\endfoot
\hbox{Mg I}  & 3829.355 & 2.709 & $-$0.231 & 4.75    & 5.60     & 5.40     & 5.15     & 5.65     & 5.65     & 5.55 \\
\hbox{Mg I}  & 3832.304 & 2.710 &    0.146 & 4.65    & 5.70     & 5.45     & 5.30     & 5.50     & 5.65     & 5.50 \\
\hbox{Mg I}  & 3838.290 & 2.720 &    0.415 & 4.85    & 5.65     & 5.40     & 5.25     & 5.50     & 5.50     & 5.60 \\
\hbox{Al I}  & 3961.520 & 0.014 & $-$0.323 & 2.68    & 3.50     & 3.20     & 3.26     & 3.30     & 3.40     & 3.25 \\
\hbox{Si I}  & 4102.936 & 1.909 & $-$2.827 & 4.80    & 5.94     & 5.64     & 5.30     & 5.45     & 5.37     & 5.48 \\
\hbox{S I}   & 9212.860 & 6.525 &    0.420 &  -----  & 5.40     & 5.25     & 4.95     & -----    & 5.05     & 5.10 \\
\hbox{S I}   & 9228.090 & 6.525 &    0.260 & 4.25    & 5.25     & -----    & -----    & -----    & 5.10     & -----\\
\hbox{S I}   & 9237.540 & 6.525 &    0.040 & 4.40    & 5.40     & -----    & -----    & 5.35     & -----    & 5.00 \\
\hbox{K I}   & 7664.911 & 0.000 &    0.130 & 2.63    & -----    & -----    & 3.00     & 3.18     & -----    & 3.30 \\
\hbox{K I}   & 7698.974 & 0.000 & $-$0.170 & 2.60    & 3.30     & 2.92     & 2.90     & -----    & 3.27     & 3.20 \\
\hbox{Ca I}  & 4226.728 & 0.000 &    0.265 & 3.25    & 4.35     & 4.20     & 3.75     & 4.20     & 3.95     & 4.10 \\
\hbox{Ca I}  & 4283.011 & 1.886 & $-$0.292 & 3.40    & 4.60     & 4.35     & 3.98     & 4.40     & 4.22     & 4.40 \\
\hbox{Ca I}  & 4289.367 & 1.879 & $-$0.388 & 3.40    & 4.40     & 4.26     & 4.00     & 4.35     & 4.08     & 4.25 \\
\hbox{Ca I}  & 4302.528 & 1.899 &    0.183 &  -----  & -----    & 4.10     & 3.78     & -----    & -----    & -----\\
\hbox{Ca I}  & 4318.652 & 1.899 & $-$0.295 & 3.40    & 4.36     & 4.20     & 3.85     & 4.32     & 4.05     & 4.22 \\
\hbox{Ca I}  & 4425.437 & 1.879 & $-$0.286 & 3.28    & 4.28     & 4.10     & 3.78     & 4.20     & 3.93     & 4.07 \\
\hbox{Ca I}  & 4434.957 & 1.886 &    0.066 &  -----  & 4.25     & 4.15     & -----    & 4.20     & -----    & 4.07 \\
\hbox{Ca I}  & 4435.679 & 1.886 & $-$0.412 &  -----  & 4.25     & 4.12     & -----    & -----    & -----    & 4.02 \\
\hbox{Ca I}  & 4454.779 & 1.899 &    0.335 & 3.22    & 4.25     & 4.08     & 3.70     & 4.12     & 3.86     & 4.03 \\
\hbox{Ca I}  & 4455.887 & 1.899 & $-$0.414 & 3.30    & 4.25     & 4.10     & 3.75     & 4.05     & 3.94     & 4.05 \\
\hbox{Sc II} & 3833.071 & 0.000 & $-$2.014 & $-$0.35 & -----    & -----    & -----    & -----    & -----    & -----\\
\hbox{Sc II} & 4246.822 & 0.315 &    0.242 & $-$0.40 & 0.80     & 0.80     & 0.38     & 0.74     & 0.40     & 0.62 \\
\hbox{Sc II} & 4294.767 & 0.605 & $-$1.391 & $-$0.25 & 0.92     & 0.85     & -----    & -----    & 0.52     & -----\\
\hbox{Sc II} & 4314.083 & 0.618 & $-$0.096 & $-$0.39 & 0.84     & 0.78     & 0.42     & 0.72     & 0.35     & 0.64 \\
\hbox{Sc II} & 4320.732 & 0.605 & $-$0.252 & $-$0.42 & 0.82     & 0.73     & 0.38     & 0.75     & 0.40     & 0.61 \\
\hbox{Sc II} & 4354.598 & 0.605 & $-$1.579 & $-$0.28 & -----    & -----    & -----    & -----    & -----    & -----\\
\hbox{Sc II} & 4374.457 & 0.618 & $-$0.418 & $-$0.42 & 0.80     & 0.77     & 0.36     & 0.76     & 0.42     & 0.62 \\
\hbox{Sc II} & 4400.389 & 0.605 & $-$0.536 & $-$0.40 & 0.82     & 0.72     & 0.36     & 0.74     & 0.43     & 0.60 \\
\hbox{Sc II} & 4415.557 & 0.595 & $-$0.668 & $-$0.36 & 0.87     & 0.76     & 0.40     & 0.77     & 0.45     & 0.62 \\
\hbox{V I}   & 3855.841 & 0.069 &    0.013 &  -----  & 1.70     & -----    & -----    & -----    & -----    & -----\\
\hbox{V I}   & 4111.774 & 0.301 &    0.408 &  -----  & 1.50     & 1.35     & 1.05     & -----    & -----    & 1.35 \\
\hbox{V I}   & 4379.230 & 0.301 &    0.580 &  0.32   & 1.60     & 1.38     & 1.18     & 1.40     & 1.22     & 1.34 \\
\hbox{V I}   & 4384.712 & 0.287 &    0.510 &  -----  & 1.48     & 1.24     & -----    & -----    & -----    & 1.30 \\
\hbox{V I}   & 4389.976 & 0.275 &    0.200 &  -----  & -----    & -----    & -----    & -----    & -----    & -----\\
\hbox{V I}   & 4408.193 & 0.275 &    0.020 &  -----  & -----    & -----    & -----    & -----    & -----    & 1.42 \\
\hbox{V II}  & 3715.466 & 1.575 & $-$0.250 & 0.50    & 1.88     & 1.68     & 1.28     & 1.58     & 1.27     & 1.50 \\
\hbox{V II}  & 3727.343 & 1.687 & $-$0.231 &  -----  & 1.78     & 1.70     & 1.30     & 1.70     & 1.33     & 1.46 \\
\hbox{V II}  & 3732.750 & 1.565 & $-$0.354 & 0.55    & 1.87     & 1.64     & 1.30     & 1.68     & 1.32     & 1.58 \\
\hbox{V II}  & 3750.870 & 1.679 & $-$0.409 &  -----  & -----    & -----    & 1.32     & -----    & 1.25     & 1.38 \\
\hbox{V II}  & 3899.129 & 1.805 & $-$0.784 &  -----  & 1.92     & -----    & -----    & -----    & -----    & -----\\
\hbox{V II}  & 3916.411 & 1.428 & $-$1.053 &  -----  & 1.90     & 1.72     & 1.40     & 1.75     & 1.38     & 1.52 \\
\hbox{V II}  & 3951.960 & 1.476 & $-$0.784 & 0.45    & 1.80     & 1.60     & 1.26     & 1.60     & 1.30     & 1.45 \\
\hbox{V II}  & 4002.936 & 1.428 & $-$1.447 &  -----  & 1.90     & -----    & 1.38     & -----    & 1.40     & 1.57 \\
\hbox{V II}  & 4005.705 & 1.817 & $-$0.522 & 0.62    & 1.90     & 1.65     & 1.45     & -----    & 1.37     & 1.60 \\
\hbox{V II}  & 4023.378 & 1.805 & $-$0.689 &  -----  & 1.86     & 1.55     & 1.50     & 1.80     & 1.40     & -----\\
\hbox{V II}  & 4035.622 & 1.793 & $-$0.767 &  -----  & -----    & -----    & 1.47     & -----    & -----    & -----\\
\hbox{Cr I}  & 4254.332 & 0.000 & $-$0.114 & 1.78    & 3.05     & 2.90     & 2.22     & 3.05     & 2.35     & 2.75 \\
\hbox{Cr I}  & 4274.796 & 0.000 & $-$0.231 & 1.77    & 3.10     & 2.78     & 2.35     & 3.05     & 2.50     & 2.78 \\
\hbox{Cr I}  & 4289.716 & 0.000 & $-$0.361 & 1.80    & 3.04     & 2.80     & 2.32     & 3.05     & 2.46     & 2.73 \\
\hbox{Mn I}  & 4030.753 & 0.000 & $-$0.470 & 1.15    & 2.40     & 2.00     & 1.67     & 2.28     & 1.80     & 2.10 \\
\hbox{Mn I}  & 4033.062 & 0.000 & $-$0.618 & 1.17    & 2.35     & 2.00     & 1.60     & 2.25     & 1.83     & 2.15 \\
\hbox{Mn I}  & 4034.483 & 0.000 & $-$0.811 & 1.25    & 2.50     & 2.20     & 1.70     & 2.35     & 1.85     & 2.17 \\
\hbox{Mn I}  & 4041.355 & 2.114 &    0.285 & 1.58    & 2.78     & 2.37     & 2.07     & 2.73     & 2.30     & 2.42 \\
\hbox{Mn I}  & 4082.939 & 2.178 & $-$0.354 & -----   & -----    & -----    & -----    & -----    & 2.22     & -----\\
\hbox{Co I}  & 3845.461 & 0.923 &    0.010 & 1.62    & 2.72     & 2.85     & 2.20     & 2.70     & 2.30     & 2.50 \\
\hbox{Co I}  & 3995.302 & 0.923 & $-$0.220 & 1.66    & 2.71     & 2.68     & 2.18     & 2.74     & 2.28     & 2.55 \\
\hbox{Co I}  & 4118.767 & 1.049 & $-$0.490 & 1.80    & 2.95     & 2.70     & 2.35     & 2.76     & 2.60     & 2.65 \\
\hbox{Co I}  & 4121.311 & 0.923 & $-$0.320 & 1.88    & 3.05     & 2.95     & 2.42     & 3.05     & 2.60     & 2.82 \\
\hbox{Ni I}  & 3807.138 & 0.423 & $-$1.205 & 2.85    & 3.80     & 3.95     & 3.36     & 3.85     & 3.40     & 3.70 \\
\hbox{Ni I}  & 3858.292 & 0.423 & $-$0.936 & 2.70    & 3.78     & 3.55     & 3.25     & 3.85     & 3.38     & 3.70 \\
\hbox{Sr II} & 4077.719 & 0.000 &    0.170 & $-$0.82 & 0.82     & 0.60     & $-$0.10  & 0.63     & 0.15     & 0.53 \\
\hbox{Sr II} & 4215.519 & 0.000 & $-$0.170 & $-$0.86 & 0.77     & 0.68     & $-$0.06  & 0.66     & 0.14     & 0.60 \\
\hbox{Y II}  & 3774.331 & 0.130 &    0.210 & $-$1.47 & $-$0.37  & $-$0.30  & $-$0.72  & -----    & $-$0.82  & $-$0.50 \\
\hbox{Y II}  & 3788.694 & 0.104 & $-$0.070 & $-$1.48 & $-$0.26  & $-$0.50  & $-$0.76  & $-$0.13  & $-$0.74  & $-$0.54 \\
\hbox{Y II}  & 3818.341 & 0.130 & $-$0.980 & $-$1.44 & -----    &  -----   & $-$0.69  & -----    & $-$0.62  & $-$0.40 \\
\hbox{Y II}  & 3950.352 & 0.104 & $-$0.490 & $-$1.43 & $-$0.26  & $-$0.38  & $-$0.74  & $-$0.14  & $-$0.69  & $-$0.40 \\
\hbox{Y II}  & 4398.013 & 0.130 & $-$1.000 & $-$1.43 & -----    &  -----   & $-$0.70  & -----    & $-$0.68  & $-$0.53 \\
\hbox{Zr II} & 3836.762 & 0.559 & $-$0.060 & $-$0.87 & 0.51     & 0.12     & $-$0.22  & 0.55     & $-$0.03  & 0.15 \\
\hbox{Zr II} & 4161.213 & 0.713 & $-$0.720 & $-$0.67 & 0.60     &  -----   &  0.15    & 0.64     & 0.25     & 0.33 \\
\hbox{Zr II} & 4208.977 & 0.713 & $-$0.460 & $-$0.76 & 0.51     & 0.30     & $-$0.02  & 0.35     & 0.03     & 0.18 \\
\hbox{Zr II} & 4317.299 & 0.713 & $-$1.380 & -----   & 0.52     &  -----   & -----    & -----    & 0.20     & -----\\
\hbox{Mo I}  & 3864.103 & 0.000 & $-$0.010 & $<-$1.15& -----    &  -----   & -----    & -----    & $<-$0.35 & -----\\
\hbox{Ru I}  & 3498.942 & 0.000 &    0.310 & $-$0.90 & -----    &  -----   & $<-$0.20 & -----    & $<-$0.40 & ----- \\
\hbox{Ru I}  & 3728.025 & 0.000 &    0.270 & $-$1.00 &  0.24    &  -----   & $<-$0.20 & $<$0.10  & $-$0.25  & $<-$0.15 \\
\hbox{Ru I}  & 3799.349 & 0.000 &    0.020 & $<-$0.60& $<$0.20  &  -----   & $<-$0.20 & -----    & $-$0.05  & $<-$0.10 \\
\hbox{Pd I}  & 3404.579 & 0.814 &    0.320 & $<-$1.00& -----    &  -----   & -----    & -----    &  -----   & $<-$0.60 \\
\hbox{Pd I}  & 3516.944 & 0.962 & $-$0.240 & $<-$1.20& -----    &  -----   & -----    & -----    &  -----   & -----\\
\hbox{Ba II} & 4554.020 & 0.000 &    0.170 & $-$1.10 & $-$0.22  & $-$0.25  & $-$0.50  & $-$0.27  & $-$0.63  & $-$0.65 \\
\hbox{La II} & 3759.080 & 0.244 & $-$0.030 & $-$2.05 & $-$1.20  &  -----   & $-$1.55  & -----    & $-$1.58  & $-$1.65 \\
\hbox{La II} & 3794.770 & 0.244 &    0.210 & $-$1.92 & $-$1.04  & $-$1.25  & $-$1.40  & $-$1.20  & $-$1.60  & $-$1.50 \\
\hbox{La II} & 3849.010 & 0.000 & $-$0.973 & $-$2.15 & $-$0.95  &  -----   & $-$1.40  & -----    & $-$1.55  & $-$1.65 \\
\hbox{La II} & 3929.210 & 0.173 & $-$0.320 & $-$2.00 & $-$0.90  &  -----   & $-$1.46  & -----    & $-$1.60  & -----\\
\hbox{La II} & 3949.100 & 0.403 &    0.490 & $-$2.10 & $-$0.98  & $-$1.25  & $-$1.50  & -----    & $-$1.55  & $-$1.50 \\
\hbox{La II} & 3988.507 & 0.403 & $-$0.969 & $-$2.15 & $-$1.03  & $-$1.13  & $-$1.40  & -----    & $-$1.60  & $-$1.60 \\
\hbox{La II} & 3995.750 & 0.173 & $-$1.084 & $-$2.03 & $-$0.93  & $-$1.03  & $-$1.50  & -----    & $-$1.57  & $-$1.62 \\
\hbox{La II} & 4077.340 & 0.235 & $-$0.060 & $-$2.07 & $-$1.03  &  -----   & $-$1.40  & -----    & $-$1.58  & -----\\
\hbox{La II} & 4086.708 & 0.000 & $-$0.696 & $-$2.01 & $-$1.03  &  -----   & $-$1.48  & -----    & $-$1.57  & $-$1.60 \\
\hbox{La II} & 4123.221 & 0.321 & $-$0.850 & $-$2.14 & $-$1.07  & $-$1.08  & $-$1.51  & $-$1.16  & $-$1.54  & $-$1.50 \\
\hbox{La II} & 4196.550 & 0.321 & $-$0.300 & $-$2.10 & $-$0.98  &  -----   & $-$1.40  & -----    & $-$1.60  & -----\\
\hbox{La II} & 4238.370 & 0.403 & $-$0.260 & $-$2.04 & -----    &  -----   & -----    & -----    &  -----   & -----\\
\hbox{La II} & 4333.753 & 0.173 & $-$0.060 & $-$2.04 & -----    &  -----   & $-$1.35  & -----    & $-$1.56  & -----\\
\hbox{La II} & 4429.910 & 0.235 & $-$0.350 & $-$2.10 & $-$1.08  &  -----   & $-$1.35  & -----    & $-$1.54  & -----\\
\hbox{Ce II} & 3942.151 & 0.000 & $-$0.180 & $-$1.60 & $<-$0.42 &  -----   & $<-$0.92 & -----    & $-$1.30  & $<-$1.00 \\
\hbox{Ce II} & 3999.237 & 0.090 &    0.060 & $-$1.95 & $<-$0.68 &  -----   & $-$1.20  & -----    & $-$1.40  & -----\\
\hbox{Ce II} & 4053.503 & 0.000 & $-$0.610 & $-$1.60 & $<-$0.40 &  -----   & $<-$0.90 & -----    &  -----   & -----\\
\hbox{Ce II} & 4073.474 & 0.478 &    0.230 & $-$1.75 & $-$0.56  &  -----   & $-$1.00  & $<-$0.40 & $-$1.20  & -----\\
\hbox{Ce II} & 4083.222 & 0.701 &    0.270 & $-$1.70 & -----    &  -----   & $<-$0.80 & -----    &  -----   & -----\\
\hbox{Ce II} & 4120.827 & 0.320 & $-$0.210 & $-$1.55 & $<-$0.20 &  -----   & $-$0.90  & -----    & $<-$1.00 & -----\\
\hbox{Ce II} & 4127.364 & 0.684 &    0.350 & $-$1.70 & $<-$0.54 & $<-$0.85 & $-$0.98  & -----    & $<-$1.20 & $<-$1.00 \\
\hbox{Ce II} & 4137.645 & 0.516 &    0.400 & $-$1.62 & $-$0.52  &  -----   & -----    & -----    & $-$1.20  & $-$1.10 \\
\hbox{Ce II} & 4142.397 & 0.696 &    0.220 & $-$1.60 & -----    &  -----   & -----    & -----    &  -----   & -----\\
\hbox{Ce II} & 4144.996 & 0.696 &    0.100 & $-$1.60 & $-$0.35  &  -----   & -----    & -----    &  -----   & -----\\
\hbox{Ce II} & 4222.597 & 0.122 &    0.020 & $-$1.65 & $-$0.50  &  -----   & $-$1.02  & $-$0.44  & $-$1.15  & $-$1.00 \\
\hbox{Ce II} & 4382.165 & 0.684 &    0.130 & $-$1.60 & $<-$0.40 &  -----   & $<-$0.88 & -----    & $<-$1.10 & -----\\
\hbox{Ce II} & 4418.780 & 0.864 &    0.280 & $-$1.45 & -----    &  -----   & -----    & -----    & $-$1.16  & -----\\
\hbox{Ce II} & 4449.330 & 0.609 &    0.040 & $-$1.90 & -----    &  -----   & $-$1.00  & $<-$0.30 &  -----   & -----\\
\hbox{Ce II} & 4486.909 & 0.295 & $-$0.260 & $-$1.55 & $<-$0.54 &  -----   & $-$0.90  & $<-$0.30 & $-$1.16  & -----\\
\hbox{Pr II} & 3964.820 & 0.055 &    0.069 & $<-$2.20& -----    &  -----   & $<-$1.60 & -----    &  -----   & -----\\
\hbox{Pr II} & 3965.260 & 0.204 &    0.204 & $<-$2.10& -----    &  -----   & $<-$1.60 & -----    &  -----   & -----\\
\hbox{Pr II} & 4038.455 & 0.000 & $-$0.510 & $-$2.10 & -----    &  -----   & -----    & -----    &  -----   & -----\\
\hbox{Pr II} & 4044.813 & 0.000 & $-$0.293 & $-$2.15 & $<-$1.00 & $<-$0.70 & $-$1.50  & -----    & $<-$1.50 & -----\\
\hbox{Pr II} & 4096.820 & 0.216 & $-$0.255 & $<-$1.90& -----    &  -----   & $<-$1.60 & -----    & $<-$1.60 & -----\\
\hbox{Pr II} & 4143.120 & 0.371 &    0.604 & $-$2.08 & $<-$1.15 & $<-$0.84 & -----    & -----    & $-$1.70  & $<-$1.40 \\
\hbox{Pr II} & 4179.400 & 0.204 &    0.459 & $-$2.16 & -----    &  -----   & $-$1.42  & $<-$0.75 & $-$1.45  & -----\\
\hbox{Pr II} & 4222.950 & 0.055 &    0.235 & $<-$2.14& -----    &  -----   & $-$1.54  & -----    &  -----   & -----\\
\hbox{Pr II} & 4408.820 & 0.000 &    0.053 & $-$2.15 & $<-$1.12 & $<-$1.00 & $-$1.60  & $<-$1.00 & $-$1.70  & -1.50 \\
\hbox{Pr II} & 4496.470 & 0.216 & $-$0.762 & $<-$2.20& -----    &  -----   & -----    & -----    &  -----   & -----\\
\hbox{Pr II} & 4510.150 & 0.422 & $-$0.007 & $<-$1.85& -----    &  -----   & -----    & -----    &  -----   & -----\\
\hbox{Nd II} & 4012.700 & 0.000 & $-$0.600 & $-$1.55 & $<-$0.50 & $<-$0.38 & $<-$1.00 & $<-$0.20 &  -----   & -----\\
\hbox{Nd II} & 4018.833 & 0.064 & $-$0.850 & $<-$1.50& -----    &  -----   & -----    & -----    &  -----   & -----\\
\hbox{Nd II} & 4021.327 & 0.321 & $-$0.100 & $-$1.60 & $-$0.60  &  -----   & $-$1.10  & -----    & $-$1.16  & -----\\
\hbox{Nd II} & 4059.950 & 0.204 & $-$0.520 & $-$1.60 & -----    & $<-$0.25 & -----    & -----    &  -----   & -----\\
\hbox{Nd II} & 4061.080 & 0.471 &    0.550 & $-$1.62 & $-$0.50  & $-$0.50  & $-$0.92  & $-$0.54  & $-$1.10  & $-$1.02 \\
\hbox{Nd II} & 4069.265 & 0.064 & $-$0.570 & $-$1.50 & -----    &  -----   & $-$0.78  & -----    & $-$0.95  & -----\\
\hbox{Nd II} & 4109.448 & 0.321 &    0.350 & $-$1.50 & $-$0.53  & $-$0.58  & $-$0.88  & $-$0.47  & $-$1.04  & $-$0.97 \\
\hbox{Nd II} & 4232.374 & 0.064 & $-$0.470 & $-$1.60 & $-$0.58  &  -----   & $-$1.00  & -----    & $-$1.05  & -----\\
\hbox{Nd II} & 4358.160 & 0.320 & $-$0.160 &  -----  & -----    & $-$0.52  & -----    & -----    & $-$1.05  & -----\\
\hbox{Nd II} & 4368.630 & 0.064 & $-$0.810 & $-$1.60 & -----    &  -----   & $-$0.90  & -----    & $-$0.98  & -----\\
\hbox{Nd II} & 4385.660 & 0.204 & $-$0.300 & $-$1.60 & -----    &  -----   & $-$0.95  & -----    & $-$1.04  & -----\\
\hbox{Nd II} & 4400.820 & 0.064 & $-$0.600 & $-$1.48 & -----    &  -----   & $-$0.92  & -----    & $-$0.94  & -----\\
\hbox{Nd II} & 4446.384 & 0.205 & $-$0.350 & $-$1.55 & $-$0.60  &  -----   & $-$0.90  & $-$0.45  & $-$1.10  & -----\\
\hbox{Nd II} & 4451.500 & 0.205 & $-$0.350 &  -----  & -----    &  -----   & $-$0.90  & $-$0.36  & $-$1.00  & -----\\
\hbox{Nd II} & 4501.810 & 0.204 & $-$0.690 & $-$1.50 & -----    &  -----   & $<-$0.80 & -----    & $-$0.95  & -----\\
\hbox{Sm II} & 3793.978 & 0.104 & $-$0.498 & $<-$2.00& -----    &  -----   & -----    & -----    &  -----   & -----\\
\hbox{Sm II} & 3896.972 & 0.041 & $-$0.578 & $<-$1.80& -----    &  -----   & -----    & -----    & $<-$1.48 & -----\\
\hbox{Sm II} & 4023.222 & 0.041 & $-$0.830 & $<-$1.73& -----    &  -----   & -----    & -----    &  -----   & -----\\
\hbox{Sm II} & 4068.324 & 0.434 & $-$0.743 & $<-$1.60& -----    &  -----   & $<-$0.86 & -----    &  -----   & -----\\
\hbox{Sm II} & 4183.767 & 0.041 & $-$0.999 & $-$1.68 & $<-$0.80 &  -----   & $-$1.20  & -----    & $<-$1.30 & -----\\
\hbox{Sm II} & 4256.391 & 0.378 & $-$0.135 & $-$1.75 & $-$0.88  &  -----   & -----    & -----    & $-$1.30  & $<-$1.20 \\
\hbox{Sm II} & 4318.927 & 0.277 & $-$0.268 & $-$1.72 & $-$1.05  & $-$0.80  & $-$1.18  & $<-$0.46 & $-$1.40  & $-$1.25 \\
\hbox{Sm II} & 4421.126 & 0.378 & $-$0.477 & $-$1.70 & -----    &  -----   & $<-$1.15 & -----    &  -----   & -----\\
\hbox{Sm II} & 4467.341 & 0.659 &    0.303 & $-$1.95 & $<-$0.90 &  -----   & $-$1.32  & -----    & $-$1.55  & $-$1.40 \\
\hbox{Sm II} & 4499.475 & 0.248 & $-$1.000 & $<-$1.60& -----    &  -----   & $-$1.04  & -----    &  -----   & -----\\
\hbox{Eu II} & 3724.931 & 0.000 & $-$0.090 & $-$2.32 & $-$1.23  &  -----   & $-$1.56  & $<-$1.10 & $-$1.75  & $<-$1.70 \\
\hbox{Eu II} & 3930.499 & 0.207 &    0.270 & $-$2.18 & $-$1.30  & $-$1.60  & $-$1.62  & $-$1.35  &  -----   & $-$1.72 \\
\hbox{Eu II} & 3971.972 & 0.207 &    0.270 & $-$2.30 & $-$1.30  &  -----   & $-$1.65  & -----    &  -----   & $-$1.70 \\
\hbox{Eu II} & 4129.725 & 0.000 &    0.220 & $-$2.20 & $-$1.25  & $-$1.55  & $-$1.53  & $-$1.37  & $-$1.75  & $-$1.74 \\
\hbox{Eu II} & 4205.042 & 0.000 &    0.210 & $-$2.22 & $-$1.25  & $-$1.45  & $-$1.52  & $-$1.30  & $-$1.74  & $-$1.72 \\
\hbox{Gd II} & 3768.396 & 0.360 &    0.210 & $-$1.90 & $-$0.90  & $<-$1.00 & $-$1.25  & $-$1.12  & $-$1.40  & $-$1.40 \\
\hbox{Gd II} & 3796.384 & 0.032 &    0.140 &  -----  & $-$0.80  & $<-$0.80 & $-$1.12  & -----    & $-$1.40  & $-$1.25 \\
\hbox{Gd II} & 3836.915 & 0.492 & $-$0.322 &  -----  & -----    &  -----   & $<-$1.12 & -----    &  -----   & -----\\
\hbox{Gd II} & 3844.578 & 0.144 & $-$0.400 & $-$1.80 & $<-$0.65 &  -----   & $-$0.90  & -----    & $-$1.35  & $<-$1.10 \\
\hbox{Gd II} & 4085.558 & 0.731 &    0.070 &  -----  & -----    &  -----   & $-$1.10  & -----    & $<-$1.35 & -----\\
\hbox{Gd II} & 4130.366 & 0.731 &    0.160 & $<-$1.75& -----    &  -----   & $<-$1.10 & -----    & $<-$1.40 & -----\\
\hbox{Gd II} & 4191.075 & 0.427 & $-$0.570 &  -----  & -----    &  -----   & -----    & -----    & $<-$1.10 & -----\\
\hbox{Tb II} & 3702.845 & 0.126 & $-$0.147 & $<-$2.40& -----    &  -----   & $<-$1.75 & -----    & $<-$1.90 & -----\\
\hbox{Tb II} & 3874.180 & 0.000 & $-$0.317 & $<-$2.60& -----    &  -----   & $<-$1.80 & -----    & $<-$2.00 & -----\\
\hbox{Tb II} & 4005.467 & 0.126 & $-$0.020 & $-$2.40 & -----    &  -----   & -----    & -----    &  -----   & -----\\
\hbox{Dy II} & 3694.810 & 0.103 & $-$0.110 & $<-$1.70& $-$0.60  &  -----   & $-$1.03  & -----    & $-$1.17  & $-$1.32 \\
\hbox{Dy II} & 3757.368 & 0.103 & $-$0.170 & $-$1.65 & -----    &  -----   & $-$0.96  & $<-$0.45 & $-$1.14  & -----\\
\hbox{Dy II} & 3788.436 & 0.103 & $-$0.570 & $-$1.72 & -----    &  -----   & $-$1.15  & $<-$0.10 & $-$1.11  & -----\\
\hbox{Dy II} & 3869.864 & 0.000 & $-$1.050 & $-$1.65 & -----    &  -----   & -----    & -----    &  -----   & -----\\
\hbox{Dy II} & 3944.681 & 0.000 &    0.100 & $-$1.60 & $-$0.62  & $-$0.98  & $-$1.05  & $-$0.70  & $-$1.16  & $-$1.18 \\
\hbox{Dy II} & 3996.689 & 0.590 & $-$0.200 & $-$1.74 & -----    &  -----   & $-$1.02  & -----    &  -----   & -----\\
\hbox{Dy II} & 4077.966 & 0.103 &    0.210 & $-$1.60 & $-$0.65  & $<-$0.90 & $-$0.99  & $-$0.75  & $-$1.13  & $-$1.16 \\
\hbox{Dy II} & 4103.306 & 0.103 & $-$0.390 & $-$1.48 & $<-$0.50 &  -----   & $-$0.98  & $<-$0.30 & $-$1.10  & $-$1.20 \\
\hbox{Dy II} & 4111.343 & 0.000 & $-$0.850 & $-$1.52 & $<-$0.45 &  -----   & $-$1.06  & -----    & $-$1.05  & $-$0.90 \\
\hbox{Ho II} & 3796.745 & 0.000 & $-$0.745 & $-$2.35 & $-$1.32  &  -----   & $-$1.60  & $<-$1.00 & $-$1.80  & $-$1.90 \\
\hbox{Ho II} & 3810.734 & 0.000 & $-$0.678 & $-$2.40 & -----    &  -----   & $-$1.65  & -----    &  -----   & -----\\
\hbox{Er II} & 3692.649 & 0.055 &    0.138 & $-$1.80 & $-$0.80  &  -----   & $-$1.25  & $<-$0.78 & $-$1.30  & $-$1.24 \\
\hbox{Er II} & 3786.836 & 0.000 & $-$0.644 & $-$1.60 & $<-$0.60 &  -----   & $-$1.15  & -----    & $-$1.18  & $-$1.10 \\
\hbox{Er II} & 3830.482 & 0.000 & $-$0.365 & $-$1.85 & $-$0.75  &  -----   & $-$1.26  & -----    & $-$1.30  & $<-$1.20 \\
\hbox{Er II} & 3896.234 & 0.055 & $-$0.241 &  -----  & -----    &  -----   & $-$1.23  & -----    &  -----   & -----\\
\hbox{Er II} & 3938.626 & 0.000 & $-$0.520 & $<-$1.82& $<-$0.70 &  -----   & $-$1.22  & -----    & $-$1.30  & -----\\
\hbox{Tm II} & 3701.363 & 0.000 & $-$0.540 & $<-$2.40& -----    &  -----   & -----    & -----    &  -----   & -----\\
\hbox{Tm II} & 3795.760 & 0.029 & $-$0.230 & $-$2.65 & $<-$1.46 &  -----   & $<-$1.78 & -----    &  -----   & $<-$1.95 \\
\hbox{Tm II} & 3848.020 & 0.000 & $-$0.130 & $-$2.78 & $<-$1.45 &  -----   & $<-$1.94 & -----    &  -----   & $<-$1.90 \\
\hbox{Yb II} & 3694.199 & 0.000 & $-$0.300 & $-$1.98 & $-$0.95  & $-$1.22  & $-$1.25  & $-$0.97  & $-$1.58  & $-$1.53 \\
\hbox{Os I}  & 4260.849 & 0.000 & $-$1.470 & $<-$1.00& -----    &  -----   & $<-$0.40 & -----    &  -----   & -----\\
\hbox{Os I}  & 4420.468 & 0.000 & $-$1.530 &  -----  & -----    &  -----   & $<-$0.45 & -----    &  -----   & -----\\
\hbox{Ir I}  & 3513.648 & 0.000 & $-$1.260 & $<-$0.74& $<$0.12  &  -----   & $<-$0.20 & -----    & $<-$0.45 & -----\\
\hbox{Ir I}  & 3800.120 & 0.000 & $-$1.450 &  -----  & $<-$0.10 &  -----   & -----    & -----    &  -----   & -----\\
\hbox{Th II} & 4019.129 & 0.000 & $-$0.228 & $-$2.45 & $<-$1.40 &  -----   & $<-$1.92 & -----    & $<-$2.12 & -----\\
\end{longtable}

\newpage

\end{appendix}

\end{document}